\documentclass[twocolumn]{aastex61}

\usepackage{amsmath}
\newcommand{\angstrom}{\text{\normalfont\AA}}

\usepackage{graphicx}
\usepackage{enumitem}

\shorttitle{star-formation and stellar masses in dark matter halos}
\shortauthors{Tacchella, Bose, Conroy, Eisenstein \& Johnson}

\begin{document}

\title{A redshift-independent efficiency model: \\ star formation and stellar masses in dark matter halos at $\MakeLowercase{z}\ga4$}

\correspondingauthor{Sandro Tacchella}
\email{sandro.tacchella@cfa.harvard.edu}

\author[0000-0002-8224-4505]{Sandro Tacchella}
\affil{Harvard-Smithsonian Center for Astrophysics, 60 Garden St, Cambridge, MA 02138, USA}
\author[0000-0002-0974-5266]{Sownak Bose}
\affil{Harvard-Smithsonian Center for Astrophysics, 60 Garden St, Cambridge, MA 02138, USA}
\author[0000-0002-1590-8551]{Charlie Conroy}
\affil{Harvard-Smithsonian Center for Astrophysics, 60 Garden St, Cambridge, MA 02138, USA}
\author{Daniel J. Eisenstein}
\affil{Harvard-Smithsonian Center for Astrophysics, 60 Garden St, Cambridge, MA 02138, USA}
\author[0000-0002-9280-7594]{Benjamin D. Johnson}
\affil{Harvard-Smithsonian Center for Astrophysics, 60 Garden St, Cambridge, MA 02138, USA}


\begin{abstract}

We explore the connection between the UV luminosity functions (LFs) of high-$z$ galaxies and the distribution of stellar masses and star-formation histories (SFHs) in their host dark matter halos. We provide a baseline for a redshift-independent star-formation efficiency model to which observations and models can be compared. Our model assigns a star-formation rate (SFR) to each dark matter halo based on the growth rate of the halo and a redshift-independent star-formation efficiency. The dark matter halo accretion rate is obtained from a high-resolution $N$-body simulation in order to capture the stochasticity in accretion histories and to obtain spatial information for the distribution of galaxies. The halo mass dependence of the star-formation efficiency is calibrated at $z=4$ by requiring a match to the observed UV LF at this redshift. The model then correctly predicts the observed UV LF at $z=5-10$. We present predictions for the UV luminosity and stellar mass functions, \textit{JWST} number counts, and SFHs. In particular, we find a stellar-to-halo mass relation at $z=4-10$ that scales with halo mass at $M_{\rm h}<10^{11}~M_{\odot}$ as $M_{\star}\propto M_{\rm h}^2$, with a normalization that is higher than the relation inferred at $z=0$. The average SFRs increase as a function of time to $z=4$, although there is significant scatter around the average: about 6\% of the $z=4$ galaxies show no significant mass growth. Using these SFHs, we present redshift-dependent UV-to-SFR conversion factors, mass return fractions, and mass-to-light ratios for different intial mass functions and metallicities, finding that current estimates of the cosmic SFR density at $z\sim10$ may be overestimated by $\sim0.1-0.2~\mathrm{dex}$.

\end{abstract}



\section{Introduction} \label{sec:intro}

In recent years, there has been a rapid improvement in our understanding of the stellar mass assembly of galaxies at the peak of cosmic star-formation rate density (SFRD) at $z=1-3$ \citep[e.g.,][]{ilbert13,muzzin13}. However, stellar masses of galaxies at $z\ga4$ have only been measured poorly, mainly because of insufficient data quality: the sensitivity and resolution of present observations are too low to probe the light of old stars at wavelengths longer than the age-sensitive Balmer break, which moves into the mid-IR at these redshifts \citep[e.g.,][]{stark16}. On the other hand, the UV luminosity function (LF) and its evolution with cosmic time are well constrained observationally out to redshifts of $z\simeq8$ \citep[e.g.,][]{finkelstein15,bouwens15}. In this paper, we investigate the information content of the UV LF as a proxy for the stellar mass assembly of galaxies by coupling the evolution of the UV LF to the dark matter halo population from a high-resolution, $N$-body simulation. In particular, this paper attempts to derive the star-formation efficiency of dark matter halos at $z=4$ using the UV LF. We then use this efficiency, assuming that it is redshift independent, to make predictions for the stellar mass growth of galaxies, which we expect to be measurable with the upcoming \textit{James Webb Space Telescope} (\textit{JWST}). 

The $\Lambda$CDM cosmological model \citep{blumenthal84} provides a theory for predicting early structure formation and the general properties of dark matter halos in which galaxies form. However, a fundamental theory for determining the stellar content associated with a given dark matter halo is still lacking. In most modern numerical/semianalytical models of galaxy formation, star formation in small halos at high redshifts is suppressed by means of `feedback' mechanisms that inhibit star formation by heating and/or removing gas from galaxies. Proposed mechanisms include photoionization by the UV background \citep{barkana99}, stellar feedback by supernovae \citep{dekel86}, radiation pressure from stars \citep{thompson05,murray10,hopkins10}, and suppression of the formation of molecular hydrogen in low-metallicity environments  \citep{krumholz12b}. However, present models can only achieve moderate resolution, leading to the use of simplified recipes (so-called `sub-grid' models) that are designed to capture the overall effects of complex feedback processes (see \citealt{somerville15} for a review). As a result, the effects of these processes then become tunable via free parameters, limiting the predictive power of such models. 

An alternative approach is to connect halos to observed galaxies in a statistical way \citep[see, e.g.,][for a comprehensive review]{wechsler18}, thereby bypassing the explicit modeling of baryonic physics. Such empirical models are useful for interpreting observations and making predictions for upcoming surveys. Furthermore, they provide useful scaling relations that can be used to constrain physical processes incorporated in numerical models. For example, feedback schemes in many hydrodynamical simulations are adjusted to reproduce the empirically determined stellar-to-halo mass relation. Empirical and numerical models are, therefore, complementary approaches in studying the physical processes driving galaxy evolution \citep[e.g.,][]{moster18}.

The link between galaxies and halos can also be used to infer the evolution of galaxy properties from the evolution of dark matter halos \citep[e.g.,][]{conroy07, white07, zheng07, firmani10, wang13b, tinker13, birrer14, sun16,cohn17,mitra17}. \citet{conroy09b} employ this approach to constrain the average stellar mass growth of galaxies in halos since $z=2$. \citet{moster13} and \citet[][see also \citealt{rodriguez-puebla17}]{behroozi13b} developed this method further using a semiempirical technique to infer observed galaxy properties from dark matter merger trees out to $z\sim8$. While this approach successfully describes the average evolution of galaxy properties, it does not self-consistently track the growth history of individual galaxies. Hence, in such an approach, galaxy properties depend only on halo mass, and not on the unique formation history of the halo. This could potentially be a limitation for understanding the properties of galaxies, since it is known, for example, that the spatial distribution of dark matter halos depends on their formation time \citep[e.g.,][]{gao05, paranjape15}. Recently, \citet[][see also \citealt{rodriguez-puebla16}]{moster18} addressed this issue by presenting an empirical model for galaxies, assigning a star-formation rate (SFR) to each dark matter halo based on its growth rate, following \citet{mutch13}. The resulting model is in good agreement with key observations, particularly for the clustering of star-forming and quenched galaxies \citep{moster18}, indicating a realistic assignment of galaxies to halos. 

The approach of linking the SFR to the growth rate of the dark matter halo has also been used to study the evolution of the UV LF at high redshifts. The connection between the high-redshift galaxies and their dark matter halos has been studied mainly through clustering since the first Lyman break galaxies were discovered \citep{giavalisco01, adelberger05, lee06}. In \citet{tacchella13}, we presented a simple model to predict the evolution of the UV LF. Specifically, we extended an earlier model by \citet{trenti10} by making the more realistic assumption that, at any epoch, all massive dark matter halos host a galaxy with a star-formation history (SFH) that is related to the time of halo assembly as predicted by Extended Press-Schechter (EPS) formalism \citep{press74, bond91}. The model is calibrated by constructing a galaxy luminosity versus halo mass relation at $z=4$ via abundance matching. After the initial calibration, the model correctly predicts the evolution of the LFs from $z=0$ to $z=8$. While the details of star-formation efficiency (defined as $\mathrm{SFR}/\dot{M_{\rm h}}$) and feedback are implicitly modeled within the calibration, our study highlights that the primary driver of cosmic SFRD across cosmic time is the buildup of dark matter halos, without needing to invoke a redshift-dependent efficiency in converting gas into stars. This model has been developed further and used in \citet{trenti15} to study the galaxies hosting gamma-ray bursts, in \citet{mason15} to constrain the UV LF before the epoch of reionization (EoR), and in \citet{ren18} to study the cosmic web around the brightest galaxies during the EoR. Similarly, \citet{trac15} use abundance matching to estimate the luminosity-mass relation and the luminosity-accretion rate relation, finding a universal luminosity-accretion-rate relation. Consistent with these studies, \citet{harikane18} analyze the clustering of $\sim600,000$ Lyman break galaxies at $z\sim4-6$, finding that a model where the star-formation efficiency does not evolve with redshift largely fits UV luminosity functions from $z=10$ to $z=0$. These findings are consistent with the key assumption of this paper (and of \citet{tacchella13}) of a redshift-independent efficiency of converting gas into stars. 

Our goal in this paper is to extend beyond predictions of the UV properties of the high-redshift galaxy population and focus on the assembly of stellar mass in these galaxies. We present a simple empirical model that can be used as a baseline for a comparison to observations and numerical models of galaxies at early cosmic epochs ($z\ga4$). The simplicity of the model stems from the assumption that the star-formation efficiency does not evolve with redshift and that each dark matter halo hosts only a single galaxy. These assumptions are fundamentally different from those in more elaborate empirical models \citep[e.g.][]{behroozi13b, rodriguez-puebla17, moster18} that are constructed to describe the evolution of the galaxy population over a wider range of redshifts (down to $z\sim0$). In such models, the efficiency of star-formation depends on halo mass, halo accretion rate, and redshift and includes treatments for satellite galaxies (i.e. multiple halo occupation). We use our redshift-independent efficiency model to make predictions for the SFHs of galaxies at $z\ga4$, which will be measurable by \textit{JWST}. We further quantify the evolution of the stellar mass function and the stellar-to-halo mass relation for the galaxy population at $z\ga4$. 

An important distinction between our present model and the one in \citet{tacchella13} is that the growth history of dark matter halos is now computed using merger trees obtained from an $N$-body simulation, rather than with the EPS formalism. This has the advantage that the merger history of halos has been fully and self-consistently evolved in a cosmological setting, while also allowing us to predict the spatial distribution of galaxies (e.g. clustering). In our model, we assume that the SFR of each dark matter halo is proportional to its accretion rate, multiplied by a redshift-independent efficiency. The assumption that the star-formation efficiency is redshift independent allows us to calibrate the efficiency at a single redshift ($z=4$ in this work) via the UV LF (Section~\ref{sec:SF_model}). After calibration, our model is able to reproduce the evolution of the observed UV LF in the range of $z=4-10$. We also make predictions for the stellar mass function and stellar-to-halo mass relation at $z=4-14$ (Section~\ref{sec:results}). Finally, we highlight that current analyses of observations at $z=10$ may overestimate the cosmic SFRD because the UV-to-SFR conversions typically used are not appropriate for increasing SFRs (Section~\ref{sec:discussion}). 

The framework in this analysis has been implemented in a flexible manner, allowing us to run the model on any arbitrary cosmological model. Throughout this paper, we assume the cosmological parameters derived from the 7-year {\it Wilkinson Microwave Anistropy Microwave Probe} \citep[WMAP-7,][]{komatsu11}: $\Omega_m = 0.272$, $\Omega_\Lambda = 0.728$, $h=0.704$, $n_s=0.967$, and $\sigma_8 = 0.81$. All observational data that we compare our model to are adjusted to match this cosmology. Furthermore, all magnitudes are quoted in the AB system. Finally, we denote the UV magnitude as the magnitude measured at rest frame $1500~\angstrom$.


\section{Model Description}
\label{sec:SF_model}

In this section we describe how our model relates the growth of galaxies to the growth of their host dark matter halos. We first extract dark mater halo merger trees from an $N$-body simulation (Section~\ref{subsec:DM_framework}). These are then populated with galaxies by assuming that the SFR of a halo is proportional to its accretion rate, normalized by a redshift-independent efficiency in converting gas into stars (Section~\ref{subsec:SF_model}). Spectral energy distributions (SEDs) of the galaxies are then calculated from a stellar population synthesis model (Section~\ref{subsec:SED}). Finally, we calibrate the star-formation efficiency by the observed UV LF at $z=4$ (Section~\ref{subsec:calibration}).

\subsection{Dark Matter Framework}
\label{subsec:DM_framework}

\begin{figure}
\includegraphics[width=\linewidth]{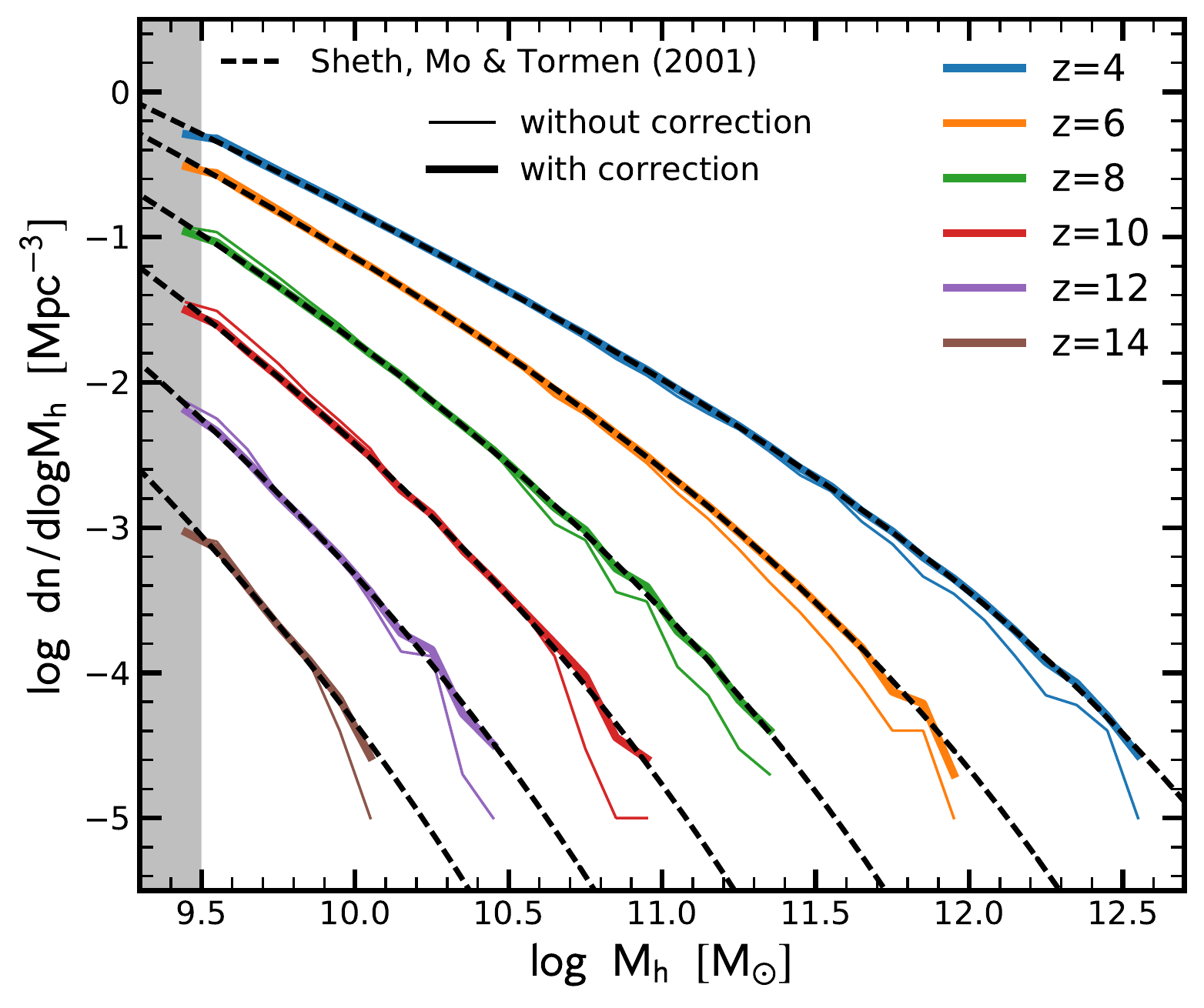}
\caption{Evolution of the dark matter halo mass function in the {\sc color} simulation. We plot the halo mass function with and without completeness corrections with thick and thin lines, respectively. The dashed black lines show the analytical halo mass function from \citet{sheth01}. The simulated halo mass functions are in good agreement with the analytical estimates. At the high-mass end, where the simulations are volume limited, the completeness corrections bring the results of the simulation in agreement with the analytic mass functions. The gray region starting at $10^{9.5}~M_{\odot}$ marks the nominal convergence limit of the simulation (300 dark matter particles).}
\label{fig:HMF}
\end{figure}

In semianalytical models, halo merger trees are typically constructed in one of two ways: ($i$) sampling analytic halo mass functions and generating realizations of merger histories using a Monte Carlo approach following the EPS model \citep[e.g.][]{kauffmann93,somerville00,cole08,yung18}, or ($ii$) directly extracting the merger history of halos from an $N$-body simulation \citep[e.g.][]{helly03,guo13,lacey16}. 

Each of these approaches has its advantages and drawbacks. A great benefit of the Monte Carlo approach is that it is possible, in principle, to achieve arbitrary resolution with relatively little computational cost. Furthermore, it is also possible to fully sample the halo mass function at any given redshift, as Monte Carlo trees do not suffer from finite volume. $N$-body trees, on the other hand, are limited by both the resolution and volume of the simulations they are extracted from, but have the advantage that the growth history of halos has been fully and self-consistently evolved in a cosmological setting, taking into account tidal forces, dynamical friction, tidal stripping, etc. Furthermore, trees extracted from $N$-body simulations also allow us to predict the spatial distribution of galaxies, enabling us to study their clustering. A quantitative comparison of $N$-body vs. Monte Carlo merger trees is presented in Appendix~\ref{app_sec:tree_compare}.

To this end, we make use of merger trees obtained from the {\it Copernicus complexio Low Resolution} ({\sc color}) simulations \citep{hellwing16,sawala16}. {\sc color} follows the evolution of $1620^3$ dark matter particles within a periodic box with volume $(70.4 \, {\rm Mpc}/h)^3$, resulting in an effective dark matter particle mass of $6.196\times10^6~M_{\odot}/h$. The gravitational softening corresponds to $1~\mathrm{kpc}/h$. The mass and temporal resolution of these simulations are particularly suited for tracing the progenitors of $z=4$ galaxies to higher redshifts. {\sc color} assumes cosmological parameters derived from WMAP-7. Zoom-in simulations based on {\sc color} have appeared as part of the {\sc coco} \citep{hellwing16,bose16} and {\sc apostle} \citep{fattahi16,sawala16} suite of simulations. 

The {\sc color} volume was evolved from $z=127$ to $z=0$ using {\sc p-gadget-3} \citep{springel08}, an updated version of the publicly available {\sc gadget-2} code \citep{springel01b,springel05c}. Halos are first constructed using the friends-of-friends algorithm \citep{davis85}, while the gravitationally bound substructures associated with them are identified with the {\sc subfind} algorithm \citep{springel01c}. {\sc subfind} entities are then connected between snapshots by identifying sets of objects that share some fraction of their most bound particles between outputs, using the formalism outlined in \cite{jiang14}. New branches in the merger tree are created whenever a new {\sc subfind} object is identified in the halo catalog. A total of 160 simulation snapshots (regularly spaced by $\Delta \ln(a) = 0.0239$) were used to construct the merger trees from {\sc color}. In what follows, we will be primarily concerned with merger trees of central (i.e. independent) halos rather than their substructures. A pathology with merger tree construction is the misidentification of subhalos as centrals when they orbit a region of low-density contrast (e.g., near the center of a larger halo); we are careful to eliminate these instances from our halo catalogs. The fraction of contaminants is small, typically composing 8-10\% of the halo population at each redshift. In total, we count 294569, 51332, and 2287 halos at $z=4, 8$, and 12, respectively, above our mass resolution limit of $10^{9.5}~M_{\odot}$

Since the {\sc color} simulation probes only a finite volume, our halo catalogs are devoid of some of the rarest and most massive halos. This is apparent in the dark matter halo mass function, shown in Figure~\ref{fig:HMF}. At $z=4$, the number density of halos with $M_{\rm h}=10^{12.5}~M_{\odot}$ is underpredicted by about 0.4 dex with respect to the analytical prediction of \citet{sheth01}. Similarly, at higher redshifts, we also miss the highest-mass objects that have number densities of $\approx10^{-5}~\mathrm{Mpc}^{-3}$. In order to correct for this, we apply a completeness correction: we estimate the magnitude of the correction from the difference between the analytical halo mass function of \citet{sheth01} and our measured halo mass function at each snapshot. As visible in Figure~\ref{fig:HMF}, the correction is up to 0.4 dex. Further details, in particular the effect of the completeness correction on the UV LF, are outlined in Appendix~\ref{app_sec:comp_corr}.

We note, however, that halos with $M_{\rm h}>10^{13}~M_{\odot}$, which are absent in our simulations, are rare (number densities of $\la10^{-6}~\mathrm{Mpc}^{-3}$). As a result, we do not expect that the absence of these halos will have a significant impact on any of our results. In the context of the \textit{JWST} mission, our model probes the bulk of the galaxy population at $z=4-12$ since \textit{JWST} has a rather small field of view, probing a rather limited volume of $\sim10^4-10^5~\mathrm{Mpc}^{3}$ at $z\sim10$. We postpone a more detailed discussion of the impact of cosmic variance on our results to future work. 

\subsection{Star Formation in Dark Matter Halos}
\label{subsec:SF_model}

Most numerical and (semi)analytical schemes model star-formation so as to reproduce empirical scaling relations such as the Kennicutt-Schmidt law, i.e., by correlating the SFR in the halo to the total (volume or surface) gas mass (density): $\mathrm{SFR}\propto M_{\rm gas, dense}$, where $M_{\rm gas, dense}$ is the mass contained in cold, dense gas. At low redshifts ($z<2$), the gas reservoirs of galaxies are large and the typical star-formation timescales are long (often quantified in terms of the gas depletion time, $t_{\rm dep}=M_{\rm gas}/\mathrm{SFR}>10^9~\mathrm{yr}$; e.g. \citealt{genzel15,tacconi18}). \citet{semenov17} attribute this longer timescale for star-formation to multiple cycles of gas into and out of a dense, star-forming phase. In particular, SFR is limited by the fact that only a small fraction of star-forming gas is converted into stars before star-forming regions get dispersed by feedback and dynamical processes. The SFR is determined by the production and disruption of dense, star-forming gas, which depends on gravity, gas compression, and the regulating nature of feedback \citep{thompson05,ostriker11,faucher-giguere13}. The gas accretion rate in low-$z$ galaxies, therefore, has little to do with the SFR itself, which is set by the availability of cold, dense gas \cite[see, e.g., the discussion on the self-regulating nature of star-formation and stellar feedback in][]{schaye15}. The star-formation mode is \textit{reservoir limited}.

At higher redshifts, the newly accreted gas is expected to transition to the dense gas phase quicker because of the overall higher density at these early times. As a further consequence, feedback also becomes less effective in disrupting the star-forming, dense gas: \citet{faucher-giguere18} argues that, at high redshifts, the characteristic galactic dynamical timescales become too short for supernova feedback to effectively respond to gravitational collapse in galactic disks (see also \citealt{lagos13} for a semianalytical model for the evolution of the mass loading in supernova feedback in the presence of higher gas densities and molecular gas fractions). This is consistent with current observations of high molecular gas fractions in $z\sim1-3$ galaxies \citep{tacconi13,tacconi18,genzel15} and predicts a high molecular-to-neutral fraction in these early galaxies. Along similar lines, \citet{krumholz12a} argue that in high-$z$ galaxies, star-forming regions are unable to decouple from the ambient interstellar medium (ISM), the result being that the free-fall times are then set by the large-scale properties of the ISM. We therefore expect that the formation of dense, star-forming gas in high-$z$ galaxies is more closely related to the gas accretion rate onto the galaxies themselves, i.e. star formation is \textit{accretion limited}. Thus, we adopt a star-formation law where $\mathrm{SFR}\propto \dot{M}_{\rm gas}$. 

Specifically, we link the SFR of a galaxy to the growth rate of its halo. We assume that ($i$) the rate of infalling baryonic mass is proportional to the mass accretion rate of the halo, rescaled by the universal baryon fraction $f_{\rm b}=\Omega_{\rm b}/\Omega_{m}=0.167$; and ($ii$) the star-formation efficiency, $\varepsilon(M_{\rm h})$, depends solely on halo mass. The SFR of a galaxy at redshift $z$ in a dark matter halo of mass $M_{\rm h}$ can then be written as the product of the baryonic growth rate times the star-formation efficiency in the following way:

\begin{equation}
\begin{split}
\mathrm{SFR}(M_{\rm h},z) & = \varepsilon(M_{\rm h}) \times \dot{M}_{\rm gas} \\
 & = \varepsilon(M_{\rm h}) \times f_b \times \widetilde{\frac{\mathrm{d}M_{\rm h}}{\mathrm{d}t}}(M_{\rm h},z),
\end{split}
\label{eq:SFR}
\end{equation}

\noindent where $\widetilde{\frac{\mathrm{d}M_{\rm h}}{\mathrm{d}t}}$ is the delayed and smoothed accretion of dark matter onto its halo. The dark matter accretion is delayed by the dynamical time of the halo in order to take into account dynamical as well as dissipative effects within the halo. The dynamical time at the virial radius of the dark matter halo can be written as

\begin{equation}
\tau_{\rm DM, dyn} = \left( \frac{3\pi}{32G\rho_{\rm 200, crit}} \right)^{1/2} \sim 0.1\tau_{\rm H},
\label{eq:dyn}
\end{equation}

\noindent where $\tau_{\rm H}$ is the Hubble time. Additionally, we smooth the dark matter accretion by $0.05\tau_{\rm H}$ in order to mitigate against sharp features that arise from the discrete snapshot sampling. This smoothing scale is smaller than (at $z>4$) or comparable to (at $z\sim4$) the lifetime of UV-bright stars, ensuring that the impact on the inferred UV luminosity is negligible. 

In this framework, the only function that then needs to be constrained is the star-formation efficiency $\varepsilon(M_{\rm h})$. This star-formation efficiency describes how efficiently gas is converted into stars, encapsulating complicated baryonic processes such as gas cooling, star formation, and various feedback processes into a single parameter. We assume -- for simplicity -- that it depends only on halo mass and is redshift independent, allowing us to calibrate $\varepsilon(M_{\rm h})$ at a single redshift. We choose to calibrate $\varepsilon(M_{\rm h})$ at $z=4$ by requiring that the model reproduces the observed UV LF at this redshift. Further details of the calibration are given in Section~\ref{subsec:calibration}.

\subsection{Predicting the SED}
\label{subsec:SED}

With the formalism introduced in the previous section (Equation~\ref{eq:SFR}), we are able to construct the SFHs for individual galaxies. We then predict the SED for each galaxy by using the Flexible Stellar Population Synthesis code \citep[\texttt{FSPS}\footnote{\url{https://github.com/cconroy20/fsps}};][]{conroy09a, foreman_mackey14}. \texttt{FSPS} has been extensively calibrated against a suite of observational data \citep[for details see][]{conroy10}. Throughout this work, we adopt the \texttt{MILES} stellar library and the \texttt{MIST} isochrones. We do not consider first-generation (Population III) stars since their contributions to the UV LF and the cosmic SFRD are minor at $z\la14$ compared to the second generation of stars \citep[e.g.][]{pallottini14, jaacks18}.

Other important parameters include the initial mass function (IMF), the stellar metallicity ($Z$), and dust attenuation. Our fiducial choice for the IMF is the one by \citet{salpeter55}, but we also investigate the implications of adopting a \citet{chabrier03} IMF. In the subsequent subsections, we discuss the treatment of metallicity and dust attenuation in the derivation of the SEDs.

\subsubsection{Metallicity}
\label{subsubsec:metallicity}

The metallicity and the abundance pattern of galaxies at $z\ga4$ are still unconstrained observationally. Throughout this work, we assume a solar abundance pattern and $Z_{\odot}=0.0142$ \citep{asplund09}. We adopt two different metallicity implementations in our model. In the first case, we assume a constant metallicity of $0.02~Z_{\odot}$ for galaxies at all redshifts. In the second, we adopt simple mass conservation to calculate the metallicity from star formation, outflows, and gas accretion. Although these two assumptions produce rather different metallicity distributions in the galaxy population, the impact on our main results is negligible. For simplicity, our default model is the one that assumes a constant metallicity. 

\citet{troncoso14} and \citet{onodera16} measured the oxygen abundance for star-forming galaxies at $z=3-4$ (see also \citealt{maiolino08}), finding $12+\log(\mathrm{O}/\mathrm{H})\approx7.5-8.5$ with a weak trend in mass (higher-mass galaxies have a higher metallicity). These observations have large uncertainties, as the metallicity calibration itself is uncertain. Using this, the \citet{troncoso14} and \citet{onodera16} observations correspond to $Z\sim0.1-0.3~Z_{\odot}$ for galaxies with $M_{\star}\approx10^{10}~M_{\odot}$. We expect the stellar metallicity to be comparable to or slightly lower than this gas-phase estimate. Hence, our fiducial model, which we denote as `Z-const', assumes a constant metallicity of $0.02~Z_{\odot}$ for galaxies at all redshifts. As shown in Section~\ref{subsec:UVLF} and Figure~\ref{fig:UVLF}, varying the metallicity from 0.001 to 0.1 has no measurable effect (change in UV magnitude of $\la0.1$ mag at $1500~\angstrom$ on average) on the UV LF and therefore our calibration. Changing it to solar metallicity ($Z=1.0~Z_{\odot}$) has an impact, making the magnitudes fainter by 0.5 mag on average. 

Although the derived UV magnitudes do not depend significantly on metallicity at $Z\la0.1~Z_{\odot}$, we explore a mass- and redshift-dependent evolution of the metallicity content of our model galaxies. Specifically, we track the evolution of the metallicity of individual galaxies by solving the equation of mass conservation \citep{bouche10, dave12, lilly13_bathtube, dekel14_bathtube}. Details are described in Appendix~\ref{app_sec:Z_implementation}. We call this version of the model `Z-evo'. We find a rather large diversity of metallicity (scatter of 0.5 dex at a given stellar mass) in the galaxy population, with massive galaxies reaching $Z\approx0.1-0.3~Z_{\odot}$, roughly consistent with observed values (Figure~\ref{fig:MZ_relation}). Since metallicities in this model are overall higher than in the default model (`Z-const'), the star-formation efficiency of this model needs to be higher in order to reproduce the same UV LF (see Section~\ref{subsec:calibration}). However, the increase in the efficiency is only $\sim0.1$ dex, leading to only a small increase ($\sim0.1$ dex) in the stellar content of the galaxies (see Section~\ref{subsec:MF} for implied change in the mass functions).

\subsubsection{Dust Attenuation}
\label{subsubsec:dust}

Since the star-formation efficiency in our model, $\varepsilon(M_{\rm h})$, is calibrated by requiring a match with the observed UV LF, it is important that dust attenuation is properly taken into account. Hence, there is -- in addition to the assumed stellar population properties (e.g. metallicity and IMF) -- also a degeneracy between the assumed dust attenuation prescription and $\varepsilon(M_{\rm h})$. The dust attenuation mainly affects UV-bright galaxies, i.e. halos with high accretion rates and SFRs. We discuss the derivation of $\varepsilon(M_{\rm h})$ and its degeneracies further in the next subsection.

We account for dust attenuation in our model by following the procedure adopted in observations by \citet{smit12}, as we did in \citet{tacchella13}. We assume that the rest-frame UV part of the spectrum can be described as a power law, $f_{\lambda} \propto \lambda^{\beta}$, where $\beta$ is the UV continuum slope. We estimate $\beta$ from the observed $\mathrm{M}_{\rm UV}-\beta$ relation: $\langle\beta\rangle=a\cdot(\mathrm{M}_{\rm UV}-19.5)+b$. The values for $a$ and $b$ are taken from \citet[][Table 3]{bouwens14}. There have also been other comprehensive investigations of the UV slopes at high redshifts \citep[e.g.,][]{finkelstein12,rogers14}. They are overall in rough agreement with each other after taking into account different measurement methods and biases as shown by \citet{bouwens14}. Specifically, at $z=5$, \citet{rogers14} find a slope of $-0.12\pm0.02$, while \citet{bouwens14} find a consistent value with $-0.14\pm0.02$. Important to note is that the scatter is in this relationship is large. We assume a Gaussian distribution for $\beta$ at each $\mathrm{M}_{\rm UV}$ value with a dispersion of $\sigma_{\beta}=0.34$ at all redshifts, which is roughly the scatter of the $\beta-\mathrm{M}_{\rm UV}$ relationship for bright $z\sim4$ and $z\sim5$ galaxies \citep{bouwens09,bouwens12a,bouwens14,castellano12,rogers14}.

In order to estimate the UV attenuation, we assume that the UV attenuation depends on the UV continuum slope $\beta$ (i.e., IRX-$\beta$ relation). The shape and normalization of the IRX-$\beta$ relation at high redshifts are still debated \citep{capak15,mclure18,bouwens16,reddy18,koprowski18,narayanan18_IRXb}. For typical little or modestly obscured systems, the results are broadly consistent with the \citet{meurer99} IRX-$\beta$ relationship. Therefore, following \citet{tacchella13}, we adopt in our fiducial model the \citet{meurer99} IRX-$\beta$ relation ($\mathrm{IRX}=2.07\times[10^{0.4(4.43+1.99\beta)}-1]$), which leads to -- after incorporating the scatter -- an average attenuation of $\langle \mathrm{A}_{\rm UV} \rangle=4.43+0.79\ln(10)\sigma_{\beta}^2+1.99\langle\beta\rangle$. This fiducial dust attenuation prescription gives $\mathrm{A}_{\rm UV}=[1.51, 1.05, 0.60]$ mag for a galaxy with an observed magnitude of $M_{\rm UV}=[-22.0, -20.0, -18.0]$ mag at $z\sim4$. For the same magnitudes at $z\sim8$, we obtain $\mathrm{A}_{\rm UV}=[1.40, 0.60, 0.0]$ mag. In order to explore how different IRX-$\beta$ relations influence our results, we also adopt the IRX-$\beta$ ($\mathrm{IRX}=1.79\times[10^{0.4(1.07+2.79\beta)}-1]$, see \citealt{gordon03}) from the Small Magellanic Cloud (SMC), which is favored by the $z=5$ observations of \citet{capak15}. We call this model version `SMC'.

A concern with this dust attenuation prescription is that it is computed solely using UV light. It seems that IR-luminous, highly obscured galaxies deviate from the \citet{meurer99} IRX-$\beta$ relation, such as high-$z$ submillimeter sources with SFR of up to $1000~M_{\odot}~\mathrm{yr}^{-1}$ \citep{casey14}. In our model, we are unable to reproduce these high SFRs. We find a maximum of $\mathrm{SFR}\approx200~M_{\odot}~\mathrm{yr}^{-1}$. One possibility is that these high SFRs are extremely rare and the volume of the {\sc color} simulation is not enough to contain the corresponding halos. Some authors have also suggested that a top-heavy IMF in starbursts may be needed to produce highly star-forming submillimeter galaxies in a $\Lambda$CDM cosmology \citep[e.g.][]{baugh05,zhang18}. Another possibility is that we have not implemented any enhancement of the star-formation efficiency that could be induced by mergers, which could result in these high SFRs \citep[e.g.][]{sargent15}.

Summarizing this section, our dust attenuation prescription closely follows observations. We are therefore confident that it describes the bulk of the galaxy population at $z\ga4$ well. Our fiducial model (`Z-const') assumes the \citet{meurer99} IRX-$\beta$ relation, while we also explore the SMC IRX-$\beta$ relation in the model version `SMC'. In a future publication we will use our model to predict the infrared galaxy LF and the infrared background in order to test our model and our assumed dust prescription. However, our model clearly has limited predictive power concerning star-forming galaxies with extreme SFRs, in which the star formation is so heavily enshrouded in dust that no UV photons can leave the star-forming region.

\subsection{Calibration of the Model}
\label{subsec:calibration}

\begin{figure*}
\centering
\begin{tabular}{lr}
\includegraphics[width=0.48\textwidth]{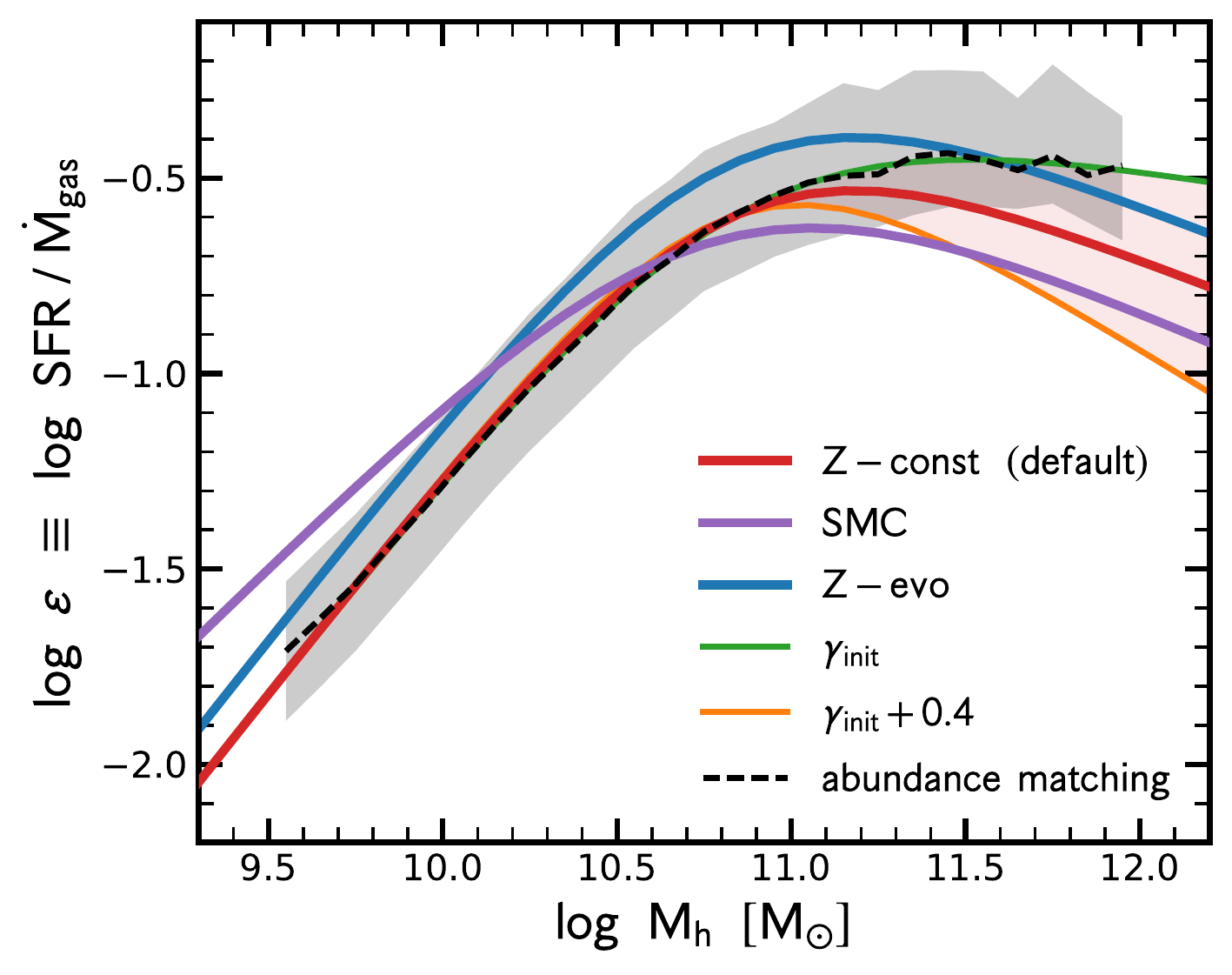} & 
\includegraphics[width=0.48\textwidth]{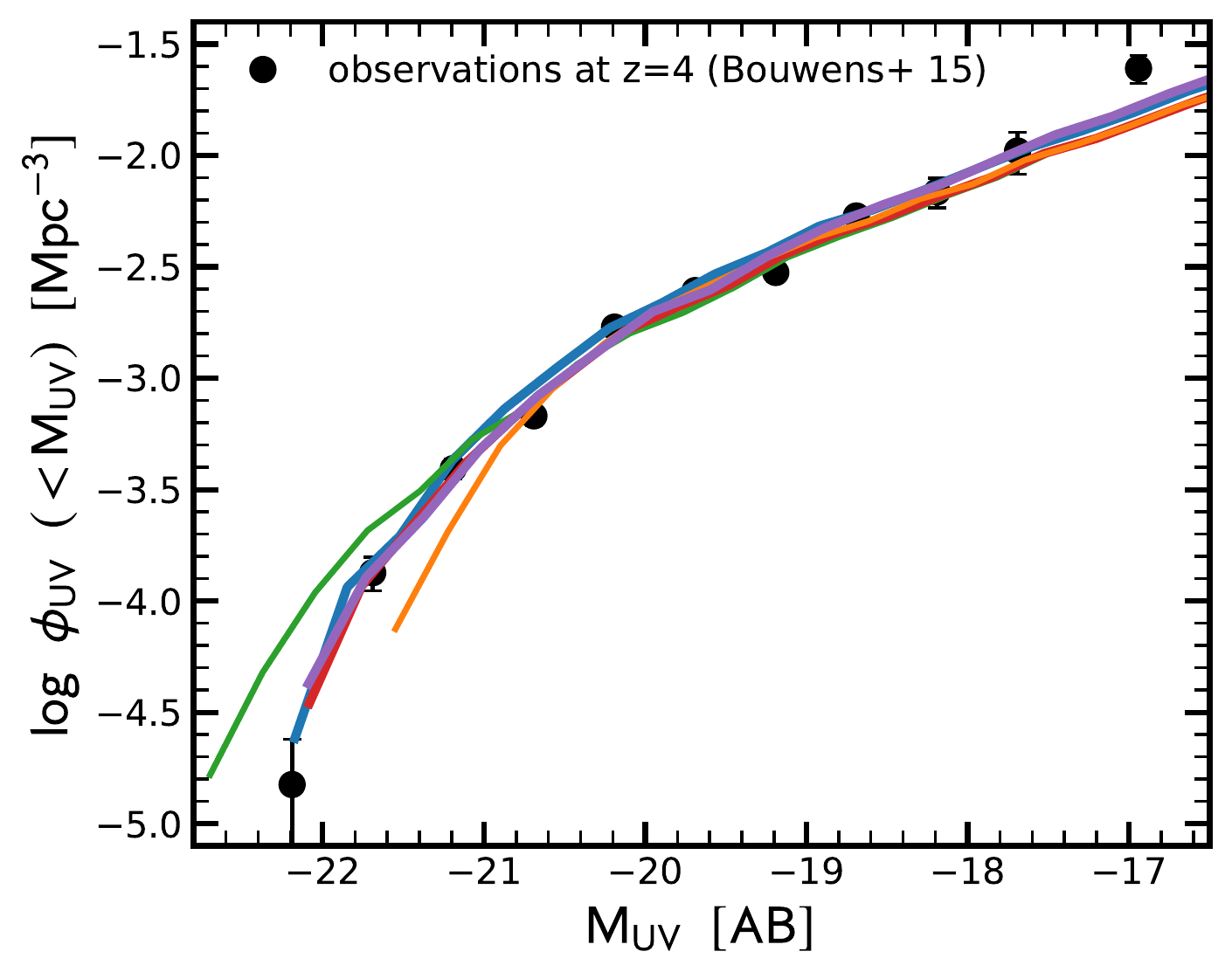}
\end{tabular}
\caption{Calibration of the model. \textit{Left panel:} star-formation efficiency as a function of halo mass, which is calibrated to reproduce the observed UV LF at $z=4$. The dashed line with the gray shaded region shows the expected relation from abundance matching and its scatter. The solid green line shows the best fit to Equation~\ref{eq:efficiency} assuming $(\varepsilon_{\rm 0, init},~M_{\rm c, init},~\beta_{\rm init},~\gamma_{\rm init}) = (0.26, 7.10\times10^{10}, 1.09, 0.16)$. In order to match the observed UV LF, we need to modify the turnover in this relation at high mass. Our best fit for the default `Z-const' model is plotted as the solid red line ($\gamma=\gamma_{\rm init}+0.2$), while the orange line indicates an extreme case of a low star-formation efficiency at high halo masses ($\gamma=\gamma_{\rm init}+0.4$). The solid blue and purple lines show the best-fit efficiency for the `Z-evo' model and the `SMC' model, respectively. \textit{Right panel:} the UV LF obtained from our model compared to the observed UF LF at $z=4$ \citep{bouwens15}. The different lines correspond to the different efficiencies plotted on the left; the solid red line, the solid blue line, and the purple line indicate the `Z-const' (fiducial) model, the `Z-evo' model, and the `SMC' model, respectively.}
\label{fig:calibration}
\end{figure*}

The only function that needs to be calibrated against observations is the star-formation efficiency, $\varepsilon(M_{\rm h})$. We achieve this by adjusting $\varepsilon(M_{\rm h})$ until the observed UV LF $\phi_{\rm UV}$ at $z=4$ is reproduced.

If the halo dark matter accretion rate is proportional to dark matter halo mass (as in EPS), one can simply perform abundance matching and equate the UV-brightest galaxy (galaxy with highest SFR) to the most massive halo. However, as shown in $N$-body simulations, dark matter halos show a range of accretion rates at a given halo mass, and the most massive halo is not necessarily the one with the highest accretion rate (see Appendix~\ref{app_sec:tree_compare}). We therefore need to go beyond halo abundance matching. 

Since each realization of our model takes several hours to run (the bottleneck being the derivation of the SEDs with \texttt{FSPS}), we cannot simply fit an arbitrary $\varepsilon(M_{\rm h})$. We use abundance matching to compute an initial guess for $\varepsilon_{\rm init}(M_{\rm h})$, which we call $\varepsilon_{\rm init}(M_{\rm h})$, following the same approach as in \citet{tacchella13}. In particular, we first run a first iteration of the model assuming $\varepsilon(M_{\rm h}) = 1.0$. We then derive a UV luminosity versus halo mass relation at redshift 4, $L_{\rm UV}(M_{\rm h}, z=4)$, by equating the number of galaxies with a UV luminosity greater than $L_{\rm UV}$ (after dust correction) to the number of halos with mass greater than $M_{\rm h}$. From the $L_{\rm UV}(M_{\rm h}, z=4)$ relation, we can solve for $\varepsilon_{\rm init}(M_{\rm h})$. At each $M_{\rm h}$ we find a distribution of $\varepsilon_{\rm init}(M_{\rm h})$,  reflecting the diversity of UV luminosities that stems from a diversity of halo accretion histories. The left panel of Figure~\ref{fig:calibration} shows $\varepsilon_{\rm init}(M_{\rm h})$ and its 16th/84th percentile. The shape of $\varepsilon_{\rm init}(M_{\rm h})$ can be well parametrized with a double power law following \citet{moster10} \citep[see also][]{moster18,behroozi13b}:

\begin{equation}
\varepsilon(M_{\rm h})=2\varepsilon_0\left[\left(\frac{M_{\rm h}}{M_{\rm c}}\right)^{-\beta} + \left(\frac{M_{\rm h}}{M_{\rm c}}\right)^{\gamma}\right]^{-1},
\label{eq:efficiency}
\end{equation}

\noindent where $\varepsilon_0$ is a normalization constant, $M_{\rm c}$ is the characteristic mass where the efficiency is equal to $\varepsilon_0$, and $\beta$ and $\gamma$ are slopes that determine the decrease at low and high masses, respectively. We find the following values: $(\varepsilon_{\rm 0, init},~M_{\rm c, init},~\beta_{\rm init},~\gamma_{\rm init}) = (0.26, 7.10\times10^{10}, 1.09, 0.16)$. 

As shown in the right panel of Figure~\ref{fig:calibration}, this $\varepsilon_{\rm init}(M_{\rm h})$ overproduces the abundance of UV-bright galaxies. In order to match the UV LF at $z=4$, we modify the high-mass slope $\gamma$. We find that the best value is $\gamma_{\rm final}=\gamma_{\rm init}+0.2=0.35$. Our best-fit $\varepsilon_{\rm init}(M_{\rm h})$ for our default model with constant metallicity (`Z-const') is then ($\varepsilon_{\rm 0, final},~M_{\rm c, final},~\beta_{\rm final},~\gamma_{\rm final})=(0.26, 7.10\times10^{10}, 1.09, 0.36)$. In order to calibrate our model `Z-evo', we use the best-fit values from above and only change the normalization. Doing so, we find ($\varepsilon_{\rm 0, final},~M_{\rm c, final},~\beta_{\rm final},$ $\gamma_{\rm final}) = (0.37, 7.10\times10^{10}, 1.09, 0.36)$. The `Z-evo' model has a higher star-formation efficiency because the metallicities in this model are overall higher than in the default model (`Z-const'), which in turn leads to a lower UV flux per unit star formation. As we will show below (Section~\ref{subsec:MF}), the increase in normalization by 0.1 dex has only a minor impact on the stellar masses of the galaxies, shifting the galaxy stellar mass function by $\sim0.1$ dex. For the `SMC' model, where we assume the SMC IRX-$\beta$ relation instead of the \citet{meurer99} relation, we obtain ($\varepsilon_{\rm 0, final},~M_{\rm c, final},~\beta_{\rm final},$ $\gamma_{\rm final}) = (0.22, 6.30\times10^{10}, 0.89, 0.40)$. With the SMC model, we find a higher star-formation efficiency at low halo masses and a lower efficiency at high halo masses than with our fiducial model. This is because at low $\beta$ values ($<-1.9$), i.e. at faint magnitudes and hence low halo masses, SMC predicts a higher IRX ratio (more dust attenuation) than \citet{meurer99}, while the opposite holds at higher $\beta$ values ($>-1.9$).

As shown in Figure~\ref{fig:calibration}, the star-formation efficiency in our model depends strongly on halo mass, peaking at a characteristic halo mass of $10^{11}-10^{12}~M_{\odot}$. A similar relation is found at lower redshifts \citep[e.g.][]{conroy09b,moster10,behroozi15,moster18}. At these times, feedback from massive stars (supernovae and stellar winds) and feedback from active galactic nuclei are thought to suppress the star formation at the low- and high-mass end, respectively \citep[e.g.][]{mo10, silk12}. At high redshifts, these feedback mechanisms are likely to be less efficient (see Section~\ref{subsec:SF_model}), but their effects on galaxies at $z>4$ are unknown. Ab initio models, such as numerical simulations, we will help to understand the baryonic processes that shape the star-formation efficiency.

\subsection{Contribution of Mergers to the Stellar Mass Growth}
\label{subsec:mergers}

In our model, stellar mass growth is solely a result of star formation. However, another contribution to the gain in stellar mass comes from mergers, which can be calculated by convolving the halo merger rate and the stellar-to-halo mass relation. \citet{behroozi15} use the the halo merger rate and stellar-to-halo mass relation from \citet{behroozi13b} to gauge the importance of mergers. Their main finding is that mergers contribute about $12-18\%$ of the total stellar mass in most galaxies at $4<z<8$, rising to $30\%$ for galaxies in halos with $M_{\rm h}\ga10^{12}~M_{\odot}$. This small fraction arises from the fact that the stellar-to-halo mass ratio declines toward lower halo mass, so most of the incoming mass will come from major mergers. Since the star-formation timescale of these high-$z$ galaxies is short ($\la 150$ Myr) compared to the typical timescale of major mergers ($\ga300$ Myr; see \citealt{fakhouri08,fakhouri10,behroozi13b}), mergers provide a minor contribution to the stellar mass growth. We therefore neglect this mode of mass growth in what follows.

\section{Results}
\label{sec:results}

In this section, we present predictions from our redshift-independent efficiency model. These predictions are intended to serve as a baseline against which observations and numerical models can be compared. 

\subsection{UV Luminosity Functions}
\label{subsec:UVLF}

\begin{figure*}
\includegraphics[width=\textwidth]{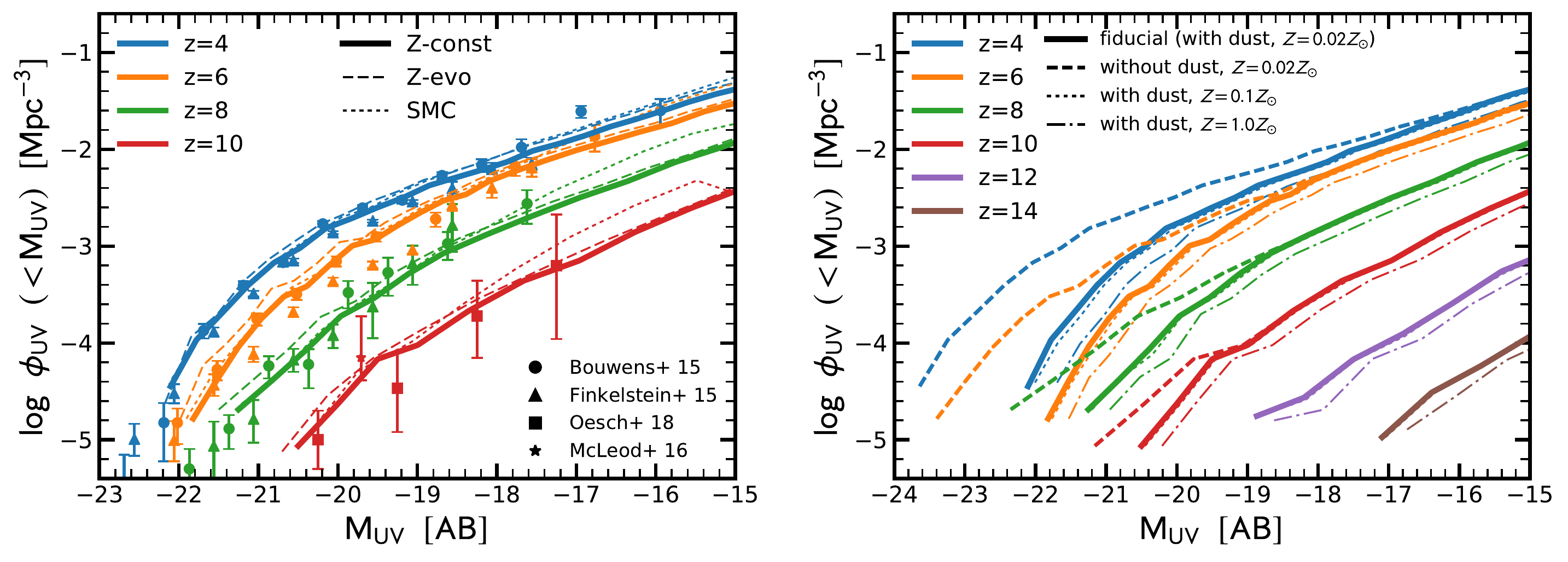}
\caption{Evolution of the UV LF. \textit{Left panel:} The thick solid, thin dashed, and thin dotted lines show the UV LFs at $z=4, 6, 8,$ and 10 in our `Z-const' (fiducial), `Z-evo', and `SMC' models, respectively. The observational data are taken from \citet{bouwens15,finkelstein15,oesch18} and \citet{mcleod16}. At $z=6-10$, there is good agreement between the predictions of our model and the observed data. \textit{Right panel:} the $z=4-14$ evolution of the UV LF for our fiducial model with dust attenuation (constant metallicity of $Z=0.02~Z_{\odot}$), our fiducial model without dust attenuation, and our model with two different metallicities ($Z=0.1~Z_{\odot}$ and $Z=1.0~Z_{\odot}$) is shown as solid, dashed, dotted, and dot-dashed lines, respectively. Reducing the metallicity to $Z\la0.1~Z_{\odot}$ has negligible impact on the UV LF, while the correction for dust attenuation clearly dominates the bright end at $z<10$.}
\label{fig:UVLF}
\end{figure*}

\begin{deluxetable}{cccc}
\tablecaption{Best-fit Schechter function (Equation~\ref{eq:LF}) parameters of our galaxy UV LF.\label{tab:LF}}
\tablecolumns{4}
\tablewidth{0pt}
\tablehead{
\colhead{redshift} &
\colhead{$\phi_{\rm UV}^{*}$} &
\colhead{$M_{\rm UV}^{*}$} & 
\colhead{$\alpha_{\rm UV}$}  \\
\colhead{} &
\colhead{[$10^{-5}~\mathrm{Mpc}^{-3}$]} &
\colhead{$[\mathrm{mag}]$} & 
\colhead{}
}
\startdata
$ z=4.0 $ & $ 137.3 \pm 24.2 $ & $ -21.09 \pm 0.19 $ & $ -1.63 \pm 0.02 $ \\
$ z=5.0 $ & $ 93.0 \pm 20.9 $ & $ -21.07 \pm 0.23 $ & $ -1.72 \pm 0.02 $ \\
$ z=6.0 $ & $ 119.9 \pm 25.7 $ & $ -20.41 \pm 0.19 $ & $ -1.72 \pm 0.03 $ \\
$ z=7.0 $ & $ 108.0 \pm 20.5 $ & $ -19.99 \pm 0.16 $ & $ -1.68 \pm 0.04 $ \\
$ z=8.0 $ & $ 56.7 \pm 10.5 $ & $ -19.9 \pm 0.15 $ & $ -1.69 \pm 0.04 $ \\
$ z=9.0 $ & $ 23.5 \pm 8.3 $ & $ -19.92 \pm 0.28 $ & $ -1.81 \pm 0.05 $ \\
$ z=10.0 $ & $ 12.4 \pm 1.2 $ & $ -19.6 (fixed) $ & $ -1.84 \pm 0.06 $ \\
$ z=12.0 $ & $ 1.2 \pm 0.5 $ & $ -19.6 (fixed) $ & $ -1.99 \pm 0.24 $ \\
\enddata
\tablecomments{The errors indicate the one standard deviation errors on the parameters. Since the UV LFs at $z\ga10$ do not show a clear turnover from the power law to the exponential cutoff, we fix $M_{\rm UV}^{*}$ to $-19.6$ mag.}
\end{deluxetable}

After calibration, our model is able to reproduce the UV LF at $z=4$ by construction. In a first step, we use our model to predict the UV LF at higher redshifts and compare it with observations. The measurements of the UV LF have improved over the past 10 years, mainly thanks to the installation of the Wide-Field Camera 3 on board the \textit{Hubble Space Telescope} \citep[\textit{HST};][]{bunker04, beckwith06, bouwens07, bouwens11, finkelstein10, mclure10, schenker13a}. The most recent $z=4-8$ measurements of the UV LF from \textit{HST} imaging \citep{bouwens15,finkelstein15} are based on $4000-6000$ $z\simeq4$ galaxies, $2000-3000$ $z\simeq5$ galaxies, $700-900$ $z\simeq6$ galaxies, $300-500$ $z\simeq7$ galaxies, and $100-200$ $z\simeq8$ galaxies. The \citet{bouwens15} study is currently the largest effort, including galaxies from all five CANDELS fields \citep{grogin11, koekemoer11}, the BoRG/HIPPIES fields \citep{trenti11,yan11}, and the HUDF/XDF and its associated parallels \citep{illingworth13}. Typically, these photometric samples are expected to have contamination levels of $\sim10\%$ at $z\simeq6-8$ owing to uncertainties in the estimation of the photometric redshifts. The current observational frontier lies at $z\simeq9-11$, where the Hubble Frontier Field dataset \citep{lotz17} recently provided an additional search volume and larger samples of galaxies at $z\sim10$ \citep{zitrin14, oesch15, oesch18, infante15,ishigaki15,ishigaki18}. 

Figure~\ref{fig:UVLF} shows the predicted evolution of the UV LF at $z=4-14$. We fit the UV LF of our model with a single Schechter function:

\begin{equation}
\phi(M_{\rm UV})dM_{\rm UV}=\phi^*\frac{\ln(10)}{2.5}10^{-0.4(M_{\rm UV}-M^*_{\rm UV})}e^{-10^{-0.4(M_{\rm UV}-M^*_{\rm UV})}}.
\label{eq:LF}
\end{equation}

\noindent
The best-fit parameters are listed in Table~\ref{tab:LF}. We find that the knee of the LF decreases from $M_{\rm UV}=-21$ at $z=4$ to $-19.6$ at $z=10$. Furthermore, we find a steepening of the faint-end slope from $ -1.61 \pm 0.02 $ at $z=4$ to  $ -2.1 \pm 0.05 $ at $z=12$. 

In the left panel, we compare our predicted UV LF to observations at $z=4-10$ \citep{bouwens15,finkelstein15,oesch18,mcleod16}. All three versions of the model, our fiducial one with a constant metallicity of $Z=0.02~Z_{\odot}$ (Z-const), the one with an evolving metallicity (Z-evo), and the one where we replace the fiducial \citet{meurer99} IRX-$\beta$ relation with the SMC relation (SMC), are remarkably consistent with the observed data, despite our simplifying assumption of a redshift-independent star-formation efficiency. This agreement is perhaps to be expected, given the success of previous implementations of this class of models that evolve according to accretion-limited growth \citep{trenti10,trenti15,tacchella13,mason15}. We find that our model naturally predicts the rather fast evolution of the UV LF from $z\sim6-8$ to $z\sim10$, although observationally this trend is still uncertain because of the small sample size and survey volumes of current observations at $z\ga8$. Since dust attenuation and metallicity only play a minor role for the evolution of the UV LF at $z>8$ (see below), accommodating a weaker redshift evolution of the UV LF is difficult with our model. In particular, in order to increase the UV LF at $z>8$, one would have to either change the star-formation efficiency in our model (making low-mass halos more efficient at higher redshifts) or change the gas accretion prescription (i.e., decoupling the gas accretion rate from the dark matter accretion rate). The SMC model increases the star-formation efficiency for low-mass halos, resulting in a small increase of the faint-end slope of the UV LF at $z\sim10$, but not in an overall weaker evolution of the UV LF. In order to achieve a weaker $z$-evolution of the UV LF, one would need to increase the efficiency with redshift. Furthermore, studying the impact of changing the gas accretion prescription is of interest, but beyond the scope of this work.

We evaluate the impact of dust attenuation and metallicity in the right panel of Figure~\ref{fig:UVLF}. The dashed lines show the dust-free UV LF. By construction (see Section~\ref{subsec:SED}), the dust attenuation mainly affects the bright end of the UV LF. At $z\la8$, the bright end is completely dominated by the dust attenuation prescription. At $z>8$, the UV LF appears to be less affected by the dust attenuation, although it must be kept in mind that, indirectly, the impact of dust attenuation depends on the specific dust prescription used in the calibration at $z=4$. This can be seen when comparing the Z-const with the SMC model: the SMC model has an underlying efficiency that is higher for low-mass halos. At the calibration redshift $z=4$, the SMC model agrees well with the fiducial model. Toward higher redshifts, in particular $z>8$, the SMC model predicts more galaxies with $M_{\rm UV}>-18$ (steeper faint-end slope) than the fiducial model.

With regard to the metallicity, we find that there is little difference between the two versions of our model (Z-const versus Z-evo; left panel). Additionally, in the right panel of Figure~\ref{fig:UVLF}, we show how the UV LF of the Z-const model is affected by a change in metallicity, while keeping the initial calibration fixed. At low, subsolar metallicity ($Z\la0.1~Z_{\odot}$), a change in metallicity has a negligible effect on the UV LF: the dotted lines ($Z=0.1~Z_{\odot}$) are barely differentiable from our fiducial model ($Z=0.02~Z_{\odot}$). Changing the metallicity to solar metallicity ($Z=1.0~Z_{\odot}$, dot-dashed lines) has a measurable effect: at $z\la8$, the impact is smaller than the one from dust, but it is comparable to or larger than the dust correction at $z>8$. 

Finally, we look into the faint-end turnover of the UV LF. Observationally, the best constraints are obtained from the Hubble Frontier Fields \citep{lotz17}, which can probe a possible turnover in the LF at faint luminosities thanks to lensing. In particular, the UV LF can be reliably measured down $\sim3$ mag fainter than the HUDF. Below that, systematics due to lensing increase significantly \citep[e.g.][]{bouwens17}. Current observations show that the UV LF continues down to $M_{\rm UV}\approx-15$, with a possible turnover below that magnitude \citep[e.g.][]{atek18}. Our model is consistent with these observations (Figure~\ref{fig:UVLF}). We find a turnover at a magnitude of $M_{\rm UV}\approx-14$ (corresponding to an SFR of $\sim0.02~M_{\odot}/\mathrm{yr}$), evolving only weakly with redshift (increasing to brighter luminosities with increasing redshift). The turnover in our model arises because of the resolution limit of the {\sc color} $N$-body simulation, which is at $M_{\rm h}=10^{9.5}~M_{\odot}$ (see Section~\ref{subsec:DM_framework}). There is only a weak evolution of the turnover with redshift because the $M_{\rm h}-M_{\star}$ and $M_{\rm UV}-M_{\star}$ relations evolve only weakly with redshift. Changing the efficiency for low-mass halos leads to a change in the turnover. Specifically, increasing the efficiency leads to a turnover at brighter magnitudes, as visible when going from the Z-const to the SMC model. Since the turnover is a resolution effect, we do not interpret this further.

\subsection{Cosmic SFRD}

\begin{figure}
\includegraphics[width=\linewidth]{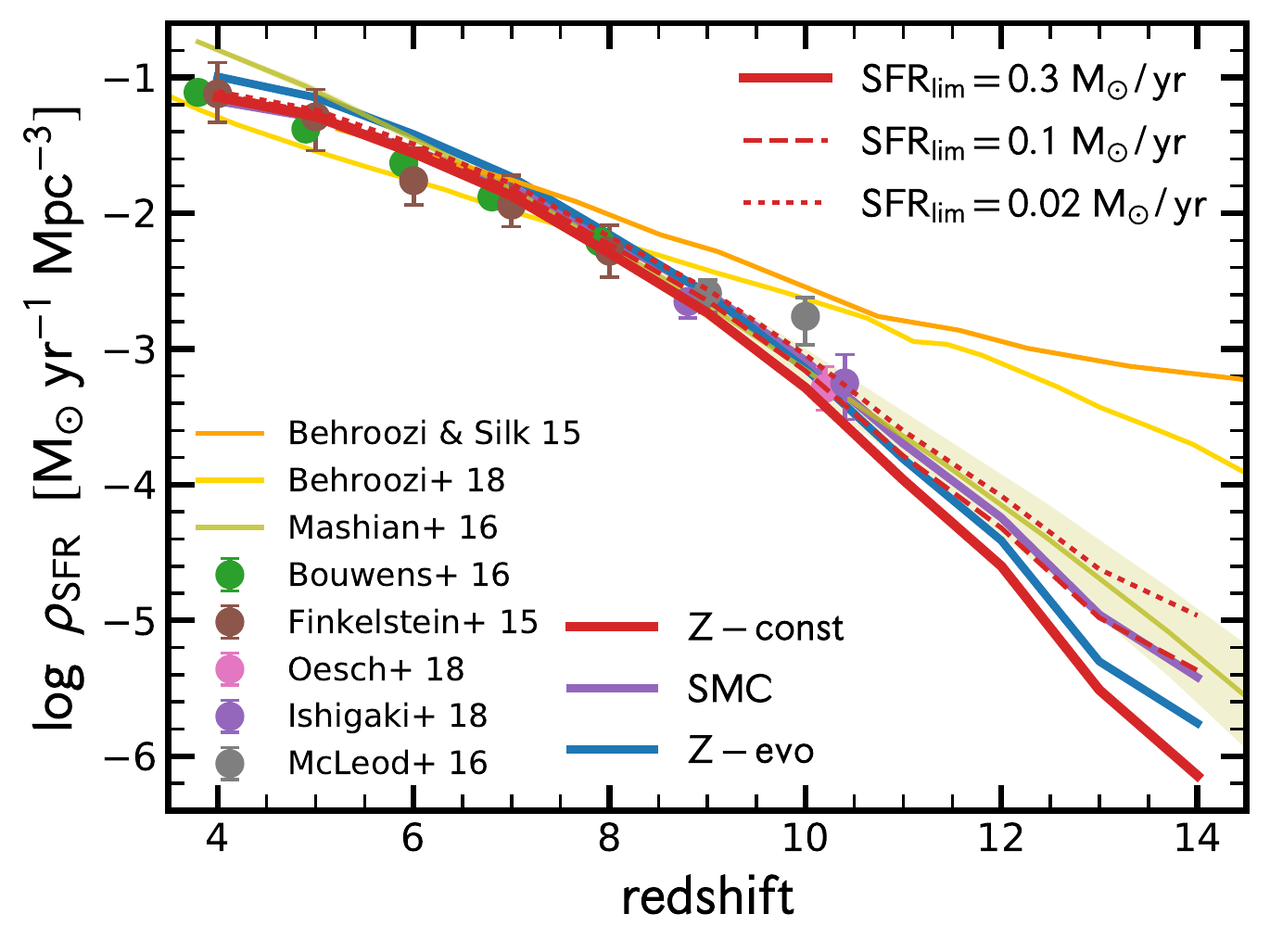}
\caption{Cosmic star-formation rate density $\rho_{\rm SFR}$. The solid, dashed, and dotted red lines show the SFR density predicted by the fiducial model, with the SFR limited to $0.3~M_{\odot}/\mathrm{yr}$, $0.1~M_{\odot}/\mathrm{yr}$ and $0.02~M_{\odot}/\mathrm{yr}$, respectively. The solid purple line shows the SMC model, which produces a higher $\rho_{\rm SFR}$ than our fiducial model owing to the higher star-formation efficiency for low-mass halos. The solid blue line shows the Z-evo model. The observational measurements are compiled from \citet{finkelstein15,bouwens15,oesch18,ishigaki18} and \citet{mcleod16}, and are obtained by integrating the UV LF down to a magnitude of $-17.0$ AB, corresponding to an $\mathrm{SFR}=0.3~M_{\odot}/\mathrm{yr}$. Overall, we find good agreement with the observations. Our model prefers a rather sharp decline of the cosmic SFRD at $z>8$, roughly consistent with the model by \citet{mashian16}, but inconsistent with the \citet{behroozi15} and \citet{behroozi18} models. We find that the cosmic SFRD declines from $z\simeq4$ to $z\simeq14$ by about 5 orders of magnitude.}
\label{fig:cSFRD}
\end{figure}

In observations, the measurement of the UV LF at $z>4$ is used to constrain the history of star formation and stellar mass growth in the first billion years \citep[see][for a detailed review]{madau14}. Calculations of the SFRD from the UV LF require a correction for dust and a conversion between the UV luminosity and the SFR. As discussed in Section~\ref{subsubsec:dust}, the UV LF is typically corrected for dust using the \citet{meurer99} relation. The dust-corrected UV luminosity is then transformed to an SFR via a conversion factor ($L_{\rm UV}/\mathrm{SFR}$) that is sensitive to stellar population properties such as age, SFH, and metallicity. Most SFR measurements use a value in the range of $L_{\rm UV}/\mathrm{SFR}= 0.6-1.0\times10^{28}~\mathrm{erg}~\mathrm{s}^{-1}~/~(M_{\odot}~\mathrm{yr}^{-1})$, which assumes a Salpeter IMF in the mass range $0.1-100~M_{\odot}$, continuous star formation for more than 100 Myr and a metallicity in the range of $\log~Z/Z_{\odot}=[+0.2,-1.0]$ \citep{madau14}. For an increasing SFH, a stellar population will produce more UV luminosity for a given average SFR, leading to a larger conversion factor by up to $0.2$ dex. This is discussed further in Section~\ref{subsec:lum_SFR_conversion}.

The SFRD estimated from the observed UV LFs is shown in Figure~\ref{fig:cSFRD}. When computing the SFRD, a lower-luminosity limit of $M_{\rm UV}=-17.0$ (which corresponds to the $\mathrm{SFR}=0.3~M_{\odot}~\mathrm{yr}^{-1}$) when integrating the UV LFs and a conversion factor of $L_{\rm UV}/\mathrm{SFR}=0.87\times10^{28}~\mathrm{erg}~\mathrm{s}^{-1}~/~(M_{\odot}~\mathrm{yr}^{-1})$ has been assumed \citep{oesch18}. The different observational datasets are in good agreement with each other in the range $z\sim4-8$: a power-law fit to the $z\simeq4-8$ values results in an SFRD evolution $\propto(1+z)^{4.2}$ \citep{bouwens15,finkelstein15}. At $z>8$, the evolution of the SFRD is more controversial. \citet[][consistent with \citealt{oesch14,bouwens16,ishigaki18}]{oesch18} find a steep decline of the SFRD from $z\simeq8$ to $z\simeq10$: the extrapolation of the power law from $z\simeq4-8$ to $z\simeq10$ lies a factor of $5-6$ above the Oesch et al. measurement, which is in contrast with other measurements \citep[e.g.][]{mcleod16}.

Our model predicts a rather steep decline with redshift. This strong decline is a direct consequence of our assumption that the star-formation efficiency is redshift independent: the characteristic mass of the dark matter halo mass function decreases with increasing redshift, far below the peak of the star-formation efficiency of $M_{\rm h}\sim10^{11}-10^{12}~M_{\odot}$. We estimate that the cosmic SFRD declines from $7.2\times10^{-2}~M_{\odot}~\mathrm{yr}^{-1}~\mathrm{Mpc}^{-3}$ at $z\sim4$ to $7.2\times10^{-7}~M_{\odot}~\mathrm{yr}^{-1}~\mathrm{Mpc}^{-3}$ at $z\sim14$, i.e. by 5 orders of magnitude. This is in excellent agreement with observational estimates at $z\la8$, and it is also consistent with the uncertainties in the observations at $z\sim10$ by \citet{oesch18} and \citet{ishigaki18}. As discussed in Section~\ref{subsec:lum_SFR_conversion}, the UV-to-SFR conversion factors adopted in these observational works at $z\sim10$ neglect the fact that most SFHs are increasing, which overestimates the cosmic SFRD at $z\sim10$ by $0.1-0.2$ dex (depending on metallicity). Furthermore, our SMC model declines slightly more weakly than our fiducial model, because the star-formation efficiency is higher for low-mass halos.

The model by \citet{mashian16} uses abundance matching at $z=4-8$ to construct the $\mathrm{SFR}-M_{\rm h}$ relation in each redshift bin. The authors find that the resulting $\mathrm{SFR}-M_{\rm h}$ scaling law remains roughly constant over this redshift range. This is, to first order, consistent with our main assumption here, namely, that the star-formation efficiency remains more or less constant with cosmic time. Mashian at al. then use the average $\mathrm{SFR}-M_{\rm h}$ relation to make predictions at $z>8$. Similar to our model, they find a rather steep decline of the SFRD. On the other hand, \citet{behroozi15} assume a constant relation between the sSFR of the galaxy and the specific mass accretion rate of the halo, i.e. $\mathrm{sSFR}\propto\dot{M}_{\rm h}/M_{\rm h}$, while we assume $\mathrm{SFR}\propto\dot{M}_{\rm h}$. The \citet{behroozi18} model solves for the SFR as a function of halo mass and redshift for satellites and central galaxies using the observed stellar mass functions, SFRs, quenched fractions, UV LF, UV magnitude versus stellar mass, autocorrelation functions, and quenching dependence on environment at $z=0-10$. Both these models predict a relatively slow decline in the SFRD at $z>8$, which is in contrast to our prediction. Furthermore, as shown in Section~\ref{subsec:MsMh}, while our model implies a rather constant stellar-to-halo mass relation with cosmic time, these models lead to a strong increase of the median galaxy mass at a fixed halo mass at $z>4$.

\subsection{M$_{\rm UV}$-$M_{\star}$ relations}
\label{subsec:MsMuv}

\begin{figure*}
\includegraphics[width=\textwidth]{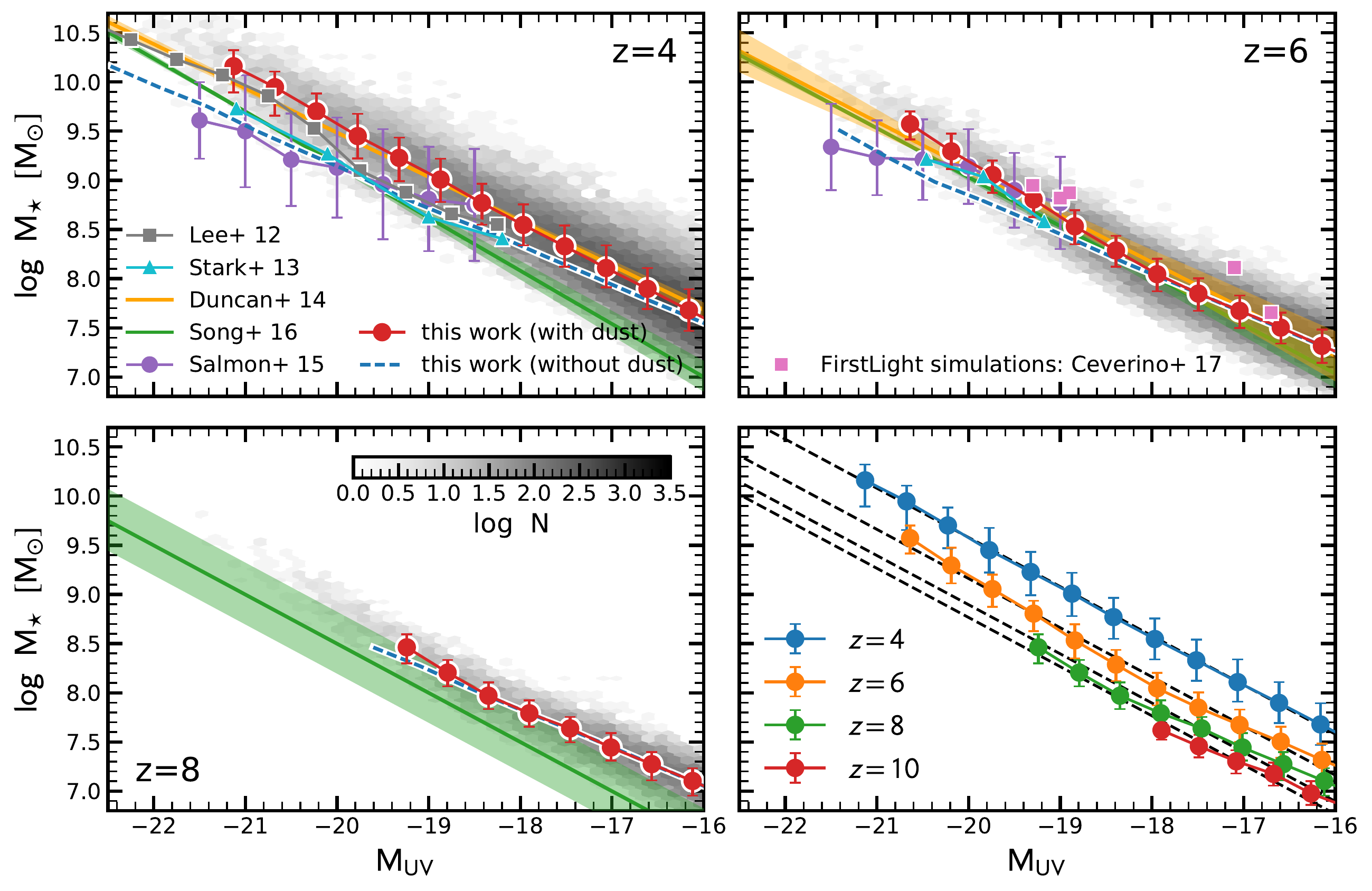}
\caption{Rest-frame UV magnitude (M$_{\rm UV}$) versus stellar mass ($M_{\star}$). \textit{Top left panel:} M$_{\rm UV}$-$M_{\star}$ relation at $z=4$ in our model compared to observations. The gray 2D histogram shows the distribution of galaxies from our model (fiducial model with dust). The red points and blue line show the median relations from our model with dust and without dust. The observational data is taken from \citet{lee12,stark13,duncan14,song16} and \citet{salmon15}.  \textit{Top right and bottom left panels:} same as the top left panel, but for $z=6$ and $z=8$, respectively. The filled pink points are from the `FirstLight' simulations of \citet{ceverino17}. \textit{Bottom right panel:} the M$_{\rm UV}$-$M_{\star}$ relation for $z=4-10$. At fixed stellar mass, galaxies have a brighter UV magnitude at earlier cosmic time. The  dashed black lines indicate the best fit, Equation~\ref{eq:MsMuv}. Overall, we find good agreement with the observations by \citet{duncan14}, but our predictions lie systematically above the relation found by \citet{song16}, particularly toward faint magnitudes at $z=4$ and $z=8$.}
\label{fig:MuvMstar}
\end{figure*}

The stellar masses, and in particular the stellar mass functions, offer a more direct probe of the mass assembly of galaxies than the UV LF. First, we study the relation between the observed (not corrected for dust) UV magnitude and stellar mass ($M_{\rm UV}-M_{\star}$ relation). While the observed $M_{\rm UV}$ can be obtained in a straightforward manner, the stellar mass, $M_{\star}$, remains poorly constrained because of insufficient data quality (both the sensitivity and resolution of current observations are too low to probe the light of old stars at wavelengths longer than the age-sensitive Balmer break, which moves into the mid-IR at $z\ga4$). 

There are two common ways to define the stellar mass of a galaxy: ($i$) the integral of the past SFR, which we denote in this work with $M_{\rm int}$; and ($ii$) the mass in stars and remnants, i.e. after subtracting stellar mass loss due to winds and supernovae, which we label $M_{\star}$. These two mass definitions are related to each other via the mass return fraction $R$:

\begin{equation}
M_{\star}=M_{\rm int}(1-R).
\label{eq:Mdef} 
\end{equation}

\noindent
We use \texttt{FSPS} to calculate the mass in stars and remnants. Specifically, \texttt{FSPS} follows the mass loss of stars due to winds and follows the prescription of \citet{renzini93} in assigning remnant masses to dead stars. Most observational and theoretical works\footnote{There are some scientific questions, in particular related to the evolution of quiescent galaxies, where it is more useful to adopt $M_{\rm int}$ as the stellar mass, since with this definition the stellar mass of a quiescent galaxy remains constant with time; see \citet{carollo13a,fagioli16,tacchella17_S1}.} use the definition of $M_{\star}$ in Equation~\ref{eq:Mdef} when quoting the stellar mass of a galaxy. We therefore also adopt this as the fiducial definition of the stellar mass of a galaxy. An extended discussion of the return fraction is presented in Section~\ref{subsec:return}.

Figure~\ref{fig:MuvMstar} plots the $M_{\rm UV}-M_{\star}$ relation predicted by our model at $z=4-10$. We find that the slope of this relation stays constant, while the normalization decreases with increasing redshift. Fitting the $M_{\rm UV}-M_{\star}$ relation with
\begin{equation}
\log M_{\star}=a(M_{\rm UV}+19.5)+\log M_{\star, 0},
\label{eq:MsMuv}
\end{equation}

\noindent
we find a redshift-independent slope of $a=-0.5$ and a zero-point that decreases with redshift following $\log M_{\star, 0}/M_{\odot}=-2.4\log(1+z)+11.0$. 

Figure~\ref{fig:MuvMstar} also compares our predicted $M_{\rm UV}-M_{\star}$ relation to observed and simulated relations from the literature. We plot the observational measurements of \citet{lee12}, \citet{stark13}, \citet{duncan14}, \citet{salmon15}, and \citet{song16}. There are discrepancies of $0.3-0.7$ dex between different studies in the measured median mass at a given UV magnitude even at $z\sim4$, in particular toward the fainter magnitude bins. This may reflect a number of systematic uncertainties associated with sample selection and stellar mass estimation. \citet{lee12} assume solar metallicity, an age grid from 5 Myr to the age of the universe at a given redshift, and SFHs that are either declining $\tau$-models or constant. \citet{stark13} utilize a moderately restricted grid, varying only the age, dust reddening, and normalization factor, fixing the SFH as either constant or rising with time following the $t^{1.7}$ power law and including the contribution from emission lines. \citet{salmon15} derive their stellar masses assuming a constant SFR and a fixed metallicity of $Z=0.2~Z_{\sun}$. \citet{duncan14} and \citet{song16} combine CANDELS \textit{HST} data and \textit{Spitzer}/IRAC data, though they use different IRAC data, different fields, and a different treatment of photometric redshifts. Both fit the SEDs to the \citet{bruzual03} stellar population synthesis models, assuming exponentially increasing and decreasing SFHs, metallicity ranging from 0.02 to $1~Z_{\odot}$, and a dust attenuation that is allowed to vary in the range $0\leq\mathrm{A}_{\rm V}\leq2$. These two studies assume a slightly different grid in the characteristic timescale $\tau$ of the SFH and in the allowed model ages. In particular, \citet{song16} allow the age to vary between 1 Myr and the age of the universe, while \citet{duncan14} limit the range to between 5 Myr and the age of the universe. \citet{duncan14}, consistent with \citet{lee12}, in general find higher stellar mass at fixed $M_{\rm UV}$ than \citet{song16} (after accounting for different assumptions for the IMF), which, on the other hand, is consistent with \citet{stark13}. As we will see in the next section, this difference propagates into their measurements of the stellar mass function. 

Our model predictions are in good agreement with the measurements by \citet{lee12} and \citet{duncan14}. On the other hand, we predict $0.2-0.7$ dex higher stellar masses for a given UV luminosity than \citet{song16} and \citet{stark13} at $z=4$. We find better agreement with \citet{song16} and \citet{stark13} at $z=6$, but again a rather large difference with \citet{song16} at $z=8$. The observations of \citet{salmon15} prefer a flatter slope in the $M_{\rm UV}-M_{\star}$ relation than our model. This is, in fact, more consistent with the dust-corrected relation of our model, even though Salmon et al. have not corrected their UV luminosity for dust. 

We compare our predicted $M_{\rm UV}-M_{\star}$ relation at $z=6$ to the one of the FirstLight project \citep{ceverino17,ceverino18_UV}, which is a cosmological zoom-in simulation of 290 halos with $M_{\rm h}=10^9-10^{11}~M_{\odot}$. These simulations include prescriptions for the cooling of gas through atomic and molecular hydrogen cooling (the latter, in particular, may be important in the buildup of the first generations of galaxies at high redshift). Subgrid models for photoionization of neutral gas, the input of thermal energy and radiative feedback from supernovae and stellar winds are included to regulate star-formation. As shown in the top right panel of Figure~\ref{fig:MuvMstar}, our model is in excellent agreement with their prediction. This agreement is perhaps not so surprising, as the UV luminosities in these simulations are obtained by assuming a proportional relationship to the SFR of the galaxy, which is heuristically similar to the procedure we have followed.

In summary, current observations show a large scatter in the $M_{\rm UV}-M_{\star}$ relation. The main reason for this is the uncertainty in the derivation of the stellar mass of galaxies. However, we are unable to clearly pinpoint the source of the discrepancy. A possible explanation of the discrepancy is the prior assumption that goes into the SED modeling. As highlighted by \citet{behroozi18}, $M_{\rm UV}-M_{\star}$ relation from \citet{song16} requires very low mass-to-light ratios, typical of recent burst or steeply rising SFHs. \citet{song16} indeed include SFHs with ages as short as 1 Myr, while \citet{duncan14} and \citet{lee12} assume a minimal age of 5 Myr, which could explain the difference in the derived stellar masses of these two studies. A possible other source for the disagreement could be the different treatment of the correction for emission lines. Future \textit{JWST} observations will provide a much tighter constraint on the $M_{\rm UV}-M_{\star}$ relation by measuring stellar masses more accurately.

\subsection{Stellar Mass Functions}
\label{subsec:MF}

\begin{figure*}
\includegraphics[width=\textwidth]{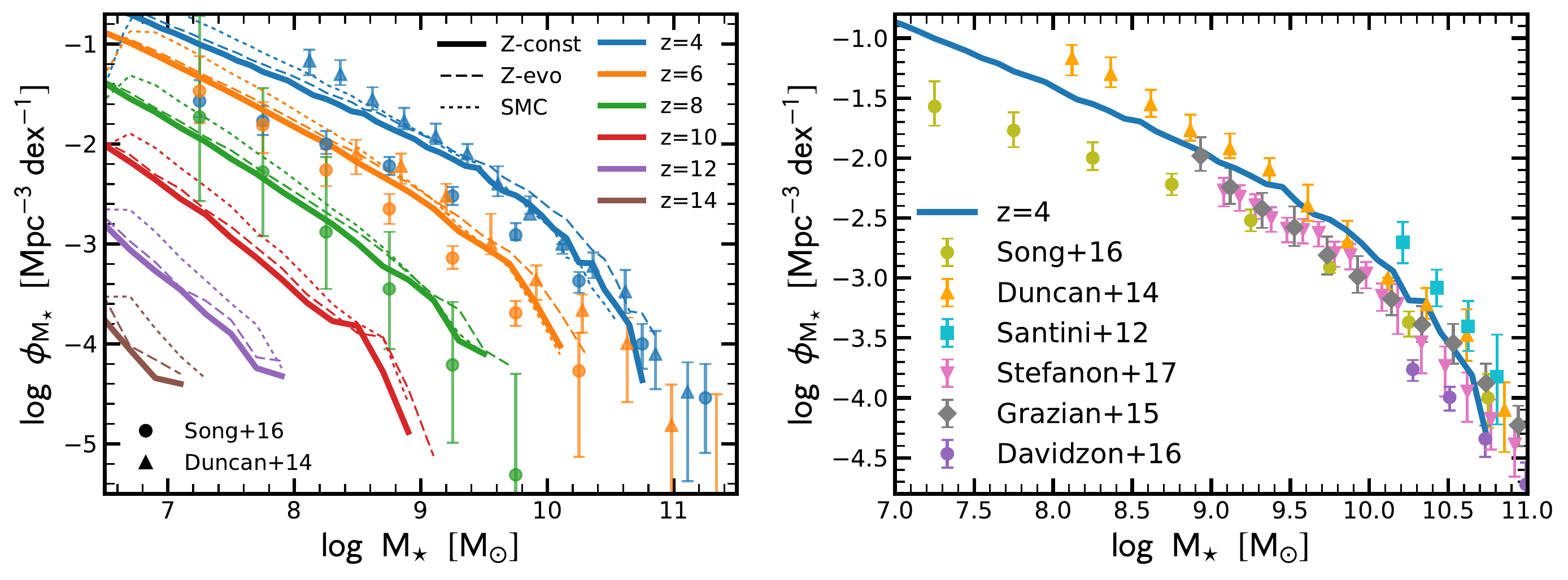}
\caption{Evolution of the stellar mass function. \textit{Left panel:} The thick solid, thin dashed, and thin dotted lines show the stellar mass functions at $z=4-14$ for our `Z-const' (fiducial), `Z-evo', and `SMC' model, respectively. The observational data are taken from \citet{song16} and \citet{duncan14}. \textit{Right panel:} Zoom-in on the comparison of the $z=4$ stellar mass function from our model and from observations \citep{duncan14,grazian15,song16,santini12,stefanon17,davidzon17}, showing good overall agreement at the high-mass end. At the low-mass end, the observed mass function of \citet{duncan14} has a significantly steeper slope than that of \citet{song16}. Our model lies between these two datasets, but prefers a higher normalization at low masses than \citet{song16}. This is consistent with our finding in Figure~\ref{fig:MuvMstar}, where we infer a higher $M_{\star}$ for a given $M_{\rm UV}$ than \citet{song16}.}
\label{fig:SMF}
\end{figure*}

\begin{deluxetable}{cccc}
\tablecaption{Best-fit Schechter function (Equation~\ref{eq:schechter}) parameters of our galaxy stellar mass function.\label{tab:MF}}
\tablecolumns{4}
\tablewidth{0pt}
\tablehead{
\colhead{redshift} &
\colhead{$\phi^{*}$} &
\colhead{$\log~M^{*}$} & 
\colhead{$\alpha$}  \\
\colhead{} &
\colhead{[$10^{-5}~\mathrm{Mpc}^{-3}$]} &
\colhead{$[M_{\odot}]$} & 
\colhead{}
}
\startdata
$ z=4 $ & $ 261.9 \pm 25.0 $ & $ 10.16 \pm 0.07 $ & $ -1.54 \pm 0.01 $ \\
$ z=5 $ & $ 201.2 \pm 22.3 $ & $ 9.89 \pm 0.08 $ & $ -1.59 \pm 0.01 $ \\
$ z=6 $ & $ 140.5 \pm 17.3 $ & $ 9.62 \pm 0.08 $ & $ -1.64 \pm 0.01 $ \\
$ z=7 $ & $ 78.0 \pm 12.9 $ & $ 9.38 \pm 0.1 $ & $ -1.70 \pm 0.01 $ \\
$ z=8 $ & $ 38.4 \pm 11.0 $ & $ 9.18 \pm 0.16 $ & $ -1.76 \pm 0.01 $ \\
$ z=9 $ & $ 37.3 \pm 18.3 $ & $ 8.74 \pm 0.26 $ & $ -1.80 \pm 0.04 $ \\
$ z=10 $ & $ 8.1 \pm 8.5 $ & $ 8.79 \pm 0.99 $ & $ -1.92 \pm 0.06 $ \\
$ z=11 $ & $ 3.9 \pm 0.8 $ & 8.50 (fixed) & $ -2.00 \pm 0.06 $ \\
$ z=12 $ & $ 1.1 \pm 0.2 $ & 8.50 (fixed) & $ -2.10 \pm 0.06 $ \\
\enddata
\tablecomments{The errors indicate the one standard deviation errors on the parameters.}
\end{deluxetable}

\begin{figure}
\includegraphics[width=\linewidth]{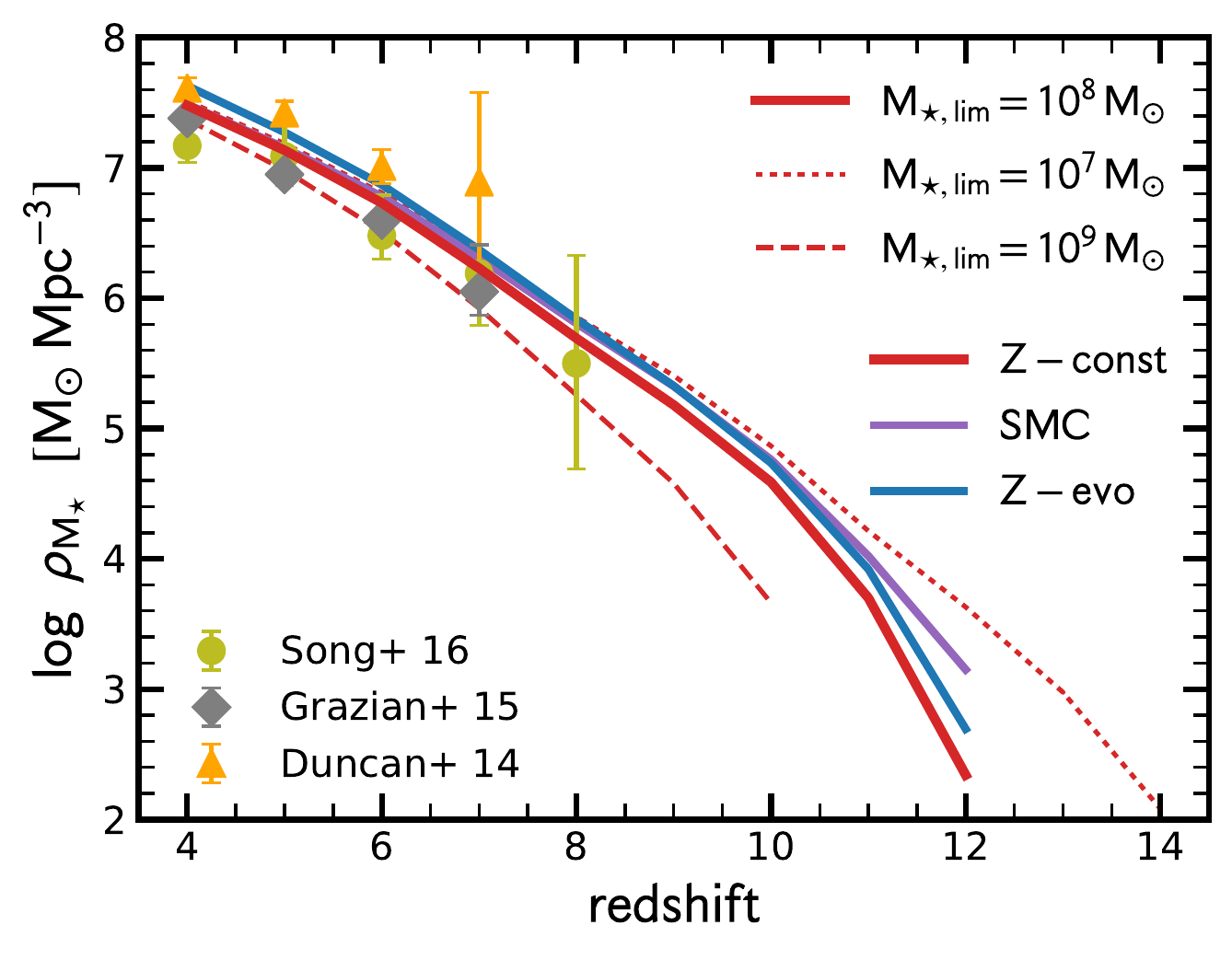}
\caption{Cosmic stellar mass density $\rho_{\rm M_{\star}}$. The dashed, solid, and dotted red lines, respectively, show the stellar mass density in our simulation box integrated down to a stellar mass limit of $M_{\star}=10^{9}~M_{\odot}, 10^{8}~M_{\odot}$ and $10^{7}~M_{\odot}$. The blue and purple lines show the Z-evo and SMC model, respectively. The observational data are taken from \citet{song16,grazian15} and \citet{duncan14}, adjusted to the Salpeter IMF and an integration limit of $10^{8}~M_{\odot}$. Our model is consistent with the current observational constraints up to $z\simeq8$. When integrating down to a limit of $M_{\star}=10^{7}~M_{\odot}$, we find that the cosmic stellar mass density declines from $z\simeq4$ to $z\simeq14$ by about 5 orders of magnitude. }
\label{fig:cSMD}
\end{figure}

The evolution of the galaxy stellar mass function over the past 10 billion years ($z=0-3$) has been extensively studied and rather well constrained \citep[e.g.][]{marchesini09, peng10_Cont, baldry12, ilbert13, muzzin13, tomczak14, weigel16}. On the other hand, in the first few billion years after the big bang, the galaxy mass function remains poorly constrained \citep[e.g.][]{song16,duncan14,grazian15}, because of limited sample size and systematic uncertainties in the stellar mass estimation, which we have highlighted in the previous subsection.

In Figure~\ref{fig:SMF}, we plot the evolution of the stellar mass function from $z=4-14$. At $z\la10$, the stellar mass function shows the well-known shape of a \citet{schechter76} function: a power law at low masses and an exponential cutoff at high masses. The knee of the mass function shifts to lower masses at higher redshifts, which is in contrast with the evolution at $z<4$ where it remains roughly constant \citep{ilbert13, muzzin13}. Furthermore, we find that the low-mass-end slope steepens with increasing redshift. More quantitatively, we fit a single Schechter function to our predicted galaxy stellar mass functions:

\begin{equation}
\phi(M_{\star})\mathrm{d}M_{\star} = \phi^{*}\left(\frac{M_{\star}}{M^{*}}\right)^{\alpha+1}\exp\left[-\frac{M_{\star}}{M^{*}}\right]\mathrm{d}M_{\star}
\label{eq:schechter}
\end{equation}

\noindent
which is characterized by a power law with a low-mass-end slope of $\alpha$, an exponential cutoff at stellar masses larger than a characteristic mass, $M^{*}$, and a normalization $\phi^{*}$. The best-fit values are provided in Table~\ref{tab:MF}. We find that the characteristic mass, $M^{*}$, indeed decreases from $\log~M^{*}/M_{\odot}=10.16\pm0.07$ at $z=4$ to $\log~M^{*}/M_{\odot}=8.79\pm0.99$ at $z=10$, while the low-mass-end slope of $\alpha$ steepens from $\alpha=-1.54\pm0.01$ at $z=4$ to $\alpha=-1.92\pm0.06$ at $z=10$. The SMC model (shown as dotted lines in Figure~\ref{fig:SMF}) produces higher-mass galaxies in low-mass halos, leading to a slightly higher normalization in the early universe ($z\ga8$) and a steeper low-mass-end slope at $z=4-8$.

We additionally compare our predicted galaxy stellar mass functions with the ones from observations in Figure~\ref{fig:SMF}. In the left panel, we compare the redshift evolution with observations of \citet{song16} and \citet{duncan14}. In the right panel, we zoom in on $z=4$, additionally including observations by \citet{santini12}, \citet{grazian15}, \citet{stefanon17}, and \citet{davidzon17}. Even after taking into account the different IMFs and cosmologies, there is a scatter of $0.3-0.8$ dex in the observational data. At $M_{\star}>10^{10.5}~M_{\odot}$, the different datasets are consistent with each other and with our model within the observational uncertainties. Our predicted mass function lies slightly below the data, even after accounting for the incompleteness in the halo mass function. Although this difference is not significant ($\sim1\sigma$ from observations), it hints that the most massive galaxies in our model are missing some stellar mass, possibly because we have neglected mass brought in through mergers.

At lower masses ($M_{\star}<10^{10}~M_{\odot}$), the observational data of \citet{song16} and \citet{duncan14} diverge significantly from each other. In particular, \citet{duncan14} measure a much steeper low-mass slope ($\alpha=-1.89^{+0.15}_{-0.13}$) than \citet{song16} ($\alpha=-1.55^{+0.08}_{-0.07}$). Our model estimate lies above the mass function of \citet{song16}, though the low-mass slope we measure ($\alpha=-1.54\pm0.01$) is in excellent agreement. The difference between \citet{song16} and \citet{duncan14} has already been seen in the M$_{\rm UV}$-$M_{\star}$ relation (Section~\ref{subsec:MsMuv}): at a given UV luminosity, \citet{song16} determine a smaller $M_{\star}$ than \citet{duncan14}, despite using similar datasets (CANDELS with \textit{Spitzer}/IRAC data) and methodologies for the derived $M_{\star}$, although \citet{song16} allow for younger ages, and hence lower mass-to-light ratios, than \citet{duncan14} (see \citealt{behroozi18} for an extended discussion). In the case of \citet{song16}, the determination of the stellar mass function depends strongly on the M$_{\rm UV}$-$M_{\star}$ relation, since they use the observed UV LF of \citet{finkelstein15} and convolve it with this relation to obtain the stellar mass function. On the other hand, \citet{duncan14} use the individual $M_{\star}$ measurements and compute the mass function using the $1/V_{\rm max}$ method of \citet{schmidt68}, where $V_{\rm max}$ is the maximum comoving volume in which a galaxy can be observed. 

Figure~\ref{fig:cSMD} shows the evolution of the cosmic stellar mass density. The inferred stellar mass density depends strongly on the stellar mass to which one integrates down the stellar mass function, which is not surprising given the steepness of the slope at the low-mass end of the mass function. Using the fiducial integration lower limit of $M_{\star}=10^8~M_{\odot}$, we find that that stellar mass density increases from $z\sim12$ to $z\sim4$ by 5 orders of magnitude from $10^2~M_{\odot}~\mathrm{Mpc}^{-3}$ to $3\times10^7~M_{\odot}~\mathrm{Mpc}^{-3}$. The redshift evolution of the stellar mass density of our model agrees qualitatively with the observed one \citep{duncan14,grazian15,song16}. 

\begin{figure*}
\includegraphics[width=\textwidth]{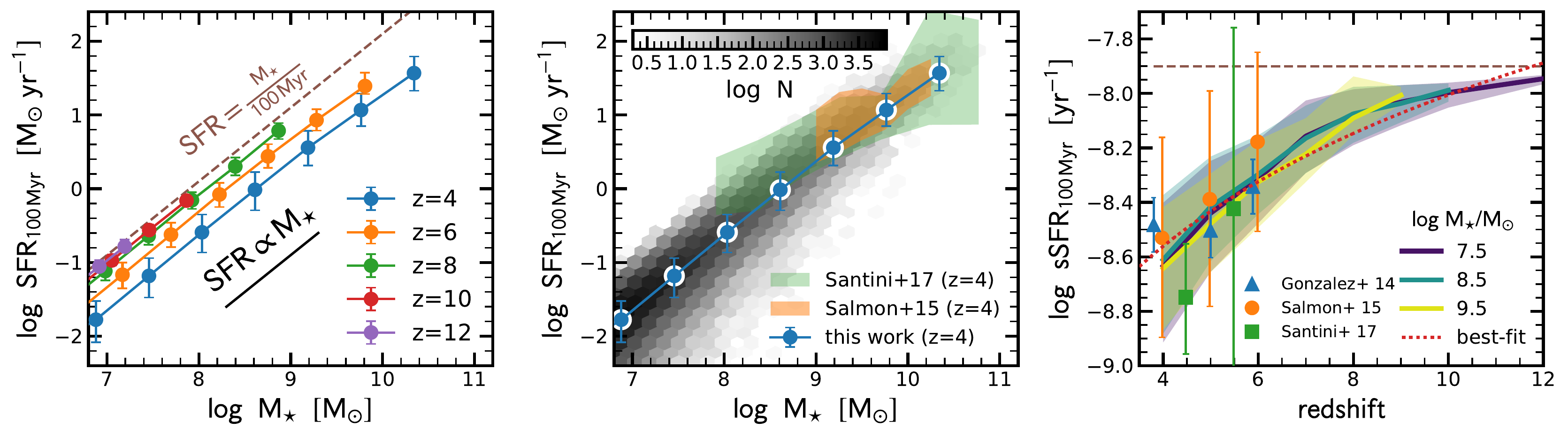}
\caption{SFR versus stellar mass ($M_{\star}$): the star-forming main sequence (MS). \textit{Left panel:} the evolution of the MS between $z=4-12$. The SFR is averaged over 100 Myr, making it comparable to UV-derived SFR estimates from observations. The  dashed brown line indicates the region where the SFR is equal to the stellar mass divided by 100 Myr (taking into account the mass returned to the ISM with $R=0.1$): by definition, no galaxy can lie above this line. At all masses, the SFR of the MS is proportional to $M_{\star}$. At earlier epochs, the normalization of the MS increases ($\propto(1+z)^{1.6}$), converging toward the dashed line, which indicates that galaxies at $z>8$ formed most of their mass in the last 100 Myr (see also Figure~\ref{fig:age}). \textit{Middle panel:} The MS in our model compared to observations by \citet{santini17} and \citet{salmon15}. The gray 2D histogram counts the number of galaxies in our model at a given value of $M_{\star}$-SFR. The median relation from our model tracks the observational trends well. \textit{Right panel:} evolution of the normalization of the MS: sSFR as a function of redshift. The different colored lines indicate the evolution of the sSFR for three different stellar masses. Since the MS in our model has a slope of 1, the evolution of the sSFR with redshift for all masses is similar. The dashed line shows again the upper limit given the averaging timescale of 100 Myr and $R=0.1$. The red dotted line indicates the best fit: $\log(\mathrm{sSFR}/\mathrm{yr})=-9.7+1.6\log(1+z)$. We find good agreement with observational estimates from  \citet{gonzalez14}, \citet{salmon15}, and \citet{santini17}.}
\label{fig:SFMS}
\end{figure*}

\subsection{Star-Formation Main Sequence}
\label{subsec:SFMS}

\begin{figure}
\includegraphics[width=\linewidth]{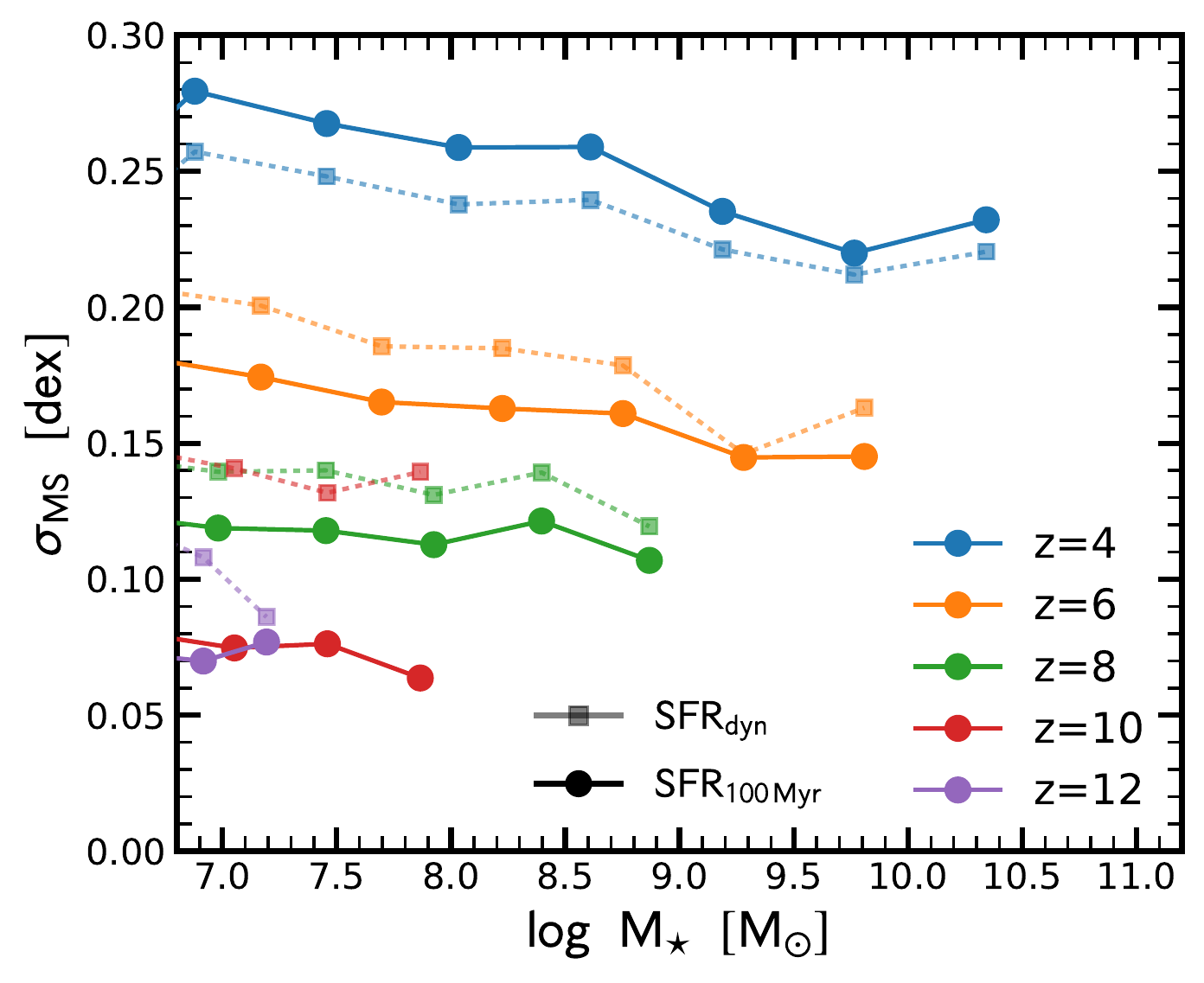}
\caption{Scatter of the MS, $\sigma_{\rm MS}$. The large filled circles and the small translucent circles show the scatter of the MS as a function of stellar mass for the SFR averaged over 100 Myr (fiducial) and the dynamical time ($\tau_{\rm dyn}\simeq0.1\tau_{\rm H}$), respectively. We find a strong trend with cosmic time: the scatter in the MS decreases from $\sim0.25$ dex at $z=4$ to 0.08 dex at $z=10$. This reflects the young stellar ages of galaxies (see also Figure~\ref{fig:age}). Furthermore, we find a weak mass dependence at $z=4$, with high-mass galaxies exhibiting a slightly smaller scatter than their lower-mass counterparts. The trend with cosmic times weakens and $\sigma_{\rm MS}$ shows a trend of assimilation when averaging the SFR over $\tau_{\rm dyn}$, which points toward the importance of dynamical effects that set $\sigma_{\rm MS}$.}
\label{fig:scatter_SFMS}
\end{figure}

\begin{figure*}
\includegraphics[width=\textwidth]{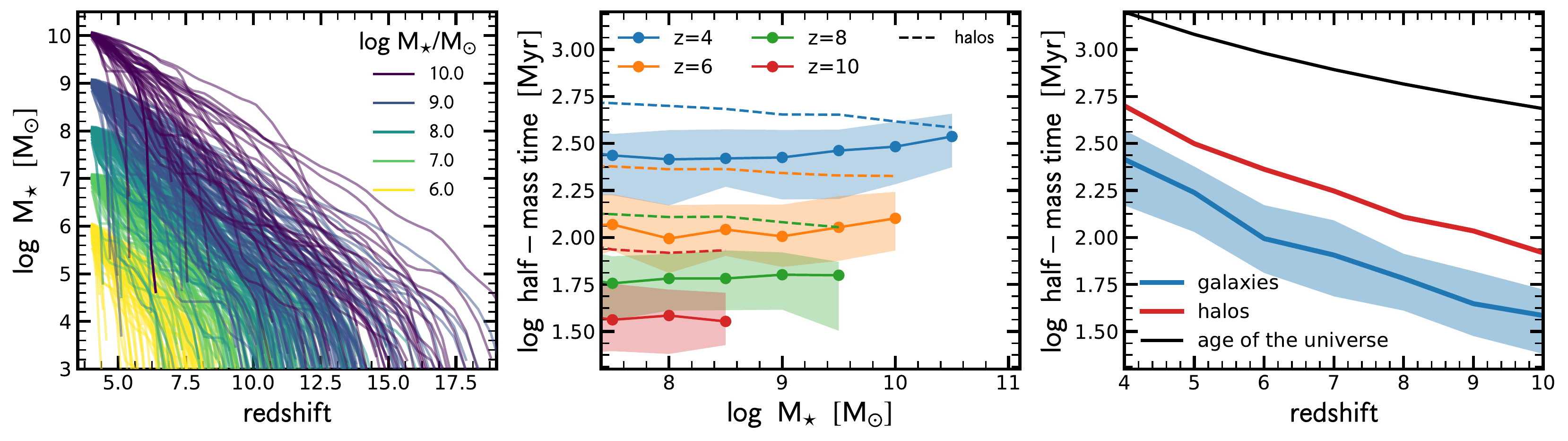}
\caption{Stellar mass growth of galaxies in our model. \textit{Left panel:} Stellar mass as a function of redshift. The individual lines show the growth in $M_{\star}$ for individual galaxies. The color-coding corresponds to the final ($z=4$) mass. Overall, we find a wide diversity of mass growth histories. \textit{Middle panel:} Half-mass time as a function of $M_{\star}$ and $z$. We define the half-mass time to be the look-back time at which half the stellar mass was assembled. We find a rather strong dependence with redshift and a comparatively weak dependence on mass. As expected from hierarchical growth, galaxies at higher redshift and lower masses are younger. The dashed lines indicate the corresponding dark matter halo assembly times (i.e., the time when the host halo had half of its final mass). \textit{Right panel:} The blue and red lines show the half-mass times of galaxies (with $M_{\star}=10^8~M_{\odot}$) and of dark matter halos (assembly time), respectively, as a function of redshift. The black solid line shows the age of the universe. At all masses and redshift, the halo assembly time is systematically longer than the half-mass time of the galaxies. }
\label{fig:age}
\end{figure*}

At $z=0-3$, a nearly linear relation between the SFR and the stellar mass of a galaxy has been found, also known as the star-forming main sequence \citep[MS; e.g.][]{brinchmann04,daddi04,elbaz07,noeske07,salim07,whitaker12, speagle14,pannella15}. An important feature of the MS is the rather small scatter of $\sigma_{\rm MS}\sim0.2-0.3$ dex. This small scatter in the MS at different redshifts indicates that most galaxies are in fact not undergoing dramatic major mergers \citep{rodighiero11,noeske07,noeske07b}, but are sustained for extended periods of time in a quasi-steady state of gas inflow, gas outflow, and gas consumption \citep{bouche10, daddi10, genzel10, tacconi10, dave12, dekel13, lilly13_bathtube, tacchella16_MS}. 

We plot the MS relation of our model in Figure~\ref{fig:SFMS}. The left panel shows the prediction for the MS for $z=4-12$. We define the SFR in the simulation to be the average SFR over the past 100 Myr. For a given $M_{\star}$, forming all stars within the last 100 Myr provides an upper limit for the SFR, which is indicated as the dashed brown line in the figure (taking into account the mass return fraction of $R=0.1$, see Section~\ref{subsec:return}). We find a linear relation between the SFR and $M_{\star}$, while the normalization increases with redshift and converges to the upper limit. More quantitatively, we find for the best fit

\begin{equation}
\mathrm{SFR}=\frac{M_{\star}}{M_{\star,0}},
\label{eq:MS}
\end{equation}

\noindent
with $\log M_{\star,0}=9.7-1.6\log(1+z)$. 

In the middle panel of Figure~\ref{fig:SFMS}, we compare the MS from the model with the one from the observations by \citet{santini17} and \citet{salmon15} at $z=4$. We find good agreement overall, with a hint of slightly lower SFRs at $M_{\star}\approx10^8-10^9~M_{\odot}$. The right panel shows the evolution of specific SFR (sSFR) as a function of redshift: since the MS in our model has a slope of 1, the sSFR evolution for all masses looks very similar. It is again important to highlight that there exists an upper limit in sSFR given the averaging timescale of 100 Myr and the mass return fraction of $R=0.1$. With the best-fit relation for the MS (Equation~\ref{eq:MS}), we put forward that the sSFR is proportional to $(1+z)^{1.6}$ at $z\ga4$, which is consistent with observational data (see also \citealt{faisst16}).

Figure~\ref{fig:scatter_SFMS} shows the scatter of the MS, $\sigma_{\rm MS}$, as a function of $M_{\star}$ and $z$. At $z=4$, we find a weak dependence on $M_{\star}$: $\sigma_{\rm MS}$ decreases weakly from 0.28 dex to 0.22 dex from $10^7~M_{\odot}$ to $10^{10}~M_{\odot}$. This trend can be explained by the fact that lower-mass galaxies have a burstier SFH. Furthermore, we find a strong trend with redshift: $\sigma_{\rm MS}$ decreases from $z=4$ to $z=10$ by $\sim0.17$ dex. The main cause for this is that the MS is already close to the upper limit in the SFR. In other words, the stellar ages of the galaxies at $z=10$ are comparable to the averaging timescale of 100 Myr (see also Figure~\ref{fig:age} for the age distribution of our model galaxies). When averaging over the dynamical time ($\tau_{\rm dyn}\simeq0.1\tau_{\rm H}$), the difference in $\sigma_{\rm MS}$ for different redshifts decreases. This assimilation of $\sigma_{\rm MS}$ points toward the importance of dynamical effects in setting $\sigma_{\rm MS}$.

\subsection{Star Formation Histories}
\label{subsec:SFH}

\begin{figure*}
\includegraphics[width=\textwidth]{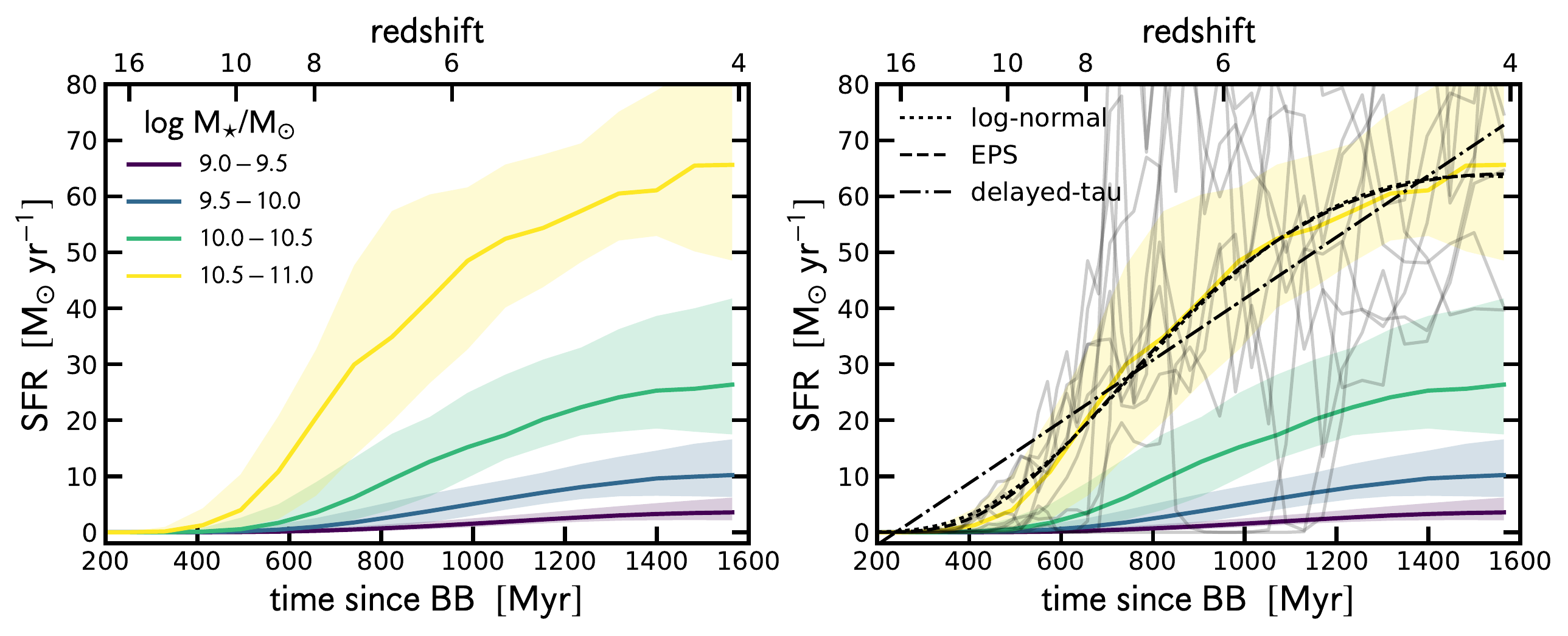}
\caption{SFHs as a function of $M_{\star}$. \textit{Left panel:} SFR as a function of time since the Big Bang (BB). The solid lines show the median (shaded regions the 16th and 84th percentiles) SFHs at $z=4$ for different mass bins. \textit{Right panel:} Same as the left panel, but we fit the median SFH of the highest-mass bin with three different parameterizations (Equations~\ref{eq:SFH_tau}-\ref{eq:SFH_EPS}). The lognormal parameterization and the one based on the EPS formalism are able to reproduce the model well; the delayed $\tau$-model, less so. Furthermore, the gray lines show SFHs of 10 individual galaxies of the most massive bin.}
\label{fig:SFH}
\end{figure*}

The SFH is one of the fundamental ingredients and outcomes of SED modeling \citep[e.g.][]{pacifici12,conroy13_rev}. Most SED-fitting analyses are based on parametric SFHs (in many cases, on a fixed grid of parameters), for which a good understanding of the shape of the SFH is fundamental. This is also true for nonparametric approaches, since these typically require a reasonable prior. 

In Figure~\ref{fig:age}, we first show the large diversity of SFHs produced by our model. The left panel shows the stellar mass growth as a function of redshift, with around 100 galaxies in each stellar mass bin. Some galaxies reach their final $M_{\star}(z=4)$ early on, while others do so later on. In order to quantify this, we look at the half-mass time distribution of galaxies and of the dark matter halos that host them. We define the half-mass time as the look-back time at which half the stellar mass was assembled. The middle panel of Figure~\ref{fig:age} shows the half-mass time of the galaxies as a function of mass for $z=4-10$. Overall, there is only a very weak trend with mass (more massive galaxies are slightly older), which can be understood by noting that, on average, our galaxies trace the MS with a slope of 1 (i.e. the sSFR is constant as a function of mass), implying that the mass doubling timescale is constant as a function of $M_{\star}$. The rather strong dependence with redshift is expected from hierarchical growth, where galaxies at higher redshift and lower masses are younger. 

The dashed line in the middle panel of Figure~\ref{fig:age} shows the dark matter halo assembly time (i.e., the time when the host halo had half of its final mass). This is larger for all galaxies at all times and masses. This can also be seen in the right panel, where the galaxy half-mass time and the halo assembly time (for galaxies of $M_{\star}=10^{7.5}-10^{8.5}~M_{\odot}$) as a function of redshift are compared to the age of the universe. Galaxies are always younger than their dark matter halo at $M_{\star}\la10^{10}~M_{\odot}$ because the star-formation efficiency $\varepsilon(M_{\rm h})$ increases with $M_{\rm h}$, making the star formation in a galaxy more efficient at late times.

We investigate the shape of the SFHs in Figure~\ref{fig:SFH}. We plot the median SFHs and their 16th and 84th percentiles as a function of time since the big bang for four different stellar mass bins at $z=4$. All median SFHs are increasing with time. At later times, the median SFRs do not increase as quickly as at early times, which can be understood by slower growth of dark matter halos at later times. Furthermore, the thin gray lines show the individual SFHs for 10 galaxies in the most massive bin ($M_{\star}=10^{10.5}-10^{11.0}~M_{\odot}$), reflecting the large scatter for individual galaxies. In particular, individual galaxies do not always have increasing SFHs with time, but they can actually have phases with declining SFRs.

We now characterize the median SFH with a parametric function. We will focus on the SFH of the most massive bin, but the results hold also for the lower-mass bins. Clearly, the rising SFH is not well fit by an increasing or decreasing $\tau$-model, or a constant SFH fit. We therefore use three other parameterizations. In particular, we fit the SFH with the following:

\begin{enumerate}[label=(\roman*)]
\item a delayed $\tau$-model, which allows linear growth at early times followed by an exponential decline at late times,

\begin{equation}
\mathrm{SFR}(t)=C \cdot (t-t_0) \cdot e^{-t/\tau};
\label{eq:SFH_tau}
\end{equation}

\item a lognormal SFH \citep{gladders13,abramson15,diemer17},

\begin{equation}
\mathrm{SFR}(t)=C/t \cdot e^{(\ln t - t_0)^2/(2\tau^2)};
\label{eq:SFH_log}
\end{equation}

\item an EPS-based SFH \citep{neistein08,dekel13},

\begin{equation}
\mathrm{SFR}(z)=C \cdot e^{-\alpha \cdot z} \cdot (1+z)^{\beta}.
\label{eq:SFH_EPS}
\end{equation}

\end{enumerate}

\begin{figure}
\includegraphics[width=\linewidth]{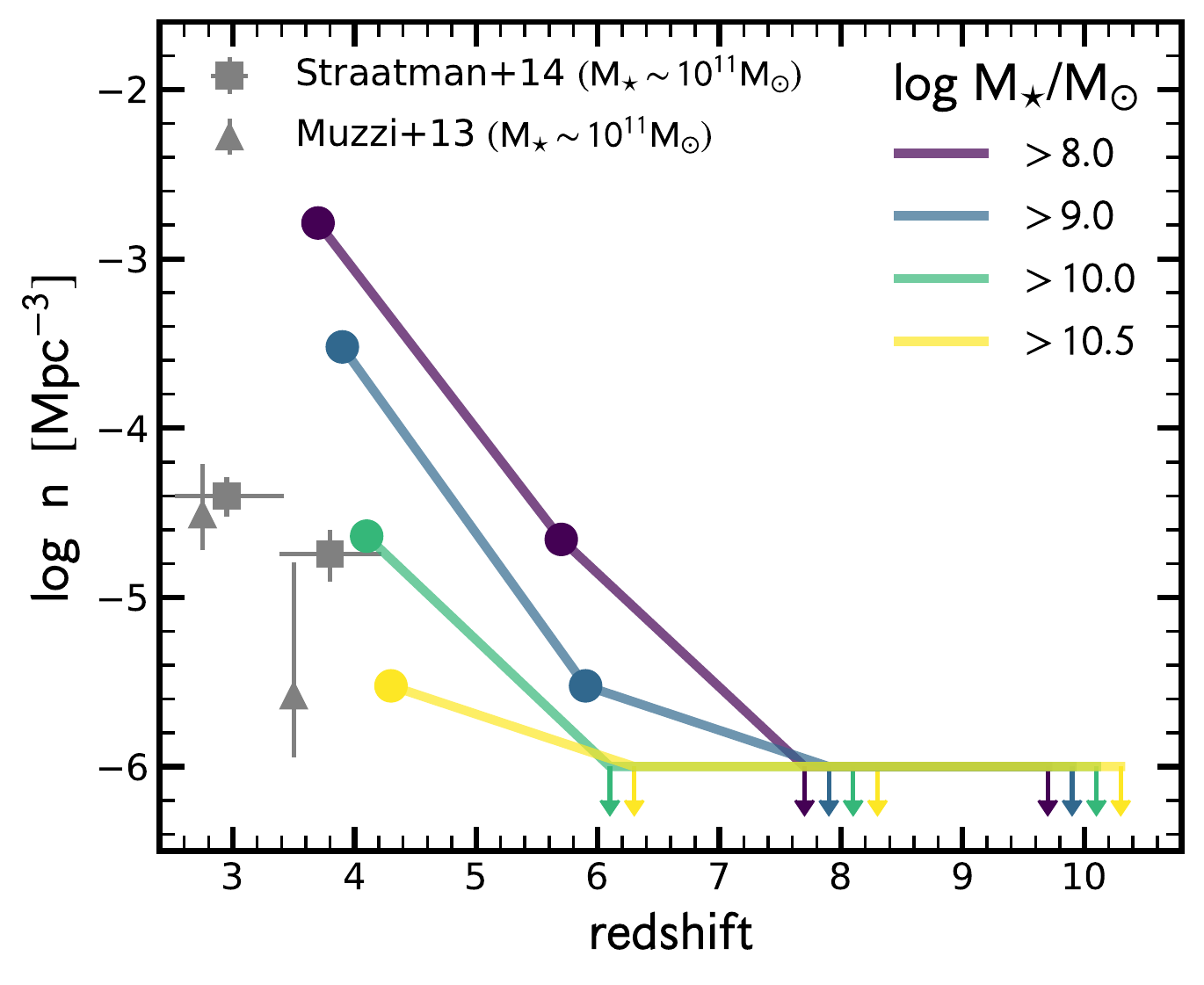}
\caption{Abundance of `slow growers'. We define slow growers as galaxies with a low SFR that have a mass $e$-folding timescale that is longer than the Hubble time ($\mathrm{sSFR}^{-1}>t_{\rm H}$). We plot the number density as a function of redshift of slow growers with mass above a certain threshold. We find such galaxies only at $z<8$. In particular, their number density increases significantly toward lower redshift, though their fraction remains rather low ($\la6\%$).}
\label{fig:QG}
\end{figure}

\noindent
All of these parameterizations have three free parameters, have a rising SFR at early times and declining SFR at later times. We find that the median SFH is well described by the lognormal and EPS-based SFH parameterizations, while the delayed $\tau$-model is a bit too high at early and late times and too low at intermediate times. It is not surprising that both the lognormal and EPS-based parameterization lead to essentially the same result, since their shapes are very similar. It is worth highlighting here that the motivation for adopting a lognormal SFH prescription is directly inspired by the growth histories of dark matter halos. Although the lognormal and EPS-based parameterizations describe well the median SFH, individual galaxies do not follow these parameterizations on a $\sim100$ Myr timescale. In particular, we find that galaxies exhibit suppressed SFR for up to a few hundred megayears, which could possibly be the first quiescent galaxies in the universe.

\subsection{Slow Growers in the Early Universe}
\label{subsec:QG}

The first quiescent galaxies in the early universe can help to better understand the physics that leads to a halt in star formation (`quenching'). Hence, finding such galaxies is of great interest. Observationally, \citet{straatman14} have identified a population of massive ($\sim10^{11}~M_{\odot}$), $z\sim4$ quiescent galaxies. Their selection is based on the ($U-V$)-($V-J$) color-color diagram (UVJ diagram) that is able to differentiate between red galaxies that are quiescent and red galaxies that are dusty and star-forming \citep{williams09}. In our model, some galaxies do show periods of a reduced SFR (see Figure~\ref{fig:SFH}, right panel), however, none of the galaxies have UVJ colors that fulfill the cut for being quiescent: all of them are in the star-forming region. This is not surprising since it takes $\sim3$ Gyr to become UVJ-quiescent for a simple stellar population (SSP) with $Z=0.02~Z_{\odot}$, which is longer than the age of the universe at $z\sim4$. For an SSP with $Z=1.0~Z_{\odot}$, this timescale reduces to $\sim0.5$ Gyr. This directly implies that any UVJ selection of quiescent galaxies at $z\ga3$ misses quiescent galaxies with such low metallicities. Therefore, a difference in metallicity could explain why we do not find any UVJ-quiescent galaxies in our model, while observationally these galaxies may indeed exist. Another possible difference could be that, with the limited volume of our simulation, we are unable to probe galaxies with masses of $\sim10^{11}~M_{\odot}$, which is the mass range where most of these quiescent galaxies are found in observations (Section~\ref{subsec:MF}). Finally, a third reason could be that we indeed miss additional physical mechanisms that stop star-formation in galaxies for an extended period, resulting in a reddening of galaxy colors.

In addition to looking at the UVJ diagram, one can look into the SFR to judge whether a galaxy is still growing as a result of star formation. Specifically, the inverse of the sSFR is the $e$-folding timescale (roughly the mass doubling timescale) for stellar mass growth. Therefore, galaxies with $\mathrm{sSFR}^{-1}>t_{\rm H}$ are no longer growing their mass significantly via star formation. We call these systems `slow growers'. As shown in Figure~\ref{fig:QG}, we identify a population of slow growers at $z<8$. At $z=6$, such galaxies make up a negligibly small fraction of the whole population. At $z=4$, the number density of slow growers with $M_{\star}>10^8~M_{\odot}$ is $10^{-3}~\mathrm{Mpc}^{-3}$, making up roughly $6\%$ of the galaxy population at this epoch. There is also a weak trend in mass: the fraction declines to $5\%$ at $>10^{10}~M_{\odot}$. The origin of these slow growers is the drop in the cosmic accretion rate of baryons into their halos compared to previous epochs.

\section{Implications}
\label{sec:discussion}

\subsection{Stellar-to-halo Mass Relation}
\label{subsec:MsMh}

\begin{figure}
\includegraphics[width=\linewidth]{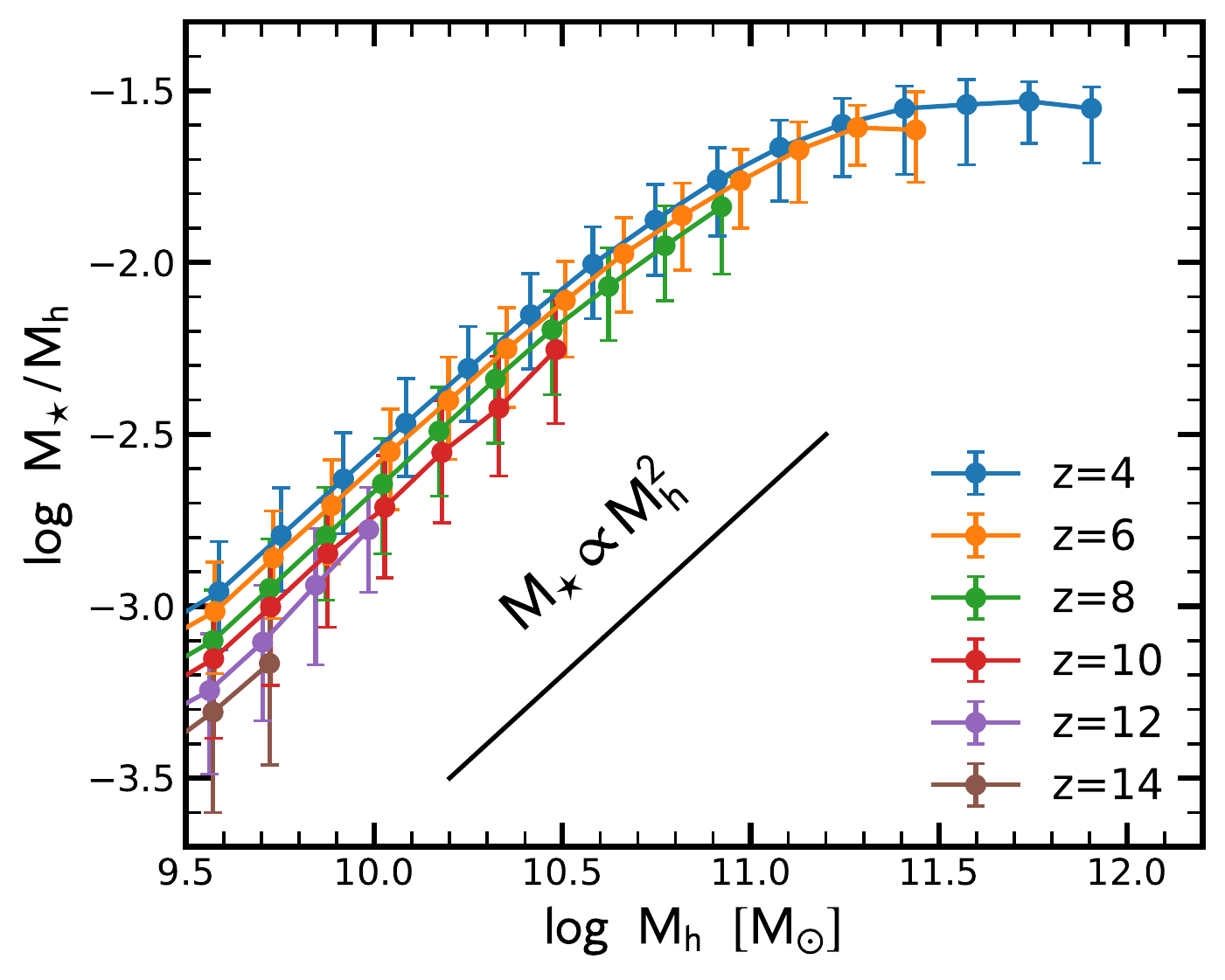}
\caption{Redshift evolution of the stellar-to-halo mass relation, which shows a weak evolution in time. The normalization of this relation decreases slightly with redshift ($\propto(1+z)^{-0.6}$). At all redshifts, the stellar-to-halo mass relationship at masses below $10^{11}~M_{\odot}$ is well described by $M_{\star}\propto M_{\rm h}^{2}$. }
\label{fig:MsMh_z}
\end{figure}

\begin{figure*}
\begin{center}
\includegraphics[width=0.8\textwidth]{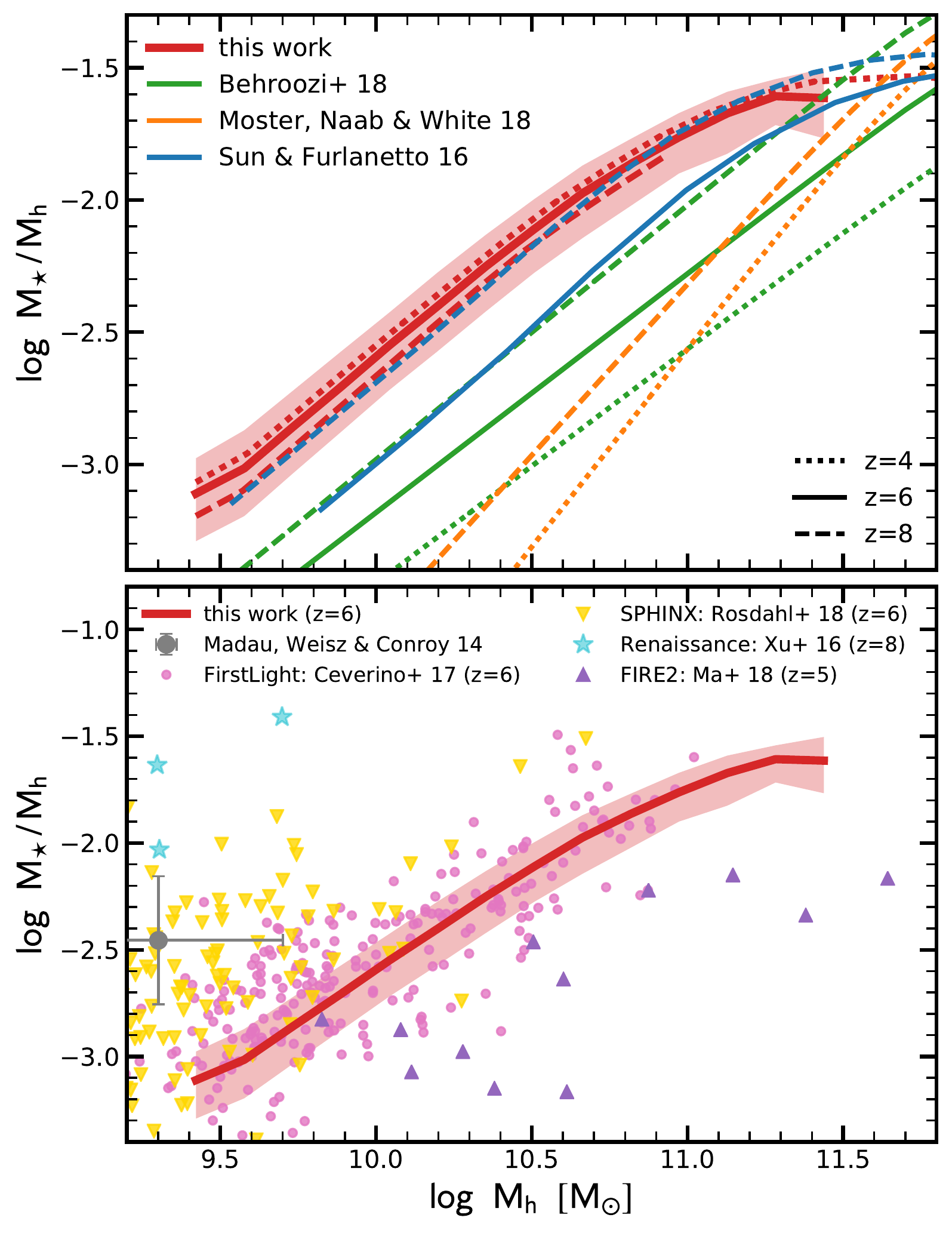}
\end{center}
\caption{Stellar-to-halo mass relation of our model in comparison with other estimates from the literature. \textit{Top panel:} Comparison of our stellar-to-halo mass relation with the ones of the empirical models of \citet{sun16}, \citet{moster18}, and \citet{behroozi18}. The dashed, solid, and dotted lines indicate the stellar-to-halo mass relation at $z=4$, $z=6$, and $z=8$, respectively. Although there are large difference between different models, the stellar-to-halo mass relation of our model lies above the others. We find good agreement with the estimate of \citet{sun16} at $z=8$. Furthermore, our relation shows little evolution, while the other empirical models show more evolution. \textit{Bottom panel:} Comparison of our stellar-to-halo mass relation with numerical simulations \citep{ceverino17,ma18,rosdahl18,xu16}, as well as a crude estimate inferred from local dwarf galaxies \citep{madau14_dwarf} at $z\sim6$. There is large scatter between different numerical simulations, some of them reproducing an even higher stellar-to-halo mass relation than our estimate.}
\label{fig:MsMh}
\end{figure*}

As highlighted in the introduction, our empirical model provides useful scaling relations that can be used to constrain the physical process of numerical models. In addition to the UV LF, stellar mass function, and the MS, an important scaling relation is the stellar-to-halo mass relation. Figure~\ref{fig:MsMh_z} shows the ratio of $M_{\star}/M_{\rm h}$ as a function of halo mass and redshift. 

The stellar-to-halo mass relation exhibits the familiar peak around $10^{11.5}~M_{\odot}$. Since our model probes a limited range in halo mass (see Section~\ref{subsec:DM_framework}), we are unable to comment on the stellar-to-halo mass relation at masses higher than $\sim10^{12}~M_{\odot}$, and we therefore focus on the range between $10^{9.5}-10^{11.5}~M_{\odot}$. We find that the stellar-to-halo mass relation stays roughly constant with redshift, which is a direct consequence of our assumption of a constant star-formation efficiency with redshift. We fit the stellar-to-halo mass relation with a double power law, following Equation~\ref{eq:efficiency}, finding

\begin{equation}
\frac{M_{\star}}{M_{\rm h}} = 0.05\cdot\left[\left(\frac{M_{\rm h}}{M_{\rm c}}\right)^{-1.0} + \left(\frac{M_{\rm h}}{M_{\rm c}}\right)^{0.3}\right]^{-1}
\end{equation}

\noindent
with $M_{\rm c}=1.6\cdot10^{11}~M_{\odot}$. The low-mass slope of $-1.0$ implies that $M_{\star}\propto M_{\rm h}^2$ at halo masses below $10^{11}~M_{\odot}$. The proportionality $M_{\star}\propto M_{\rm h}^2$ follows directly from $\varepsilon \propto M_{\rm h}$ at $M_{\rm h}<10^{11}~M_{\odot}$ and our assumed star-formation law: $M_{\star}=\int \mathrm{SFR}(t) \mathrm{d}t \propto \int \varepsilon(M_{\rm h}) \dot{M}_{\rm h} \mathrm{d}t \propto M_{\rm h}^2$. The scatter in the stellar-to-halo mass ratio is roughly constant as a function of halo mass and increases slightly with redshift: from 0.14 dex to 0.18 dex from $z=4$ to $z=10$.

The normalization of the stellar-to-halo mass relation decreases weakly with redshift as $\propto(1+z)^{-0.6}$ (Figure~\ref{fig:MsMh_z}). At first glance, this is surprising since we expect a constant or slightly rising (due to reduced stellar mass loss) relation toward earlier times. Furthermore, most other models (see next paragraph) predict a constant or rising relation. We obtain this weak decline toward earlier times because of the time delay of $0.1\tau_{\rm H}$ in our model (see Section~\ref{subsec:SF_model}). Since SFHs are more sharply increasing at earlier times, the time delay has a larger impact at higher redshifts, where it moves a larger fraction of the mass accretion (and hence star formation) beyond the epoch considered. Removing the time delay from our model leads to a constant stellar-to-halo mass relation (less than 0.05 dex difference between $z=4$ and $z=10$). 

Observationally, it is still difficult to constrain the evolution of the stellar-to-halo mass relation with redshift. Based on the abundance matching, \citet{stefanon17} find that the stellar-to-halo mass ratio at fixed cumulative number density is roughly constant with redshift for $M_{\rm h}\ga10^{12}~M_{\odot}$. \citet{harikane18} find, by combining the halo occupation distribution models and clustering measurements, that the stellar-to-halo mass relation increases from $z\sim4-7$ by a factor of 4 at $M_{\rm h}\sim10^{11}~M_{\odot}$, while the stellar-to-halo mass relation shows no strong evolution in the similar redshift range at $M_{\rm h}\sim10^{12}~M_{\odot}$.

In Figure~\ref{fig:MsMh} we compare the stellar-to-halo mass relation of our model with others in the literature. The red lines indicate our model. The dashed, solid, and dotted lines mark the $z=4$, $z=6$, and $z=8$ estimates, respectively. The blue, orange, and green lines in the top panel of Figure~\ref{fig:MsMh} show the empirical models of \citet{sun16}, \citet{moster18}, and \citet{behroozi18}, respectively. It is important to stress that these models use different cosmologies: in our model we assume WMAP-7 cosmological parameters, while the other models assume Planck or some other variation of cosmological parameters (e.g. $\Omega_m = 0.28, \Omega_\Lambda = 0.72, \sigma_8=0.82, n_s=0.95$ and $h=0.7$ in the case of \citealt{sun16}). Together with slightly different halo mass definitions, this can introduce differences of up to 0.2 dex in the stellar-to-halo mass relation. It is, however, not a straightforward task to correct for a difference in cosmology between empirical models since one needs to rerun the whole model (in addition to renormalizing halo number densities and differences in accretion rates onto halos). We therefore present the stellar-to-halo mass relation of other empirical models as presented in the literature, solely correcting for differences in the assumed IMF. However, these possible systematics should be borne in mind when comparing models in Figure~\ref{fig:MsMh}. 

\citet{sun16} use halo abundance matching (UV LF) over the redshift range $5<z<8$ and assuming smooth, continuous gas accretion to model the star-formation efficiency of dark matter halos at $z>6$. The star-formation efficiency evolves with redshift, where lower-mass halos are forming stars more efficiently at higher redshifts. Therefore, their stellar-to-halo mass relation evolves with redshift: at $z=6$, our relation lies above their relation, while at $z=8$ the relations are consistent with each other. 

As mentioned in the introduction, the \citet{moster18} model also assumed that SFR is proportional to the dark matter halo accretion rate. Their model is tailored to describe the galaxy population at $z=0-8$ and is therefore more complex than ours, including prescriptions for satellites and quiescent galaxies. In addition to the halo accretion rate, the star-formation efficiency also depends on redshift and halo mass, i.e., $\varepsilon(M_{\rm h}, \dot{M}_{\rm h}, z)$. They then constrain $\varepsilon(M_{\rm h}, \dot{M}_{\rm h}, z)$ by using the observed stellar mass functions, cosmic SFRD, sSFRs, fractions of quiescent galaxies, and projected galaxy correlation functions. Similarly, \citet{behroozi18} model the SFR distribution of halos as a function of halo mass and redshift, using stellar mass function, SFRs, quenched fraction, UV LFs, $M_{\rm UV}-M_{\star}$ relations, autocorrelation functions, and quenching dependence on environment. Since most of these observations have large uncertainties at $z\ga4$, both models are mainly constrained by low-$z$ observations. This does not imply that they are less trustworthy at high $z$.

Interestingly, the \citet{moster18} and \citet{behroozi18} models not only predict a different stellar-to-halo mass relations at $z=4-8$, but while \citet{moster18} predict only little evolution, \citet{behroozi18} predict  a significant evolution with redshift. At $M_{\rm h}\approx10^{11.5}-10^{12.0}~M_{\odot}$ our empirical model roughly agrees with the ones of \citet{moster18} and \citet{behroozi18}. However, toward lower halo masses, we find a shallower decrease than \citet{moster18}, while we are consistent with the slope of \citet{behroozi18}. Furthermore, our model predicts nearly no evolution with redshift, while the \citet{behroozi18} relation, on the other hand, evolves by $\sim0.5$ dex from $z=4$ to $z=8$.

The gray point in the bottom panel of Figure~\ref{fig:MsMh} marks a $z=6$ estimate from nearby isolated dwarf galaxies in the local universe by \citet{madau14_dwarf}, who combine resolved SFHs with simulated mass growth rates of dark matter halos. They show that these dwarfs have more old stars than predicted by assuming a constant or decreasing star-formation efficiency with redshift, which leads to a high stellar-to-halo mass ratio at early times. 

We also compare our predicted stellar-to-halo relation to relations inferred from numerical simulations. In particular, we compare it to the following:
($i$) The FirstLight project \citep{ceverino17,ceverino18_UV}, which is a cosmological zoom-in simulation of 290 halos with $M_{\rm h}=10^9-10^{11}~M_{\odot}$. This simulation includes a prescription for the thermal energy and radiative feedback \citep[as a local approximation of radiation pressure;][]{ceverino14_radfeed} for the injection of momentum coming from the (unresolved) expansion of gaseous shells from supernovae and stellar winds \citep{ostriker11}.
($ii$) SPHINX \citep{rosdahl18}, a suite of simulations that includes a series of cosmological boxes with volume (5-10 cMpc)$^3$, in which halos are well resolved down to or below the atomic cooling threshold ($3\times10^7~M_{\odot}$, resolution of 11 pc at $z=6$). The simulations are the first nonzoom radiation hydrodynamics simulations of reionization that capture the large-scale reionization process and simultaneously predict the escape fraction of ionizing radiation from thousands of galaxies.
($iii$) The Renaissance simulations \citep{xu16}, a suite of zoom-in cosmological radiation hydrodynamics simulations focusing on the first generation of galaxies. This simulation contains 3000 halos with $M_{\rm h}=10^7-10^{9.5}~M_{\odot}$ at redshifts of 15, 12.5, and 8 and incorporates the effects of radiative and supernova feedback from Population III stars.
($iv$) The FIRE-2 project \citep{hopkins18_FIRE2, ma18}, which is a suite of cosmological zoom-in simulations in the halo mass range $M_{\rm h}=2\times10^9-10^{12}~M_{\odot}$. 

As visible in Figure~\ref{fig:MsMh} (bottom panel), there is considerable scatter between the different numerical simulations that arises because of different treatments of the feedback, as well as the star-formation process. The Renaissance simulations, which target low-mass halos ($M_{\rm h}<10^{9.5}~M_{\odot}$), lie about 1 order of magnitude above our estimate. The SPHINX simulation lies slightly above our relation by about 0.3 dex. The FirstLight project is in excellent agreement with our estimate. FIRE-2 seems to lie rather low, about 0.6 dex below our estimate. Finally, both FirstLight and FIRE-2 find no strong evolution of this relation with redshift: FIRE-2 seems to not show any redshift evolution (see their Fig. 4 in \citealt{ma18}), while the median stellar-to-halo mass ratio for halos with $M_{\rm h}\sim10^{10}~M_{\odot}$ decreases with from $z=10$ to $z=6$ by less than a factor of 2 (0.24 dex). It is not clear how this relation evolves for SPHINX.

The manner in which the stellar-to-halo mass relations are derived in these simulations is quite different from the model we have presented in this work. Hydrodynamical simulations model a number of complex physical processes such as the radiative cooling of gas, the conversion of dense gas into stars, feedback from photoionization and supernovae, etc., to build up galaxy populations over cosmic time. The exact manner in which these processes are treated varies from simulation to simulation. On the other hand, the model we present in this paper has no explicit treatment of galaxy formation physics. By requiring a match to an observed relation (the $z=4$ UV LF in our case), we are able to translate the dark matter halo population into a galaxy catalog; the physics is therefore modeled {\it implicitly}. Despite this, however, the {\it mean} trend in statistics like the stellar-to-halo mass relation are comparable in simulations and in our model. This suggests that while galaxy formation is a complex network of processes, there are some derivative quantities, such as the stellar-to-halo mass relation, which can be measured by simply matching overall population statistics (e.g. the LF) with simple models. Note, however, that while the {\it mean} relations are reproduced, this may not be equally true for the {\it scatter} at fixed halo mass -- as shown, for example, in the bottom panel of Figure~\ref{fig:MsMh}. This is where the stochastic impact of physical processes on individual galaxies in hydrodynamical simulations may be particularly informative.

\subsection{Number Count Predictions for JWST}

\begin{figure}
\includegraphics[width=\linewidth]{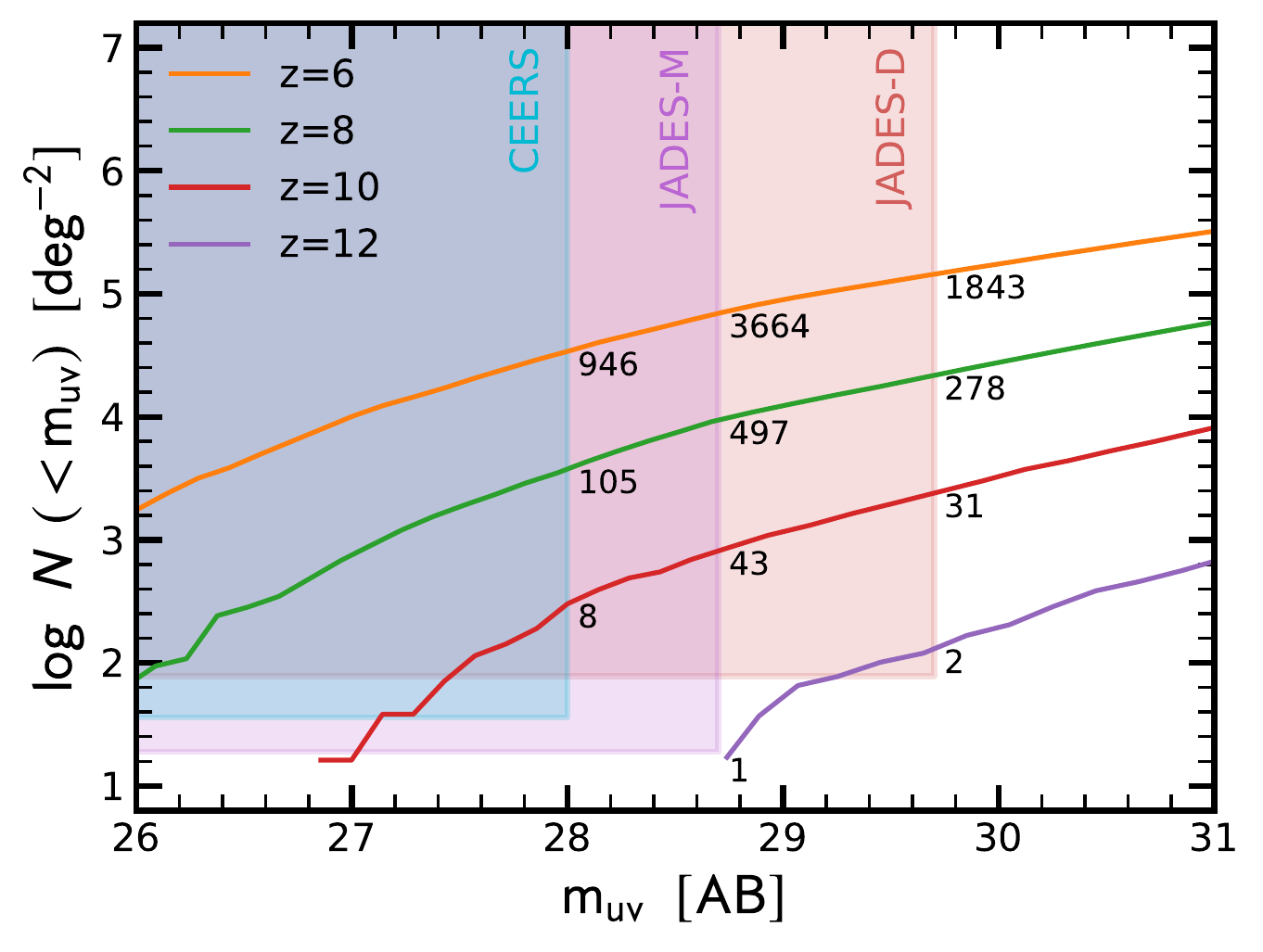}
\caption{Predicted number counts of galaxies brighter than apparent magnitude $m_{\rm UV}$ (rest-frame UV) per square degree for $z=6-12$ based on our model UV LFs. We plot the coverage of future \textit{JWST} surveys as shaded regions: the ERS program CEERS and the medium and deep GTO programs JADES-M and JADES-D, respectively. The predicted number counts are indicated by the black numbers. With these three extragalactic programs, we expect of the order of $1000$ $z\sim8$ galaxies, $100$ $z\sim10$ galaxies, and a few $z>10$ galaxies.}
\label{fig:UV_counts_predictions}
\end{figure}

In Section~\ref{sec:results}, we have shown that our new empirical model is able to reproduce current observational constraints. Furthermore, we have made predictions for the galaxy UV luminosity and stellar mass functions at $z>8$. We now use our model to make predictions for the number counts of \textit{JWST} NIRCAM high-$z$ dropout surveys. In particular, we focus on two extragalactic surveys that are currently planned with \textit{JWST}. The first is a large ($\sim720$ hr) observational program, the \textit{JWST} Advanced Deep Extragalactic Survey (JADES), a joint program of the NIRCam and NIRSpec Guaranteed Time Observations (GTO) teams. The second is the Cosmic Evolution Early Release Science Survey (CEERS; PI Finkelstein), which is an Early Science Release (ERS). 

We use our predicted UV LFs to make the number count predictions. We assume that the galaxies are detected with at least $10\sigma$ (assuming point-source detection limits) in two rest-frame UV photometric bands: that closest to $1500~\angstrom$, and the nearest band at a longer wavelength. The detection bands correspond to F115W and F150W at $6<z<7$, F150W and F200W at $7<z<9.6$, and F200W and F277W at $9.6<z<13$. JADES has a medium and a deep component: the deep component (JADES-D) covers an area of $46~\mathrm{arcmin}^{2}$, with an average depth of 29.7 AB mag ($10\sigma$ point-source limit; assuming to hold for all passbands), while the medium component (JADES-M) covers $190~\mathrm{arcmin}^{2}$ at 28.7 AB mag. The CEERS program covers an area of area of $100~\mathrm{arcmin}^{2}$ and reaches an average depth of 28.0 mag. 

Figure~\ref{fig:UV_counts_predictions} shows the predicted number counts for redshifts $6\la z\la12$ and the regions of CEERS, JADES-M, and JADES-D. The numbers in the figure indicate the expected number of dropouts. Our model predicts of the order of $1000$ $z\sim8$ galaxies, $100$ $z\sim10$ galaxies, and a few $z>10$ galaxies. This is consistent with the numbers quoted in \citet{williams18}, who extrapolate low-$z$ scaling relations and provide a mock catalog for extragalactic observations with \textit{JWST}. Our model forecasts a significant drop in number density from $z\sim6$ to $z\sim10$ compared to lower redshifts \citep[see also][]{cowley18}. Observationally, due to large uncertainties that stem from small sample sizes and survey volumes at $z\ga8$ in current observations, there remain discrepancies in the literature as to how fast the LF evolves \citep{oesch12,oesch14,zheng12,ellis13,mcleod16,bouwens15}. The most recent observations by \citet[][see also \citealt{ishigaki18}]{oesch18} are consistent with the fast evolution predicted by our model. \textit{JWST} will accurately measure the evolution at $z>8$, thereby constraining the relationship between the star-formation efficiency and the evolution of the halo mass function at early times, testing our fundamental assumption made in this work of a redshift-independent star-formation efficiency \citep{tacchella13,mason15}.

We note that these number count predictions are not significantly affected by short-term ($<30$ Myr) fluctuations of the SFH. Since this is below the time resolution of our dark matter merger tree ($\sim60~\mathrm{Myr}$), a complete treatment is beyond the scope of this work. However, to first order, the far-UV luminosity traces the SFH averaged over a timescale of $20-50$ Myr, which is comparable to the temporal resolution of our model. Therefore, short-term fluctuations will only affect our number count predictions minimally. On the other hand, one has to keep in mind that these UV-selected galaxy populations are not representative of a mass-complete galaxy sample. The importance of this effect can be directly seen from the scatter in the $M_{\rm UV}-M_{\star}$ relation (Figure~\ref{fig:MuvMstar} in Section~\ref{subsec:MsMuv}). For example, at $z\sim8$, the maximal difference in $M_{\rm UV}$ amounts to [2.5, 1.7, 1.2] mag for galaxies with $M_{\star}=[10^7, 10^8, 10^9]~M_{\odot}$.

\subsection{Luminosity-to-SFR Conversion}
\label{subsec:lum_SFR_conversion}

\begin{figure*}
\includegraphics[width=\linewidth]{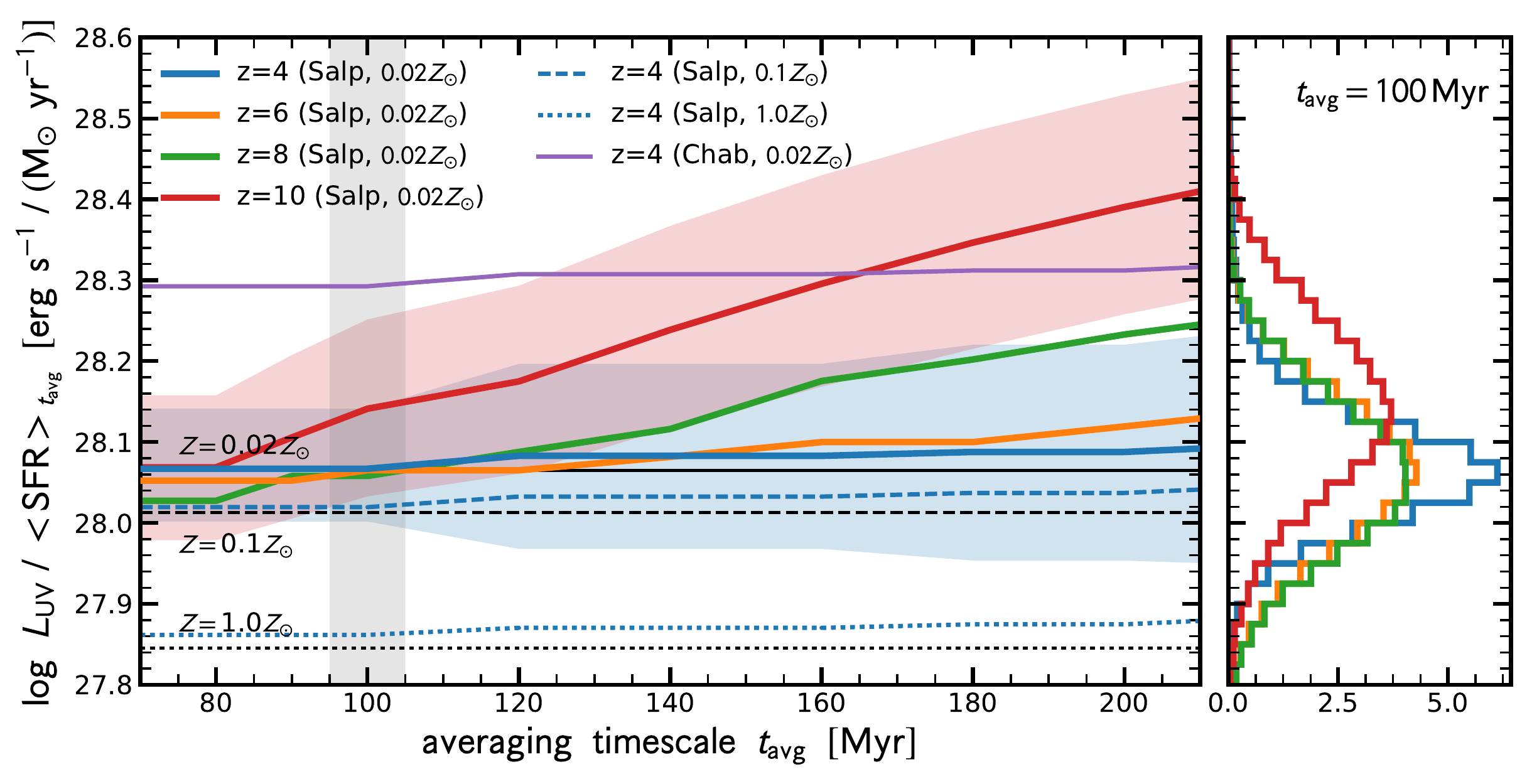}
\caption{UV-to-SFR conversion factor $L_{\rm UV}/\mathrm{<SFR>}_{t_{\rm avg}}$ as a function of the timescale over which the SFR is averaged ($t_{\rm avg}$). The UV luminosity is measured at $1500~\angstrom$. Our fiducial model with $Z=0.02~Z_{\odot}$ and a Salpeter IMF at $z=4-10$  is represented by the solid lines, with the shaded region indicating the 16th and 84th percentiles at $z=4$ and 10. The histogram on the right shows the distribution of $L_{\rm UV}/\mathrm{<SFR>}_{t_{\rm avg}=100\mathrm{Myr}}$ for an averaging timescale of 100 Myr (indicated with the vertical gray band). At $z=4-8$, we find good agreement between the conversion factor derived from our model and one obtained assuming a constant SFR (horizontal black lines; dotted, dashed, and solid lines indicate $Z=1.0,0.1,$ and $0.02~Z_{\odot}$, respectively). Deviation of the order of $\ga0.1$ dex at large $t_{\rm avg}$ and high redshift can be explained by the increasing SFHs. Changing metallicity and the IMF lead to an overall shift in the conversion factor. }
\label{fig:lum_SFR_conversion}
\end{figure*}

As highlighted above, the UV luminosity is commonly used as a tracer of the SFR. The UV luminosity output by a stellar population depends on metallicity, IMF, and SFH (for a given stellar library and isochrones). The conversion between UV luminosity and SFR is typically made assuming that the SFR is approximately constant for at least 100 Myr \citep{madau98,kennicutt98}. However, if the SFR varies on shorter timescales, the conversion between luminosity and SFR becomes more complicated \citep[e.g.,][]{wilkins12}. In particular, as discussed in \citet{faucher-giguere18} and in Section~\ref{subsec:SF_model}, bursty star formation occurs where there is an imbalance between stellar feedback and gravity, which is the case for all galaxies at high redshifts ($z>1$) and for local low-mass galaxies. For example, in the dwarf galaxy regime, \citet{johnson13} show that there is a factor of two dispersion in the ratio of UV luminosity to SFR. Furthermore, both \citet{weisz12} and \citet{kauffmann14} find that in low-mass galaxies the amplitude of star formation bursts can be up to a factor of 30. As shown in \citet{shivaei15}, various uncertainties (e.g., dust attenuation, IMF, and stellar population models) make it difficult to constrain the stochasticity of star formation in high-$z$ galaxies. Based on SFRs measured from H$\beta$ and far-UV, \citet{guo16} claim that burstiness increases toward higher redshifts and lower stellar masses.

In Figure~\ref{fig:lum_SFR_conversion} we explore the conversion factor between UV luminosity and SFR in our model at $z=4$ for different metallicities and IMFs. We plot the ratio of the UV luminosity at $1500~\angstrom$ ($L_{\rm UV}$) and the SFR average over a timescale $t_{\rm avg}$ as a function of this timescale $t_{\rm avg}$. For a Salpeter IMF in the mass range $0.1-100~M_{\odot}$, the MILES stellar library, the MIST isochrones, and a constant SFR, the \texttt{FSPS} model of \citet{conroy09a} yields $\log~L_{\rm UV}/\mathrm{SFR}=(27.85, 28.01, 28.06)$ for $\log~Z_{\star}/Z_{\odot}=(0.0, -1.0, -1.7)$; these values are shown as vertical dashed lines in Figure~\ref{fig:lum_SFR_conversion}. Our fiducial model, which assumes a Salpeter IMF and $\log~Z_{\star}/Z_{\odot}=-1.7$, is shown as solid blue line. The 16th to 86th percentile region is indicated as the shaded area. For intermediate timescales ($t_{\rm avg}\approx100~\mathrm{Myr}$), the UV-to-SFR conversion factor of our model is consistent with one obtained by assuming a constant SFR. For longer averaging timescales, we find that the conversion factor is slightly higher. This can be explained by the nature of the SFHs, which are on average increasing in our model as shown in Section~\ref{subsec:SFH}. In particular for $z=10$, where nearly all SFHs are increasing steeply, our model predicts a conversion conversion factor of $\log~L_{\rm UV}/<\mathrm{SFR}>_{100\mathrm{Myr}}=28.15^{+0.09}_{-0.08}$, which is 0.21 dex higher than the one used by \citet{oesch18}, who adopt the conversion factor of \citet{madau14}, which is a compromise of different metallicities and constant and slightly increasing SFHs.  Using our conversion factor would decrease their cosmic SFRD estimate to $10^{-3.5}~M_{\odot}~\mathrm{yr}^{-1}~\mathrm{Mpc}^{-3}$. 

The distribution of conversion factors for averaging timescales of $100$ Myr is shown in the histogram in the right panel of Figure~\ref{fig:lum_SFR_conversion}. At all redshifts, we find a scatter of $\sim0.1$ dex (higher at higher redshifts), which can be attributed to variations of the SFHs on timescales of $\sim50-100$ Myr. Furthermore, for a Chabrier IMF, we find a 0.2 dex higher conversion factor, while for $\log~Z_{\star}/Z_{\odot}=0.0$ and $-1.0$, the conversion factor is lower by 0.05 and 0.2 dex, respectively. The exact conversion factors, as a function of redshift, are listed in Table~\ref{tab:kappa}.

Summarizing the discussion above, for the same metallicity and IMF, we find a consistent UV-to-SFR ratio in our model with that when assuming a constant SFR at $z=4-8$. At $z=10$, we find a 0.1 dex higher UV-to-SFR ratio than when assuming a constant SFR, which can be explained by the increasing SFHs in our model. In addition, we find a scatter of $\sim0.05-0.1$ dex that increases with redshift. This can be attributed to variation in the SFH on timescales of $\la100$ Myr.

\begin{deluxetable*}{ccccc}
\tablecaption{UV-to-SFR conversion factor ($L_{\rm UV}/\mathrm{<SFR>}_{100\mathrm{Myr}}$) in $\mathrm{erg}~\mathrm{s}^{-1}/(M_{\odot}~\mathrm{yr}^{-1})$.\label{tab:kappa}}
\tablecolumns{5}
\tablewidth{0pt}
\tablehead{
\colhead{redshift} &
\colhead{Salp./0.02Z$_{\odot}$} &
\colhead{Chab./0.02Z$_{\odot}$} & 
\colhead{Salp./1.0Z$_{\odot}$} & 
\colhead{Salp./0.1Z$_{\odot}$}
}
\startdata
4.0 & $28.07^{+0.07}_{-0.07}$ & $28.29^{+0.07}_{-0.07}$ & $27.86^{+0.07}_{-0.09}$ & $28.02^{+0.07}_{-0.07}$ \\
5.0 & $28.06^{+0.08}_{-0.08}$ & $28.29^{+0.09}_{-0.09}$ & $27.87^{+0.10}_{-0.11}$ & $28.02^{+0.09}_{-0.09}$ \\
6.0 & $28.07^{+0.09}_{-0.10}$ & $28.29^{+0.10}_{-0.10}$ & $27.88^{+0.11}_{-0.12}$ & $28.02^{+0.10}_{-0.11}$ \\
7.0 & $28.06^{+0.09}_{-0.09}$ & $28.29^{+0.09}_{-0.09}$ & $27.88^{+0.11}_{-0.11}$ & $28.02^{+0.09}_{-0.10}$ \\
8.0 & $28.06^{+0.10}_{-0.10}$ & $28.29^{+0.10}_{-0.10}$ & $27.88^{+0.12}_{-0.12}$ & $28.02^{+0.11}_{-0.10}$ \\
9.0 & $28.10^{+0.11}_{-0.10}$ & $28.33^{+0.11}_{-0.10}$ & $27.94^{+0.13}_{-0.12}$ & $28.07^{+0.11}_{-0.11}$ \\
10.0 & $28.14^{+0.11}_{-0.11}$ & $28.37^{+0.11}_{-0.11}$ & $27.99^{+0.13}_{-0.13}$ & $28.12^{+0.12}_{-0.12}$ \\
11.0 & $28.18^{+0.12}_{-0.11}$ & $28.42^{+0.12}_{-0.12}$ & $28.05^{+0.14}_{-0.14}$ & $28.16^{+0.13}_{-0.12}$ \\
12.0 & $28.2^{+0.12}_{-0.11}$ & $28.44^{+0.12}_{-0.11}$ & $28.06^{+0.13}_{-0.13}$ & $28.19^{+0.12}_{-0.11}$ \\
13.0 & $28.28^{+0.11}_{-0.12}$ & $28.52^{+0.12}_{-0.12}$ & $28.16^{+0.13}_{-0.14}$ & $28.27^{+0.12}_{-0.12}$ \\
14.0 & $28.34^{+0.12}_{-0.12}$ & $28.57^{+0.12}_{-0.12}$ & $28.21^{+0.14}_{-0.13}$ & $28.32^{+0.12}_{-0.12}$ \\
\enddata
\tablecomments{The errors indicate the $1\sigma$ scatter.}
\end{deluxetable*}

\subsection{Mass Return Fraction}
\label{subsec:return}

\begin{figure}
\includegraphics[width=\linewidth]{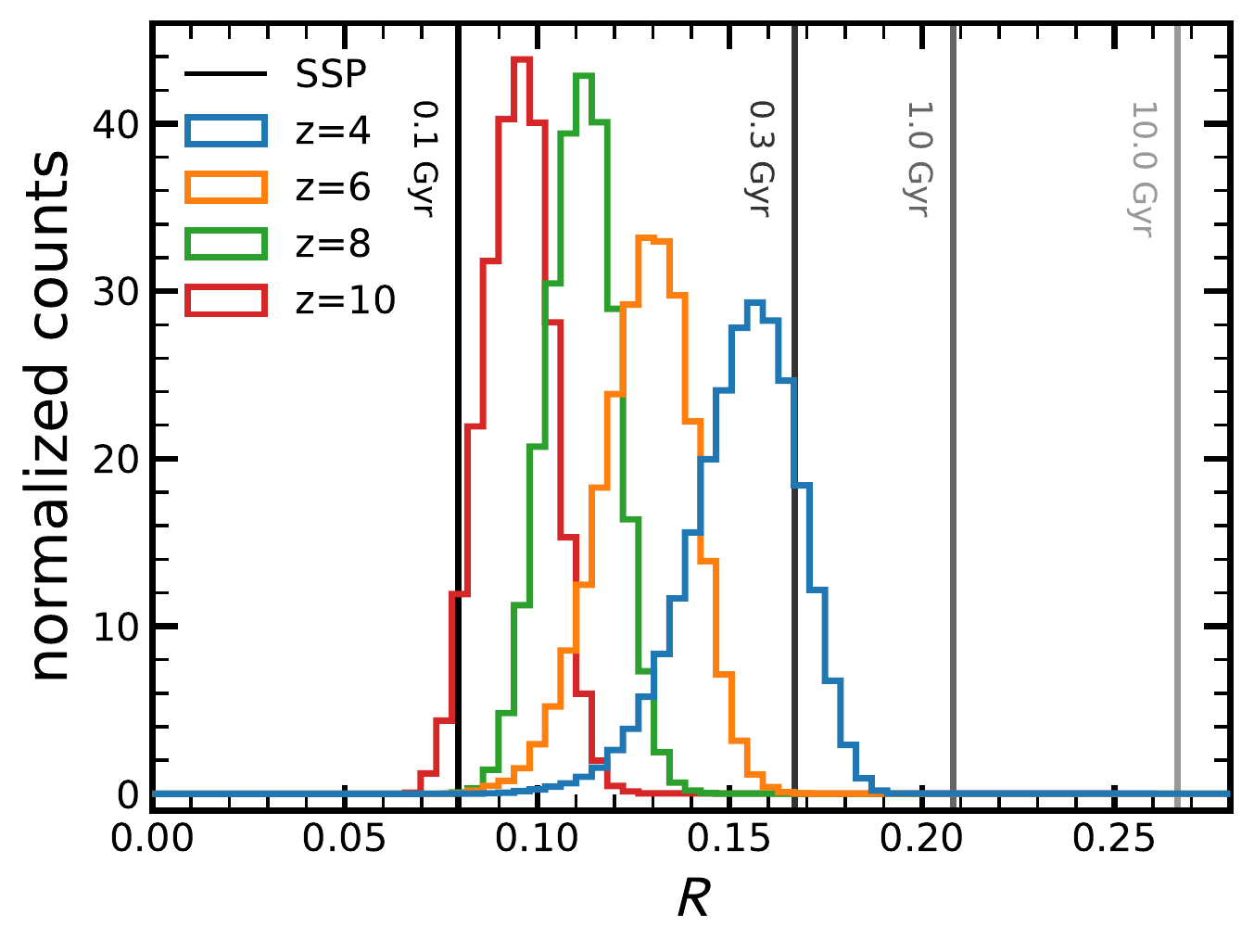}
\caption{Distribution of the mass return fraction ($R$). The mass return fraction is defined as the mass fraction of each generation of stars that is put back into the interstellar medium, i.e. $R=1-M_{\star}/M_{\rm int}$, where $M_{\star}$ is the mass in stars and remnants, and $M_{\rm int}$ is the integral of the SFH. The histograms show the distribution of $R$ for $z=4,6,8$ and 10. The vertical lines from the left to the right indicate $R$ for an SSP (Salpeter IMF) with an age of 0.1 Gyr, 0.3 Gyr, 1.0 Gyr and 10.0 Gyr. None of high-$z$ galaxies have values of $R>0.25$, because the majority of galaxies have increasing SFHs and hence a young stellar age.}
\label{fig:return}
\end{figure}

\begin{deluxetable}{ccc}
\tablecaption{Mass return fraction ($R$) as a function of redshift, metallicity and IMF.\label{tab:R}}
\tablecolumns{3}
\tablewidth{0pt}
\tablehead{
\colhead{redshift} &
\colhead{Salp./0.02Z$_{\odot}$} &
\colhead{Chab./0.02Z$_{\odot}$}
}
\startdata
$4.0$ & $0.15^{+0.01}_{-0.02}$ & $0.27^{+0.02}_{-0.02}$ \\
$5.0$ & $0.14^{+0.01}_{-0.02}$ & $0.25^{+0.02}_{-0.02}$ \\
$6.0$ & $0.13^{+0.01}_{-0.01}$ & $0.23^{+0.02}_{-0.02}$ \\
$7.0$ & $0.12^{+0.01}_{-0.01}$ & $0.21^{+0.02}_{-0.02}$ \\
$8.0$ & $0.11^{+0.01}_{-0.01}$ & $0.20^{+0.02}_{-0.02}$ \\
$9.0$ & $0.10^{+0.01}_{-0.01}$ & $0.18^{+0.02}_{-0.02}$ \\
$10.0$ & $0.10^{+0.01}_{-0.01}$ & $0.17^{+0.02}_{-0.02}$ \\
$11.0$ & $0.08^{+0.01}_{-0.01}$ & $0.15^{+0.02}_{-0.02}$ \\
$12.0$ & $0.08^{+0.01}_{-0.01}$ & $0.15^{+0.01}_{-0.02}$ \\
$13.0$ & $0.07^{+0.01}_{-0.01}$ & $0.13^{+0.02}_{-0.02}$ \\
$14.0$ & $0.07^{+0.01}_{-0.01}$ & $0.13^{+0.02}_{-0.02}$ \\
\enddata
\tablecomments{The errors indicate the $1\sigma$ scatter.}
\end{deluxetable}

In Section~\ref{subsec:MsMuv}, we discussed two different stellar mass definitions. The fiducial one ($M_{\star}$) that we adopt in this work is the mass in stars and remnants, i.e. after subtracting stellar mass loss due to winds and supernovae. The second is the integral of the past SFR ($M_{\rm int}$). These two mass definitions are related via the mass return fraction $R$ (see Equation~\ref{eq:Mdef}), which is defined as the total mass fraction returned to the ISM by a stellar generation. 

Under the assumption of `instantaneous recycling', where the release and mixing of the products of nucleosynthesis by all stars more massive than $m_0$ occur on timescales much shorter than the Hubble time, whereas stars with $m<m_0$ live forever, \citet{madau14} obtain a return fraction of $R=0.27$ for $m_0=1.0~M_{\odot}$ and a Salpeter IMF in the range of $0.1~M_{\odot}<m<100.0~M_{\odot}$. Under the same assumptions, they obtain $R=0.41$ for a Chabrier IMF, a higher value because the Chabrier IMF is more weighted toward short-lived massive stars. 

Since we follow both stellar mass definitions in our model, we are able to provide a more physically motivated estimate of the mass return fraction $R$. In particular, we can abandon the assumption of instantaneous recycling and instead use \texttt{FSPS} to follow the mass loss of stars due to winds as they evolve along the isochrones. We follow the stars until the end of their lifetime, where we then assign remnant masses according to \citet{renzini93}.

Figure~\ref{fig:return} shows the distribution of the $R$ values of the galaxies in our model. We find that for our fiducial model (Salpeter IMF, $Z=0.02~Z_{\odot}$), $R$ lies below the typically quoted value of $R\approx0.3$ for a Salpeter IMF at all redshifts. We find that $R$ monotonically increases with cosmic time from $R=0.10$ at $z=10$ to $R=0.15$ at $z=4$ (Table~\ref{tab:R}). These low values of $R$ are expected since galaxies at these early times are all young, as seen in Figure~\ref{fig:age}. In Figure~\ref{fig:return} we also plot with vertical lines the values of $R$ for different input ages for the SSP. Our model galaxies lie roughly between the ages of 0.1 Gyr and 0.3 Gyr, which is fully consistent with our age estimates (Section~\ref{subsec:SFH}). A similar analysis concerning the mass-to-light ratio conversion of a Salpeter and Chabrier IMF is shown in Appendix~\ref{app_sec:ML}.

This mass returned to the ISM can again be available for star formation or be ejected from the galaxy. Although the mass return fraction is rather small at these early cosmic times, it is a non-negligible fraction toward lower redshifts ($z<4$) that could lead to an increase in star-formation. Therefore, it should be taken into account for models that describe the galaxy population toward $z=0$.

\section{Summary and Conclusions}

In this paper, we have presented an empirical model that connects the dark matter halo population to the stellar mass and star-formation content of galaxies at $z\ga4$. We assume that the star formation of a galaxy is accretion limited, implying that we can write the SFR as the product of the star-formation efficiency and the baryon accretion rate: $\mathrm{SFR}=\varepsilon(M_{\rm h}) \times \dot{M}_{\rm gas}$. We further assume that $\varepsilon(M_{\rm h})$ is simply a function of the halo mass $M_{\rm h}$ and that the baryon accretion rate is proportional to the dark matter accretion rate, which we obtain from the {\sc color} $N$-body simulations. 

We calibrate $\varepsilon(M_{\rm h})$ at $z\simeq4$ with the UV LF rather than the stellar mass function because the observational uncertainty on the UV LF is much smaller than that on the stellar mass function. After calibration, our model correctly predicts the evolution of the UV LF at $z=4-10$. In particular, our model predicts a rather strong decline of the cosmic SFRD with redshift (5 orders of magnitude from $z=4$ to $z=14$). The main cause of this decline is that there are fewer and fewer halos that are massive enough to be able to form stars efficiently. Additionally, a second-order effect is that the dark matter accretion rates onto halos are decreasing.

Concerning the stellar content of the galaxies, the stellar mass functions of observations show a significant amount of scatter (about 0.7 dex at $M_{\star}=10^{8.5}~M_{\odot}$). Our model is in the ballpark of these observations. We predict a steepening of the low-mass-end slope and a decrease of the mass scale of the knee of the Schechter function toward higher redshifts. Our model predicts a linear relation between SFR and $M_{\star}$ (star-forming MS) with a scatter of $\sim0.25$ dex at $z\sim4$ that declines to $<0.1$ dex at $z\ga10$. The decreasing scatter can be understood by the fact that an increasing number of galaxies have assembled all their stellar mass within the timescale probed by the SFR ($\sim100$ Myr for the UV). This is consistent with the age constraints we obtain for our model galaxies: $<60$ Myr for galaxies at $z\ga10$. This implies that the SFR (or UV flux) is a good tracer of $M_{\star}$. Furthermore, our model galaxies show a large diversity of SFHs. On average, they are well described by a lognormal and an EPS-based parameterization. Since the SFHs are rising (particularly at $z>8$), this leads to a modification of the UV-to-SFR conversion factor: SFRD measurements based on a constant or only slightly increasing SFH underestimate the true SFRD. Finally, our model predicts a stellar-to-halo mass relation that evolves only little with redshift and is well described with $M_{\star}\propto M_{\rm h}^{2}$ at $M_{\rm h}<10^{11}~M_{\odot}$. 

In conclusion, the strength of our model is that it is based on an $N$-body merger tree, which has the advantage that the growth history of halos is fully and self-consistently evolved in a cosmological setting, taking into account tidal forces, dynamical friction, and tidal stripping. Furthermore, it also allows us to predict the spatial distribution of galaxies, enabling us to study their clustering, and it includes the diversity of accretion histories of halos.

A drawback of our model is the finite resolution and volume of the $N$-body simulation. With the {\sc color} simulation used in developing this model, we are currently unable to go beyond $z=14$. Furthermore, we miss some of the most massive and rarest galaxies at $z=4$. Another caveat of our model is its calibration, which is currently based on the UV LF. Since our predicted UV fluxes are dust-free, we are required to adopt a dust prescription to be able to compare our model prescriptions with observations. Although our adopted dust prescription closely follows what is adopted in observations, there is a degeneracy between the dust prescription and the star-formation efficiency. We explored this degeneracy by substituting the fiducial \citet{meurer99} IRX-$\beta$ relation with the SMC relation, which leads to an overall higher efficiency for low-mass halos ($M_{\rm h}\approx10^{9.5}$) and hence a slightly higher SFRD at $z\ga10$, but this does not affect our conclusions qualitatively. With \textit{JWST}, the stellar masses of galaxies will provide a stronger constraint, allowing future calibrations to be based on the stellar mass functions.

Finally, future high-$z$ observations will allow us to gain an understanding for how far the assumption of a redshift-independent star-formation efficiency model remains a good one. Which observations would immediately show that a more complicated model is necessary? The assumption of a redshift-independent star-formation efficiency combined with the accretion-limited star-formation law directly leads to a linear $M_{\star}-\mathrm{SFR}$ relation, a rather steep decline of the cosmic SFRD, and a stellar-to-halo mass relation that is rather constant with redshift and $M_{\star}\propto M_{\rm h}^{2}$ at low masses. A nonlinear $M_{\star}-\mathrm{SFR}$ relation would imply that the SFR does not track mass accretion as closely as assumed. Furthermore, an observed evolution in the stellar-to-halo mass relation, together with a constraint on the cosmic SFRD, would indicate that the star-formation efficiency must be changing as a function of time. In particular, a redshift-dependent efficiency that increases at low masses with cosmic time would lead to a higher cosmic SFRD at earlier times and to a stronger evolution of the stellar-to-halo mass relation, with more massive galaxies at lower halo masses at earlier cosmic times.

\acknowledgments

We thank the referee for their thorough report that helped to improve the paper. We are grateful to Peter Behroozi for sharing his paper in preparation, for making the data on the cosmic SFRD and stellar-to-halo mass relation available to us, and for giving insightful comments on our paper. Furthermore, we are thankful to Xiangcheng Ma, Joakim Rosdahl, and Kyoung-Soo Lee for providing us their data. We acknowledge Daniel Ceverino for publishing the FirstLight database DR3 at \url{http://www.ita.uni-heidelberg.de/~ceverino/FirstLight/index.html}. We also thank Mimi Song for discussing her work with us and Benedikt Diemer and Joel Leja for useful discussions. This research made use of NASA's Astrophysics Data System (ADS), the arXiv.org preprint server, the Python plotting library \texttt{matplotlib} \citep{hunter07}, \texttt{astropy}, a community-developed core Python package for Astronomy \citep{astropycollaboration13}, and the python binding of \texttt{FSPS} \citep{foreman_mackey14}. S.T. is supported by the Smithsonian Astrophysical Observatory through the CfA Fellowship. S.B. is supported by Harvard University through the ITC Fellowship. B.J. and D.J.E. are supported by NASA's JWST/NIRCam contract to the University of Arizona, NAS5-02015, and D.J.E. is further supported as a Simons Foundation Investigator.

\appendix

\section{Comparing Monte Carlo and $N$-body merger trees}
\label{app_sec:tree_compare}

\begin{figure}
\includegraphics[width=\textwidth]{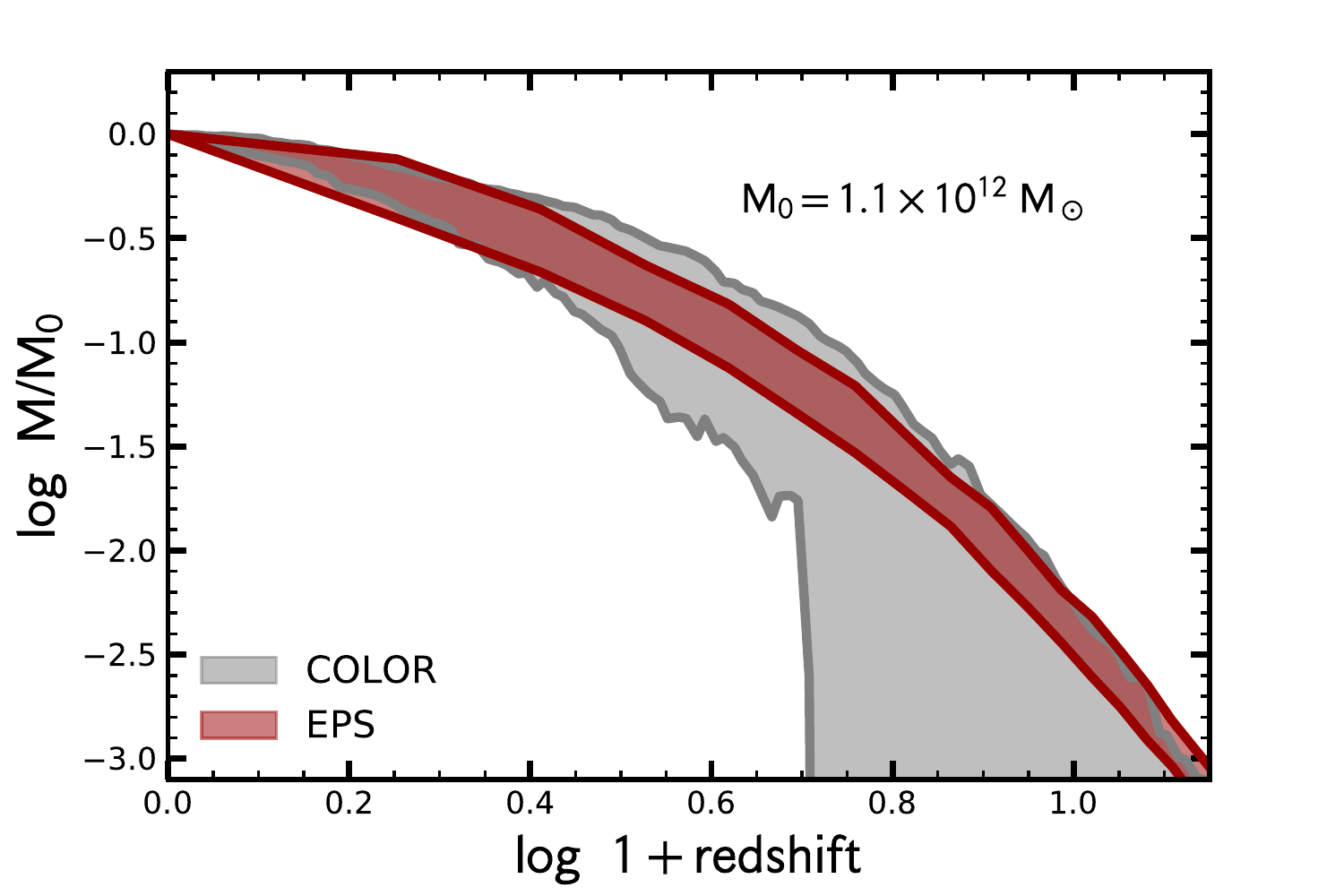}
\caption{Mass accretion histories (MAHs) of $\sim 1.1 \times 10^{12}\/\/M_\odot$ halos identified at $z=0$. In this comparison, we generate 100 realizations of halo merger trees using the EPS formalism, following the scheme outlined by \citet{parkinson08}; this is shown in crimson. Similarly, we randomly select 100 halos with a $z=0$ mass in the interval $\log \/\/M/M_\odot = [12,12.7]$ from the {\sc color} simulation (in gray). The mass resolution of the EPS-derived trees is set equal to the mass resolution in {\sc color}. The shaded regions encapsulate the 16th and 84th percentiles of the accretion histories. The MAHs plotted in this figure simply follow the main progenitor branch of each individual halo. Merger trees extracted from the $N$-body simulation exhibit a wider diversity of MAHs than those constructed via a Monte Carlo approach, particularly at $z\ga3$.}
\label{fig:epsNbody_MAH}
\end{figure}

\begin{figure*}
\includegraphics[width=\textwidth]{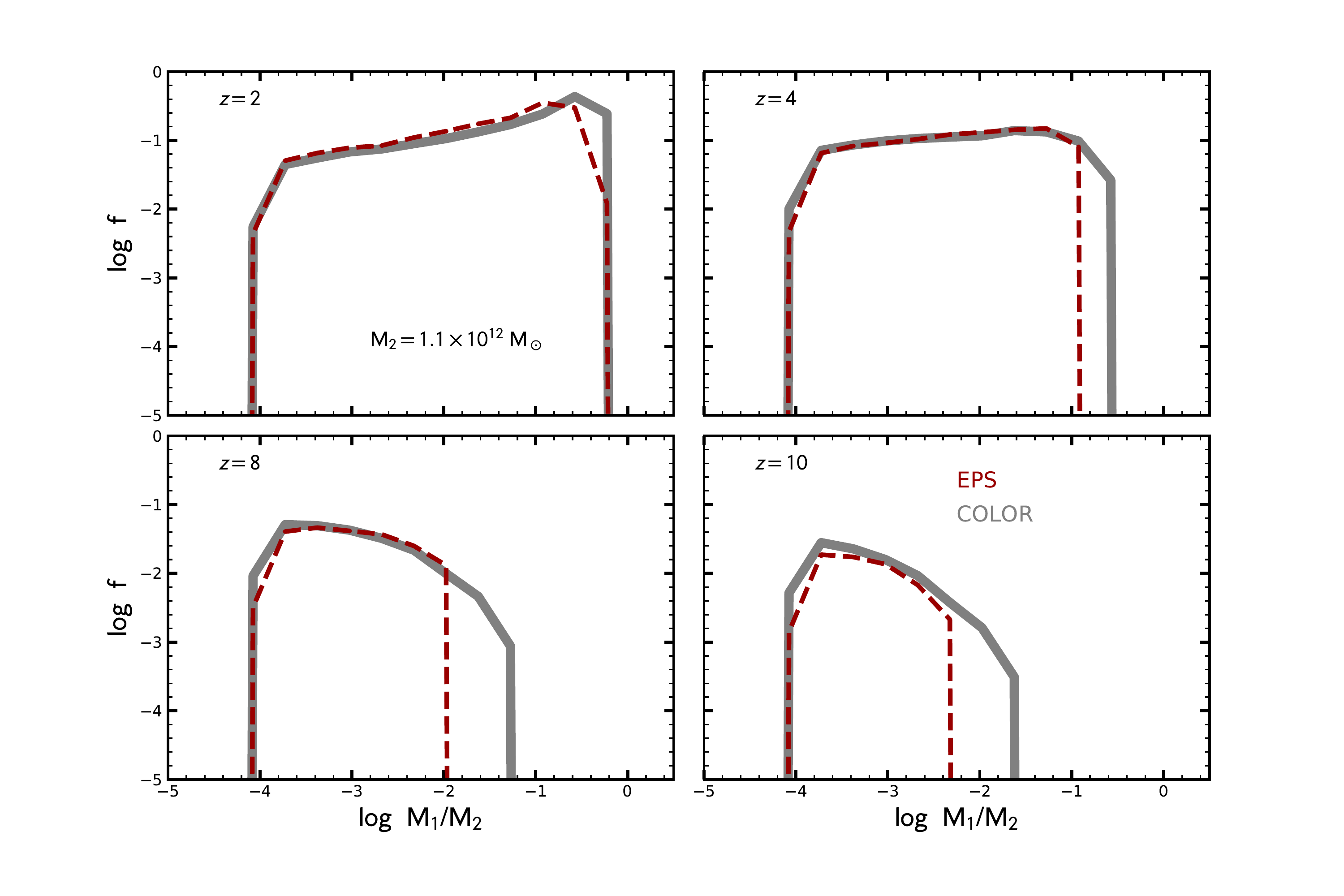}
\caption{Fraction of mass of halos with mass $M_2=1.1\times 10^{12}\/\/M_\odot$ at $z=0$ that is contained in progenitors of mass $M_1$ at $z=2,4,8,$ and 10. The gray histogram is obtained from {\sc color}; the dashed lines show the equivalent result using the \citet{parkinson08} algorithm as described in Figure~\ref{fig:epsNbody_MAH}. While there is good agreement between the two sets of merger trees at $z=2$, the consistency between the $N$-body and Monte Carlo trees worsens at higher redshift. In particular, EPS trees at $z\ga4$ tend to underestimate the mass contained in the most massive progenitors.}
\label{fig:CMF_compare}
\end{figure*}

In this appendix, we present a comparison of halo merger trees obtained from the {\sc color} $N$-body simulation with those that have been generated using the EPS formalism. In particular, the Monte Carlo trees make use of the \citet{parkinson08} algorithm, which has been specifically tuned to match the results from numerical simulations at $z\la4$. In this comparison, we select (at random) 100 merger histories of halos with $z=0$ mass $\sim 1.1\times 10^{12}\/\/M_\odot$ from {\sc color}; additionally, we generate 100 realizations of halos with the same mass using EPS. For consistency between the two methods, the mass resolution of the Monte Carlo trees is set equal to that in {\sc color}.

Figure~\ref{fig:epsNbody_MAH} compares the mass accretion histories (MAHs) of the main progenitor branch for the two sets of merger trees. In general, the two methods agree in the overall shape of the MAH, with the EPS trees showing a similar level of scatter to what is seen in {\sc color} at $z\la2$. At higher redshifts, however, there is clearly a wider diversity in the MAHs in {\sc color} than is sampled by the EPS trees.

A second metric for comparing the two sets of merger trees is shown in Figure~\ref{fig:CMF_compare}, which shows the {\it conditional mass function} (CMF) for $10^{12}\/\/M_\odot$ halos. The CMF quantifies the fraction of the final halo mass, $M_2$, that is contained in progenitors with mass $M_1$ at a particular redshift. As we can see from Figure~\ref{fig:CMF_compare}, EPS trees reproduce the CMF measured in {\sc color} at $z<4$ (where the \citealt{parkinson08} method has been calibrated), but there are systematic differences at higher redshift. In particular, the EPS-generated trees tend to underestimate the cumulative mass contained within the most massive progenitors at $z\ga4$. As this is the regime of interest, merger trees obtained from an $N$-body simulation are more appropriate for the objectives of our present work.

\section{Completeness Correction}
\label{app_sec:comp_corr}

\begin{figure}
\includegraphics[width=\textwidth]{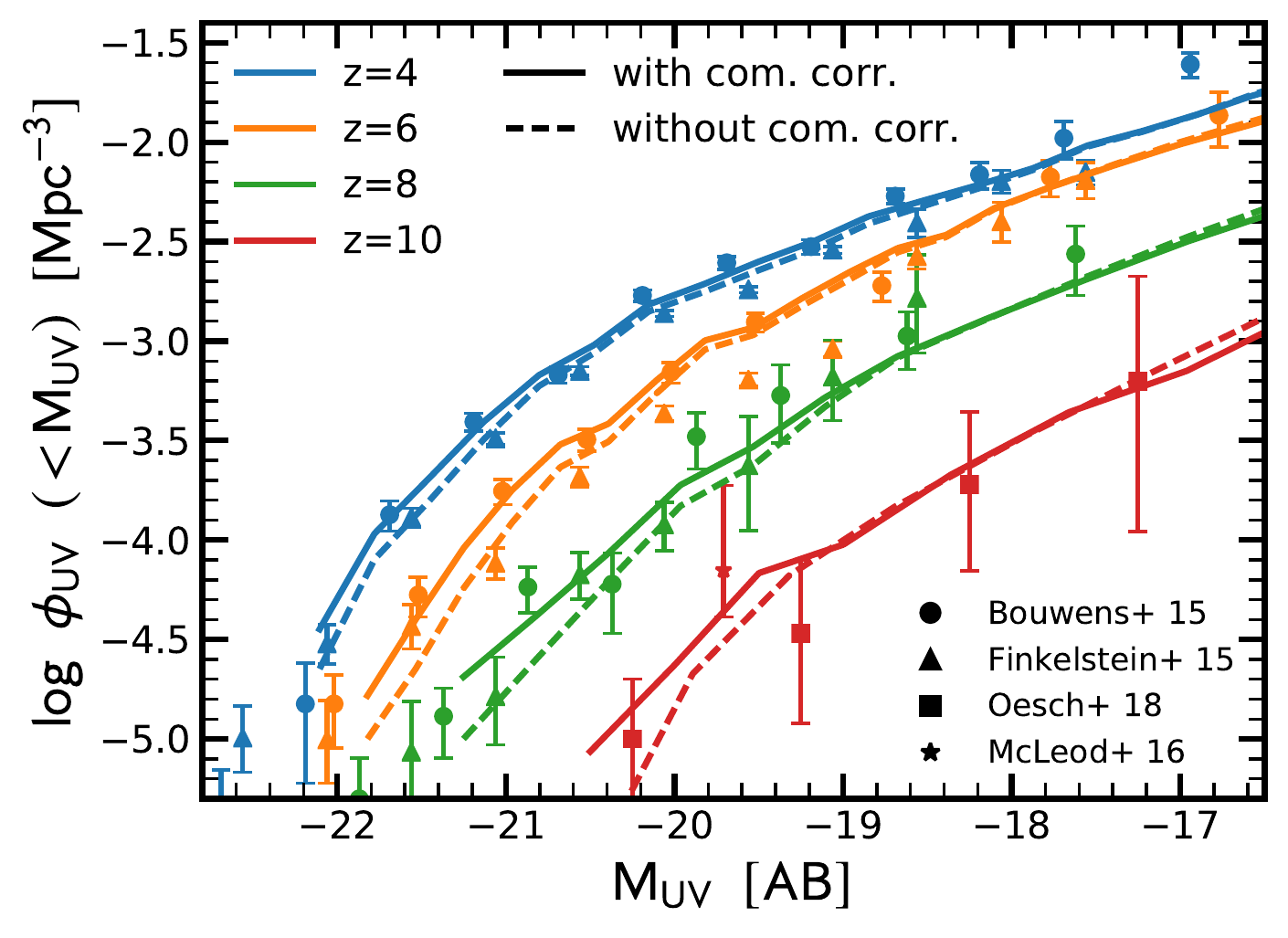}
\caption{Effect of the completeness correction in the halo mass function on the UV LF. The solid and dashed lines, respectively, show the UV LFs at $z=4,6,8,$ and 10 with and without completeness correction. The completeness correction was introduced to adjust the lack of massive halos in our simulation box as shown in Figure~\ref{fig:HMF}. The most massive halos typically exhibit the highest SFRs and are also the brightest in UV; the completeness correction therefore boosts the bright end of the UV LFs, particularly at earlier cosmic times where the size of the completeness correction is larger. }
\label{fig:comp_corr}
\end{figure}

We compare the halo mass function of the {\sc color} simulations, which we adopt in this work, to the one obtained from the analytical formalism of \citet{sheth01}. At all redshifts, we miss some of the most massive halos because the simulation only probes a finite volume. In order to correct for this, we calculate the completeness correction by taking the difference from our estimated halo mass function and the analytical one. At low masses (masses below the knee of the mass function), this correction is negligible. At high masses, the correction is up to 0.4 dex. 

Figure~\ref{fig:comp_corr} shows the effect of the completeness correction on the UV LF. At $z=4-6$, the completeness correction not only affects the brightest halos, since in our model the brightest halos have a variety of halo masses. Toward higher redshifts, the correspondence between brightness and halo mass increases, and hence the completeness correction mainly affects the bright end.

\section{Metallicity Implementation}
\label{app_sec:Z_implementation}

\begin{figure}
\includegraphics[width=\textwidth]{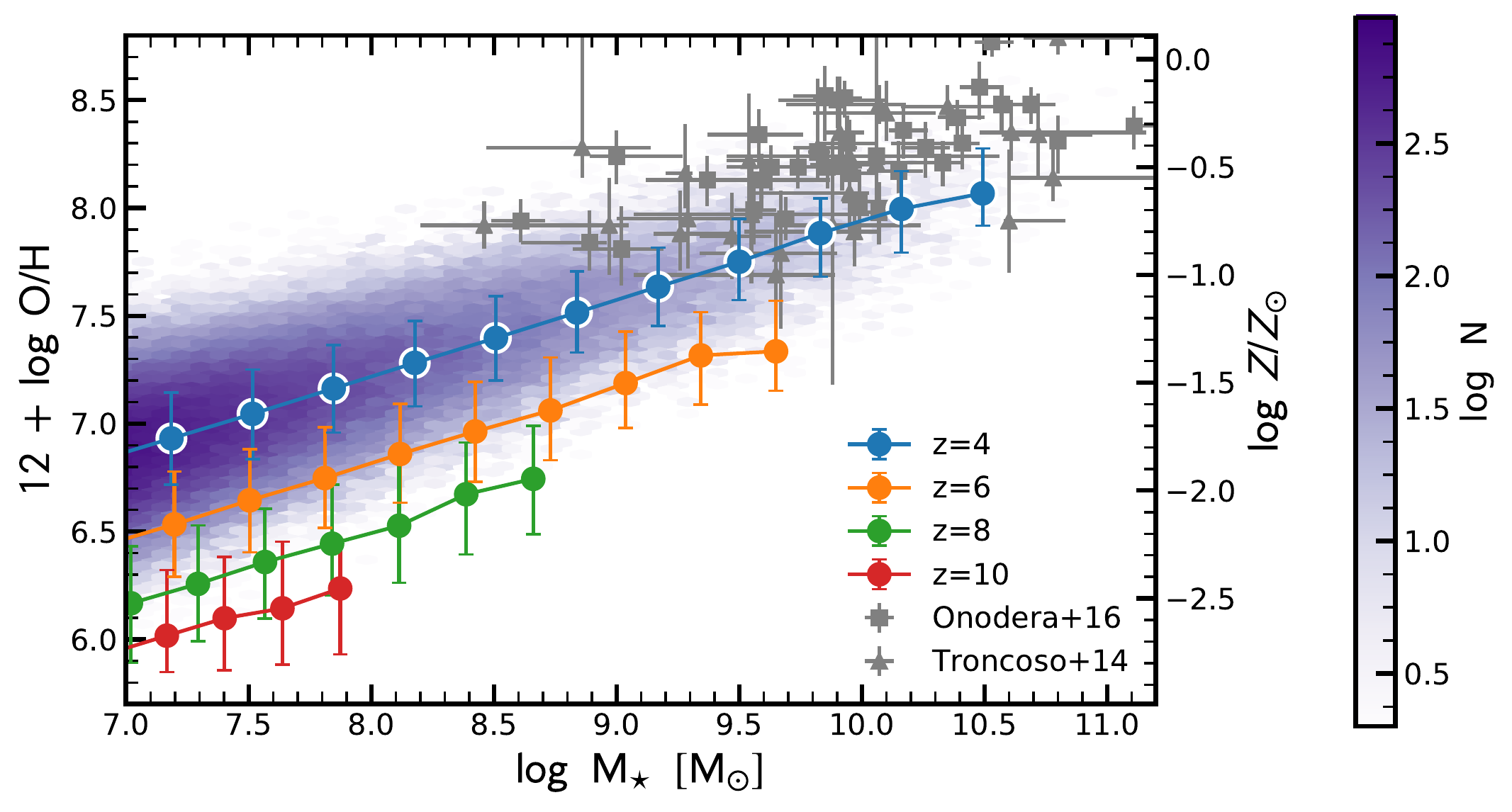}
\caption{Evolution of the mass-metallicity ($M_{\star}-Z$) relation. The metallicity is quantified as gas-phase oxygen abundance ($12+\log\mathrm{O}/\mathrm{H}$) on the left-hand $y$-axis and as the mass fraction of elements heavier than helium ($Z$) on the right-hand $y$-axis, assuming a solar abundance pattern \citep{asplund09}. The filled points show the median relations at $z=4,6,8$ and 10; the error bars show the 16th and 84th percentiles. Gas-phase metallicity measurements at $z=3-4$ made by \citet{onodera16} and \citet{troncoso14}, respectively, are shown with the gray squares and triangles. The $M_{\star}-Z$ relation declines weakly with redshift. At $z=4$, our model quantitatively reproduces the slope of the observed $M_{\star}-Z$ relation, albeit with a slightly lower normalization.}
\label{fig:MZ_relation}
\end{figure}

Our fiducial model, `Z-const', assumes a metallicity of $0.02~Z_{\odot}$ for all galaxies at all cosmic times. Undoubtedly, we expect the metallicity to evolve with cosmic time for individual galaxies, as well as the global galaxy population as a whole. We track the evolution of the metallicity of individual galaxies by solving the equations of mass conservation. We call this version of the model `Z-evo'. As shown above, the key observational predictions do not change in the model `Z-evo' with respect to the fiducial model `Z-const'. In particular, the UV luminosity for a given SFH and IMF does not strongly depend on metallicity at $Z\la0.1~Z_{\odot}$. Hence, the UV LF and the calibration only changes weakly. 

In order to calculate the redshift evolution for the individual galaxies, we follow \citet[][see also \citealt{bouche10, dave12, dekel14_bathtube}]{lilly13_bathtube} and express the change in the mass of metals $m_{\rm Z}$:

\begin{equation}
\frac{\mathrm{d}m_{\rm Z}}{\mathrm{d}t} = (y(1-R)-(Z-Z_0)(1-R+\lambda)) \cdot \mathrm{SFR} + Z_0\frac{\mathrm{d}m_{\rm gas}}{\mathrm{d}t},
\label{eq:Zevolution}
\end{equation}

\noindent where $y$ is the yield, $Z_0$ is the metallicity of the accreting gas, $\lambda$ is the mass-loading factor (assume that the mass-loss rate from the galaxy is equal to $\lambda\cdot\mathrm{SFR}$), and $R$ is the faction of mass that is converted into stars, as measured by the SFR, that is promptly (we assume instantaneously) returned to the ISM. We define the yield to be the mass of metals returned to the ISM per unit mass that is locked up into long-lived stars, i.e. $(1-R)$ times the mass of stars formed. We assume that the gas-phase metallicity and the stellar metallicity are the same, which is a reasonable assumption since the mass doubling timescale is short for these high-$z$ galaxies.

In order to solve Equation~\ref{eq:Zevolution}, we need to make several assumptions. The key input from our model is the SFR. Furthermore, from the stellar population modeling, we can self-consistently find $R\approx0.1$ (see Section~\ref{subsec:return}). All other terms in Equation~\ref{eq:Zevolution} ($y$, $Z_0$, $\lambda$, and $\mathrm{d}m_{\rm gas}/\mathrm{d}t$) need to be estimated. 

The stellar nucleosynthetic yields depend on metallicity, rotation,  and the mass limit for black hole formation $M_{\rm BH}$. By integrating over the IMF the subsolar metallicity stellar yields (where the effect of mass loss is negligible) tabulated by \citet{maeder92} from $10~M_{\odot}$ to $M_{\rm BH}=60~M_{\odot}$, \citet{madau14} obtain $y=0.023$, which lies within the observed range of $0.010-0.036$ and calculated range by \citet{vincenzo16}. Throughout this paper, we assume a fixed yield of $y=0.023$.

For the metallicity of the accreting gas, we assume $Z_0=10^{-3}~Z_{\sun}$. The main motivation for this comes from observations of Lyman limit and Ly$\alpha$ forest systems, which are typically enriched at $10^{-3.5} \la Z/Z_{\odot} \la 10^{-2}$ \citep{meiksin09}.

We assume that the mass loading follows a weak inverse relationship with mass: $\lambda = \lambda_{10}\cdot(M_{\star}/10^{10})^{-1/3}$. The motivations for this dependence come from theoretical models: the momentum-driven wind model of \citet{murray05} has $\lambda \propto M_{\star}^{-1/3}$. On the other hand, an energy-driven wind model has $\lambda \propto M_{\star}^{-2/3}$ \citep{dekel86}. We choose the normalization to be $\lambda_{10}=0.4$, motivated by the fit of Equation~\ref{eq:Zevolution} to the observed Fundamental Metallicity Relation by \citet{lilly13_bathtube}. 

In order to constrain the last term in Equation~\ref{eq:Zevolution}, $\mathrm{d}m_{\rm gas}/\mathrm{d}t$, and to be able to convert from $M_{\rm Z}$ to $Z$, we need to estimate the (total) gas mass $M_{\rm gas}$ of the galaxy. We estimate $M_{\rm gas}$ from the stellar mass $M_{\star}$ by extrapolating the observed scaling relation at $z=0-3$ of \citet{tacconi18} with an upper limit of $\log(M_{\rm gas}/M_{\star})=3.0$. 

With these assumptions, we are able to predict the metallicity evolution for individual galaxies. Figure~\ref{fig:MZ_relation} shows the $M_{\star}-Z$ relation at $z=4-10$ predicted from our model. For a comparison, we plot also the observations of \citet{troncoso14} and \citet{onodera16} of the gas-phase oxygen abundance of star-forming galaxies at $z=3-4$. Although the zero-point of the observations is rather uncertain because of the uncertainty in the metallicity calibration, we find that our model $M_{\star}-Z$ relation relation lies slightly below the observations, as we expect for galaxies at higher redshifts. 

Our model predicts a relation between metallicity and stellar mass of $Z\propto M_{\star}^{0.35}$, which is consistent with the observed slope of $0.31\pm0.05$ at lower redshifts \citep{onodera16}. Furthermore, we find that the normalization declines with redshift as $\propto(1+z)^{-2}$, which is driven by our prescription for deriving the gas mass.

\section{Implications on the mass-to-light ratio from the IMF}
\label{app_sec:ML}

\begin{figure*}
\includegraphics[width=\textwidth]{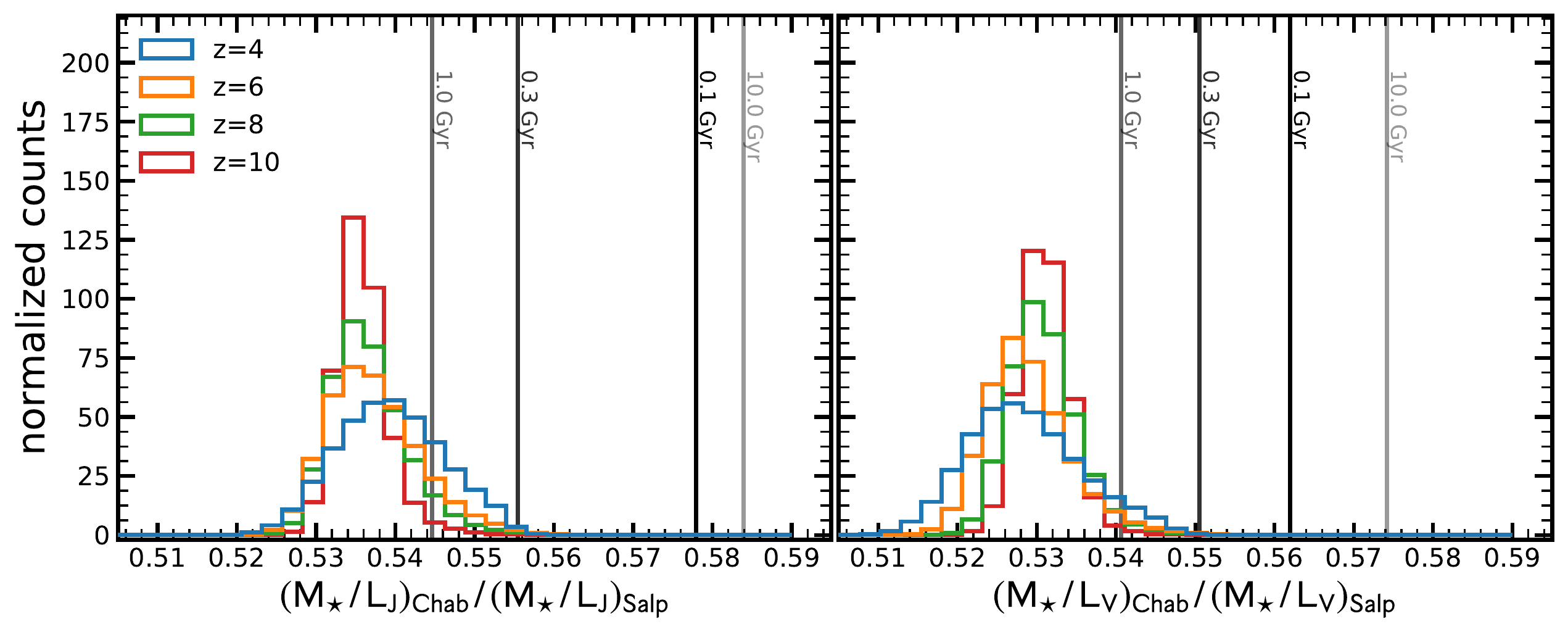}
\caption{Distribution of the mass-to-light ($M_{\star}/L$) ratio conversion factor between a Salpeter and Chabrier IMF. The $M_{\star}/L$ conversion factor for the $J$- and $V$-band is shown in the left and right panels, respectively. Our model galaxies have a conversion factor of 1.85-1.95 (or about 0.28 dex), with little redshift evolution. }
\label{fig:ML}
\end{figure*}


Different analyses assume different IMFs. In order to make different measurements (such as $M_{\star}$) comparable, we need to convert the measurements to the same IMF. Figure~\ref{fig:ML} shows the mass-to-light ($M_{\star}/L$) ratio conversion factor between Salpeter and Chabrier IMF. We use -- as in Section~\ref{subsec:return} -- our SFHs to derive stellar masses and rest-frame luminosities in the $V$- and $J$-band via \texttt{FSPS}. We find a nearly redshift-independent distribution with a median of $0.539^{+0.005}_{-0.005}$.


\end{document}